\documentclass[twocolumn,useAMS,usenatbib]{mn2e}
\usepackage{natbib}
\usepackage{amsmath,amssymb}
\usepackage{epsfig,graphicx}
\newcommand{\mytilde}{\raise.17ex\hbox{$\scriptstyle\mathtt{\sim}$}}
\def\reff@jnl#1{{\rm#1\/}}

\def\aj{\reff@jnl{AJ}}                  
\def\araa{\reff@jnl{ARA\&A}}            
\def\apj{\reff@jnl{ApJ}}                        
\def\apjl{\reff@jnl{ApJ}}               
\def\apjs{\reff@jnl{ApJS}}              
\def\apss{\reff@jnl{Ap\&SS}}            
\def\aap{\reff@jnl{A\&A}}               
\def\aapr{\reff@jnl{A\&A~Rev.}}         
\def\aaps{\reff@jnl{A\&AS}}             
\def\baas{\reff@jnl{BAAS}}              
\def\jrasc{\reff@jnl{JRASC}}            
\def\memras{\reff@jnl{MmRAS}}           
\def\mnras{\reff@jnl{MNRAS}}            
\def\physrep{\reff@jnl{Phys.~Rep.}}
\def\pra{\reff@jnl{Phys.~Rev.~A}}         
\def\prb{\reff@jnl{Phys.~Rev.~B}}         
\def\prc{\reff@jnl{Phys.~Rev.~C}}         
\def\prd{\reff@jnl{Phys.~Rev.~D}}         
\def\prl{\reff@jnl{Phys.~Rev.~Lett}}      
\def\pasp{\reff@jnl{PASP}}              
\def\pasj{\reff@jnl{PASJ}}              
\def\skytel{\reff@jnl{S\&T}}            
\def\solphys{\reff@jnl{Solar~Phys.}}    
\def\sovast{\reff@jnl{Soviet~Ast.}}     
\def\ssr{\reff@jnl{Space~Sci.~Rev.}}     
\def\nat{\reff@jnl{Nature}}             

\newcommand{\pimax}{\ensuremath{\Pi_\mathrm{max}}}

\newcommand{\invhmpc}{\ensuremath{h/\mathrm{Mpc}}}
\newcommand{\hmpc}{\ensuremath{h^{-1}\mathrm{Mpc}}}

\newcommand{\hMsun}{h^{-1}M_{\odot}}
\newcommand{\hmsun}{h^{-1}M_{\odot}}

\newcommand{\mvir}{\ensuremath{M_\mathrm{vir}}}

\newcommand{\ds}{\ensuremath{\Delta\Sigma}}
\newcommand{\dsr}{\ensuremath{\Delta\Sigma(R)}}
\newcommand{\scinv}{\ensuremath{\Sigma_c^{-1}}}

\newcommand{\rmg}{\mathrm{g}}
\newcommand{\rmm}{\mathrm{m}}

\newcommand{\rmgm}{\mathrm{gm}}
\newcommand{\rmgg}{\mathrm{gg}}

\newcommand{\beq}{\begin{equation}}
\newcommand{\eeq}{\end{equation}}
\newcommand{\beqa}{\begin{eqnarray}}
\newcommand{\eeqa}{\end{eqnarray}}
\newcommand{\refresp}[1]{#1}
\newcommand{\refrespt}[1]{#1}
\newcommand{\refrespm}[1]{#1}

\newcommand{\wgg}{\ensuremath{w_{\rmg\rmg}}}
\newcommand{\upsgg}{\ensuremath{\Upsilon_{\rmg\rmg}}}
\newcommand{\upsgm}{\ensuremath{\Upsilon_{\rmg\rmm}}}

\newcommand{\dsrand}{\ensuremath{\Delta\Sigma_\mathrm{rand}}}
\newcommand{\dsx}{\ensuremath{\Delta\Sigma_\times}}
\newcommand{\dsxrand}{\ensuremath{\Delta\Sigma_{\times,\mathrm{rand}}}}
\newcommand{\rmd}{\mathrm{d}}

\newcommand{\photoz}{photo-$z$}
\newcommand{\zphot}{\ensuremath{z_\mathrm{phot}}}

\title[Galaxy-galaxy lensing cosmology]{Cosmological parameter constraints from galaxy-galaxy lensing and galaxy clustering with the SDSS DR7}

\author[Mandelbaum et al.]
{Rachel Mandelbaum$^{1,2}$\thanks{\tt rmandelb@andrew.cmu.edu}, 
An\v{z}e Slosar$^3$, 
Tobias Baldauf$^4$, 
Uro\v{s} Seljak$^{4,5,6,7}$, 
\newauthor
Christopher~M. Hirata$^8$,
Reiko Nakajima$^9$, 
Reinabelle Reyes$^{10}$,
\newauthor
Robert E. Smith$^{4,9}$
\\$^1$Peyton Hall Observatory, Princeton University, Peyton Hall, Princeton, NJ 08544, USA
\\$^2$Department of Physics, Carnegie Mellon University, Pittsburgh, PA 15213, USA
\\$^3$Brookhaven National Laboratory, Upton NY 11375, USA
\\$^4$Institute for Theoretical Physics, University of Zurich, Zurich, Switzerland
\\$^5$Space Sciences Lab, Department of Physics and Department of Astronomy, University of California, Berkeley, CA 94720 USA
\\$^6$Lawrence Berkeley National Lab, University of California, Berkeley, CA  94720, USA
\\$^7$Institute of the Early Universe, Ewha Womans University, Seoul, Korea
\\$^8$Department of Astronomy, Caltech M/C 350-17, Pasadena, CA 91125, USA
\\$^9$Argelander-Institut f\"{u}r Astronomie, Universit\"{a}t Bonn, 53121 Bonn, Germany
\\$^{10}$Kavli Institute for Cosmological Physics and Enrico Fermi Institute, University of Chicago, Chicago, IL 60637, USA
}

\date{\today}

\begin{document}
\maketitle

\begin{abstract}
  Recent studies have shown that the cross-correlation coefficient
  between galaxies and dark matter is very
  close to unity on scales outside a few virial radii of galaxy halos,
  independent of the details of how galaxies populate dark matter
  halos. This finding makes it possible to determine the dark matter clustering from
  measurements of galaxy-galaxy weak lensing and galaxy
  clustering.  We present new cosmological parameter constraints based
  on large-scale measurements of spectroscopic galaxy samples 
 from the Sloan Digital Sky Survey (SDSS) Data Release 7
  (DR7).  We generalise the approach of
  \cite{2010PhRvD..81f3531B} to remove small scale information
  (below 2 and 4\hmpc\ for lensing and clustering measurements, respectively), where the cross-correlation
  coefficient differs from unity. 
We derive constraints for three galaxy
  samples covering 7131 deg$^2$, containing 69150, 62150, and 35088 galaxies with mean redshifts of $0.11$, $0.28$, and $0.40$.
We clearly detect scale-dependent galaxy bias for the more luminous
galaxy samples, at a level consistent with theoretical expectations. 
When
  we vary both $\sigma_8$ and $\Omega_m$ (and marginalise over
  non-linear galaxy bias) \refresp{in a flat $\Lambda$CDM model}, the best-constrained quantity is
  $\sigma_8 (\Omega_m/0.25)^{0.57}=0.80\pm 0.05$ ($1\sigma$, stat. $+$
  sys.), where statistical and systematic errors (photometric redshift and shear calibration)
  have comparable contributions\refresp{, and we have fixed  $n_s=0.96$
    and $h=0.7$.}   
These strong constraints on \refresp{the matter} clustering suggest that this
  method is competitive with cosmic shear in 
  current data, while having very complementary and in some ways less serious systematics. 
We therefore 
expect that this method will play a prominent role in 
  future weak lensing surveys.
When we combine these data with WMAP7 CMB
  data, constraints on $\sigma_8$, $\Omega_m$, $H_0$, $w_\mathrm{de}$ and $\sum m_\nu$ become 30--80 per cent tighter than with CMB data alone, since our data break several parameter degeneracies.
\end{abstract}

\begin{keywords}
gravitational lensing: weak -- cosmology: observations -- cosmological
parameters -- large-scale structure of Universe
\end{keywords}

\section{Introduction}
\label{S:intro}

The currently accepted cosmological model that is broadly consistent
with multiple observations, known as $\Lambda$CDM, is dominated by
dark ingredients: dark matter, which we observe through its
gravitational effects, and dark energy, the presence of which was
inferred due to the accelerated expansion of the universe as detected
using supernovae \citep{1998AJ....116.1009R,1999ApJ...517..565P}.  Further attempts to constrain this model,
such as those described by the Dark Energy Task Force \citep{2006astro.ph..9591A}, rely on
observational methods that can broadly be classified in two ways: {\em
  geometric} measurements such as supernovae (standard candles) and
Baryon Acoustic Oscillations (BAO, standard rulers); and measurements
of large-scale {\em structure growth}.  The latter measurements of
structure growth - particularly as a function of time - can 
constrain the initial amplitude of matter fluctuations, the matter
density, and even the nature of dark energy; the scale-dependence of
structure growth can be used to constrain the neutrino mass.

Theoretical predictions for structure growth, such as from
perturbation theory or $N$-body simulations, are cleanest when
expressed in terms of fluctuations in the density of dark matter.
Fortunately, weak gravitational lensing provides us with a way of
observing the total matter density (including dark matter), via the
deflections of light due to intervening matter along the
line-of-sight, which both magnifies and distorts galaxy shapes
\citep[for a review,
see][]{2001PhR...340..291B,2003ARA&A..41..645R,2008ARNPS..58...99H,2010RPPh...73h6901M}.
The lensing measurement that is commonly used to constrain the
amplitude and growth of matter fluctuations is `cosmic shear',
the auto-correlation of galaxy shape distortions due to intervening
matter along the line-of-sight.  Since the initial detections of
cosmic shear a decade ago \citep{2000MNRAS.318..625B,
  2000A&A...358...30V,2001ApJ...552L..85R, 2002ApJ...572...55H},
increasingly enlarging datasets and sophisticated measurement
techniques have led to steadily decreasing errors, both statistical
and systematic
\citep[e.g.,][]{2010A&A...516A..63S,2013arXiv1303.1808H}.

However, cosmic shear is, by its very nature, a difficult measurement:
in the auto-correlation of galaxy shape distortions, coherent systematic errors (such as 
those induced by seeing or distortions in the telescope) become
an additional additive term.  Moreover, 
intrinsic alignments with the local density field that 
anti-correlate with the real gravitational shear
\citep{2004PhRvD..70f3526H} can contaminate cosmic shear measurements in ways that are difficult to remove.  

\cite{2010PhRvD..81f3531B} provided an alternate approach to
constraining the growth of structure using gravitational lensing which
is less subject to the aforementioned difficulties.  This approach
involves the combination of two measurements: the auto-correlation of
galaxy positions (galaxy clustering), and the cross-correlation
between foreground galaxy positions and background galaxy shears
(galaxy-galaxy lensing, which measures the galaxy-mass
cross-correlation).  By combining these two measurements, we can
recover the matter correlation function, the quantity that is most
easily predicted by the theory.  To reduce uncertainties associated
with exactly how galaxies populate dark matter halos, 
\cite{2010PhRvD..81f3531B} construct a two-point observable that
explicitly eliminates all information below scales equal to a few
times the typical dark matter halo virial radius.  The use of these
two observations allows for a direct measurement of the galaxy bias
(the factor relating the matter and the galaxy density fluctuations,
which can be both mass- and scale-dependent), thus eliminating one of
the main systematic uncertainties in using galaxy clustering alone to
constrain the matter power spectrum, by converting it to a statistical
error over which we marginalise when constraining cosmology.  This
measurement can constrain the amplitude of matter
fluctuations at quite low redshift, which is very useful when combined
with higher-redshift measurements, providing a
measure of structure growth in the time when dark energy is most
dominant.  Also, since it relies on shear cross-correlations rather
than auto-correlations, coherent additive errors in galaxy shapes can
be removed from the analysis entirely.

This paper is a proof of concept of the method described in
\cite{2010PhRvD..81f3531B} to constrain the amplitude of matter
fluctuations at $z<0.4$, using data from the Sloan Digital Sky Survey
(SDSS).  For this measurement, we use lens samples that have
spectroscopy\refresp{: one sample of typical galaxies from the SDSS
  `Main' galaxy sample, and two samples consisting of} Luminous Red
Galaxies (LRGs)\refresp{, which are commonly used for large-scale
  structure measurements} due to their
homogeneous photometric properties, simple selection criteria, and the
large cosmological volume that they sample.    By dividing our sample
into three lens samples, we can test for consistency between the
results at different redshifts (modulo the expected amount of
evolution due to the different mean redshifts).  We will demonstrate
that even this very shallow survey can constrain the amplitude of
matter fluctuations at the $\sim 6$ per cent level, which is
especially cosmologically interesting when combined with Cosmic Microwave
Background (CMB) data.  

We begin in Sec.~\ref{S:theory} with a more detailed outline of the
theoretical background behind the observation we wish to carry out, and simulations that we use for tests of this method.  The data that we use is
described in Sec.~\ref{S:data}, and our observational methodology in
Sec.~\ref{S:observations}.  The
observational results for the galaxy-galaxy lensing are in
Sec.~\ref{S:results-lensing} and for the galaxy clustering, in
Sec.~\ref{S:results-clustering}.  We show the resulting constraints on
cosmological parameters and on galaxy bias in
Sec.~\ref{S:paramconstraints}, and conclude in Sec.~\ref{S:discussion}
with perspective on how
this method may be used in upcoming surveys that will carry out deep,
wide-field lensing observations.

Here we note the cosmological model and units used in this paper.  All
estimates of observed quantities assume a flat $\Lambda$CDM universe
with $\Omega_m=0.25$, $\Omega_{\Lambda}=0.75$; we discuss the
implications of this choice in Sec.~\ref{SSS:coscorr}.  Distances quoted for transverse
lens-source separation are comoving \hmpc,
where $H_0=100\,h$ km$\mathrm{s}^{-1}$ Mpc$^{-1}$.  Likewise,
\ds{} is computed using the expression for \scinv{} in comoving
coordinates, Eq.~\eqref{E:sigmacrit}.  In the units used, $H_0$ scales
out of everything, so our results are independent of this quantity.
Finally, 2-dimensional separations are indicated with  capital $R$,
3-dimensional radii with lower-case $r$ (occasionally $r$ may denote
$r$-band magnitude as well; this should be clear from context).

\section{Theory}\label{S:theory}

The most basic theory predictions for the growth of structure are
phrased in terms of the statistics of the matter distribution - for
example, the 2-point matter auto-correlation function $\xi_{\rmm\rmm}({\bmath r})$ or
the power spectrum $P_{\rmm\rmm}({\bmath k})$.  Here the matter auto-correlation
function is defined in terms of the matter density contrast
$\delta_\rmm = \rho_m/\bar{\rho}_m-1$ as
\beq
\xi_{\rmm\rmm}({\bmath r}) = \langle \delta_\rmm ({\bmath x})
\delta_{\rmm}^{*}({\bmath x} + {\bmath r})\rangle. 
\label{xi}
\eeq
Perturbation theory is
sufficient to predict such statistics of the matter distribution when the perturbations are
linear (density contrast $\delta_m \ll 1$);
$N$-body simulations are used to predict the non-linear power spectrum
\citep[e.g.,][]{2010ApJ...715..104H} in the absence of modifications
due to gas physics, which may be significant on the scales used for
typical weak lensing analyses
\citep{2004ApJ...616L..75Z,2006ApJ...640L.119J,2008ApJ...672...19R,2008PhRvD..77d3507Z,2011MNRAS.417.2020S}.

Galaxy redshift surveys allow us to constrain analogous
auto-correlation functions for the galaxy density field,
$\xi_{\rmg\rmg}({\bmath r})$ or $P_{\rmg\rmg}({\bmath k})$.  Unfortunately, the connection between the theory predictions for the
matter statistics 
to the two-point statistics of the galaxy density field is non-trivial.
We can define the relation as 
\beq
\xi_{\rmg\rmg}(r)=b^2(r)\xi_{\rmm\rmm}(r).
\eeq
On large scales, it is possible to use the linear bias approximation,
$b(r)= { \rm constant}$, 
where the galaxy bias depends on the mass of the dark matter
halos hosting the galaxies.  However, the bias is
also scale-dependent  on smaller scales \citep{2005MNRAS.362..505C,2005ApJ...631...41T,2007PhRvD..75f3512S,2008MNRAS.385..830S}, $\lesssim
20$\hmpc\ for galaxies in very massive halos.  The
existence of galaxy bias causes significant difficulty in inferring
the statistics of the underlying matter density field from galaxy
redshift surveys,  
additional information is needed. 

Galaxy-galaxy weak lensing provides a simple way to probe the
connection between galaxies and matter via their
cross-correlation function
\beq
\xi_{\rmg\rmm}({\bmath r}) = \langle \delta_\rmg ({\bmath x})
\delta_{\rmm}^{*}({\bmath x} + {\bmath r})\rangle.
\eeq
This cross-correlation can be related to the
projected\footnote{In Eq.~\eqref{E:sigmar} we ignore the radial
  lensing window, which is so broad as to be insignificant on all but
  the largest scales, as was demonstrated explicitly in the context of
  this method by \cite{2010PhRvD..81f3531B}.} surface density around lensing galaxies
\beq\label{E:sigmar}
\Sigma(R) = \overline{\rho} \int \left[1+\xi_{\rmg\rmm}\left(\sqrt{R^2 +
    \Pi^2}\right)\right] \rmd\Pi,
\eeq
where $\Pi$ is the line-of-sight separation measured from the lens,
and therefore $r^2=R^2+\Pi^2$.  This surface density is then related to the observable
quantity for lensing, the tangential shear distortion $\gamma_t$ of
the shapes of background galaxies, via 
\beq\label{E:ds}
\ds(R) = \gamma_t(R) \Sigma_c= \overline{\Sigma}(<R) - \Sigma(R), 
\eeq
where
\beq\label{E:means}
\overline{\Sigma}(<R) = \frac{2}{R^2}\int_{0}^{R'} R'\rmd R' \Sigma(R').
\eeq
When averaging over (`stacking') large numbers of lens galaxies to determine the average
signal around them, the resulting matter distribution is axisymmetric
about the line-of-sight. 
The observable quantity $\ds$ can
be expressed as the product of two factors, a tangential shear
$\gamma_t$ and a geometric factor
\beq\label{E:sigmacrit}
\Sigma_c = \frac{c^2}{4\pi G} \frac{D_A(z_s)}{D_A(z_l) D_A(z_l,z_s)(1+z_L)^2}
\eeq
where $D_A(z_l)$ and $D_A(z_s)$ are angular diameter distances to the lens and
source, $D_A(z_l,z_s)$ is the angular diameter distance between the lens
and source, and the factor of $(1+z_L)^{-2}$ arises due to our use of
comoving coordinates.  

Generally, for some 2-point statistic $\zeta$ (for example, the
real-space correlation function $\xi$ or Fourier-space power spectrum $P(k)$), we
can relate the three possible 2-point correlations that can be
constructed out of the matter and galaxy fields, $\zeta_{\rmm\rmm}$,
$\zeta_{\rmg\rmg}$, and $\zeta_{\rmg\rmm}$, as follows:
\beqa
\zeta_{\rmg\rmm} &=& b^{(\zeta)} r_\mathrm{cc}^{(\zeta)}
\zeta_{\rmm\rmm}, \label{E:gmstat} \\
\zeta_{\rmg\rmg} &=& b^{(\zeta)2} \zeta_{\rmm\rmm} =\frac{b^{(\zeta)}}{r_\mathrm{cc}^{(\zeta)}}
\zeta_{\rmg\rmm}. \label{E:obsbias}
\eeqa
All quantities in these equations are a function of scale, where the
scale depends on the exact statistic (e.g., 3D $r$, 2D $R$, Fourier
wavenumber $k$, multipole $\ell$).  
Here $b^{(\zeta)}$ is the galaxy bias relating the galaxy and dark
matter fluctuations, and $r_\mathrm{cc}^{(\zeta)}$,
defined as $r_\mathrm{cc}^{(\zeta)}=\zeta_{\rmg\rmm}/\sqrt{\zeta_{\rmm\rmm}\zeta_{\rmg\rmg}}$,
is the 
cross-correlation coefficient between the matter and galaxy
fluctuations\footnote{This statistic is
  often denoted $r$.  We use the subscript `cc' to avoid confusion
  with 3D length scales.}.  Generically, the galaxy bias tends to a
constant value on large scales
(`linear bias'), and the cross-correlation coefficient approaches one, but the rate at which this happens depends on the choice of statistic $\zeta$.  
In particular, if $\zeta$ is defined as a product of either a Fourier mode (i.e. the power spectrum) or of 
a count in cell (of varying size, called the smoothing size), then $| r_\mathrm{cc}^{(\zeta)}| <1$ (note that 
no shot-noise subtraction is applied here). In this case, 
the deviation of $r_\mathrm{cc}^{(\zeta)}$ from unity can be related to stochasticity \refresp{\citep[e.g., ][]{1999ApJ...520...24D}}, 
which is defined as
\beqa
\langle(\delta_\rmg-b^{(\zeta)}\delta_\rmm)^2\rangle&=&\zeta_{\rmg\rmg}-2b^{(\zeta)}\zeta_{\rmg\rmm}+(b^{(\zeta)})^2\zeta_{\rmm\rmm}\notag\\
&=&2(b^{(\zeta)})^2\refrespm{(1-r_\mathrm{cc}^{(\zeta)})\zeta_{\rmm\rmm}}. 
\eeqa
This is zero if $r_\mathrm{cc}^{(\zeta)}=1$. 
However, the
rate at which $r_\mathrm{cc}^{(\zeta)}$ approaches unity as a function of 
either the size of the cell or the wavevector of the Fourier mode is slow,  
because stochasticity (such as the shot noise caused by finite number of galaxies)
contributes to it. This rate of convergence to unity 
is even worse if compensated windows with positive and negative
weights, such as for the aperture 
mass statistic, are used \citep{1998MNRAS.296..873S}; this effect has been
observed in practice by, e.g., \cite{2009MNRAS.398..807S} and \cite{2012ApJ...750...37J}.

On the other hand, $\zeta$ can be defined as a correlation function, as in Eq.~\eqref{xi}, 
not as a product of a field with itself (or another field), 
in which case the shot noise does not explicitly contribute to it except at zero lag. A
related statistic in Fourier space is the  shot-noise-subtracted power spectrum, where stochasticity is explicitly subtracted. 
In this case, as shown in \cite{2010PhRvD..81f3531B}, $r_\mathrm{cc}^{(\zeta)}$ is much closer to unity (except at zero lag) 
and the scale dependence of $b^{(\zeta)}$ is significantly reduced (which is why shot-noise subtraction is a standard 
procedure in the analysis of the galaxy power spectrum). Moreover, even on 
scales where $b^{(\zeta)}$ is strongly scale dependent, $r_\mathrm{cc}^{(\zeta)}$ is close to unity, with deviations 
from unity of only a few per cent on scales above 3\hmpc, where the scale-dependent bias can be tens of per cent.
In this case, $r_\mathrm{cc}^{(\zeta)}$ has no relation to stochasticity, since its contribution 
does not enter or is explicitly subtracted from it, and we no longer need to have $| r_\mathrm{cc}^{(\zeta)}| <1$.

If we can ensure that we are working in a regime where the
cross-correlation $r_\mathrm{cc}^{(\zeta)} \approx 1$, or, more 
generally, if we have a 
robust model for its scale dependence, then we can 
infer the combination of the mean matter density and the
correlation statistic of matter by combining the galaxy-galaxy lensing
and galaxy clustering statistics. \refrespt{Note that the galaxy-galaxy lensing
observable is not sensitive to just $\zeta_{\rmg\rmm}$, but rather to
$\bar{\rho}_\rmm\zeta_{\rmg\rmm}$ (e.g., see Eq.~\ref{E:sigmar}), so this combination of observables gives}
\beq
\refrespm{\bar{\rho}_\rmm^2}\frac{\zeta_{\rmg\rmm}^2}{\zeta_{\rmg\rmg}} =
\refrespm{\bar{\rho}_\rmm^2} [r_\mathrm{cc}^{(\zeta)}]^2 \zeta_{\rmm\rmm}. \label{E:obsmm}
\eeq
\refrespt{As a result, on fully linear scales, g-g lensing and
  clustering together would constrain the product $\sigma_8 \Omega_m$;
since the majority of analyses (including ours) have substantial
constraining power in the nonlinear regime, this changes the
best-constrained parameter combination to $\sim \sigma_8 \Omega_m^{0.6}$.}

So far, this discussion has been fairly general.
\cite{2010PhRvD..81f3531B} carried out a detailed exploration of the
behaviour of $r_\mathrm{cc}^{(\zeta)}$ for a variety of statistics
$\zeta$, using a simulated sample of Luminous Red Galaxies residing in
dark matter halos with masses $\gtrsim 3\times 10^{13}\hMsun$, at
$z=0.23$.  As shown there, a key point in determining the optimal
statistic $\zeta$ is that we want to avoid information from 
within the halo virial radius, because those are the scales
for which the correlation coefficient is intrinsically quite different from
unity in a way that cannot be predicted from first principles (without 
some detailed model for how galaxies populate dark-matter halos).  The observed lensing signal $\ds$ is therefore quite
non-optimal from the perspective of wanting to do cosmology using
large scales only, because as shown in Eqs.~\eqref{E:ds}
and~\eqref{E:means}, at a given $R$ it depends on the surface density of matter
around galaxies all the way from $R=0$.

The statistic that was proposed by \cite{2010PhRvD..81f3531B} and
\cite{2010MNRAS.405.2078M} to remove 
small-scale information is known as the annular differential surface
density (ADSD) $\Upsilon$, defined as 
\beqa
\Upsilon(R;R_0) &=& \ds(R) - \left(\frac{R_0}{R}\right)^2
\ds(R_0) \label{E:upsilon} \\
 &=& \frac{2}{R^2} \int_{R_0}^{R} \rmd R' R' \Sigma(R') \label{E:upsfromw}\\
 && \qquad -\frac{1}{R^2}\left[R^2\Sigma(R) - R_0^2 \Sigma(R_0)\right].\notag
\eeqa
This statistic depends not only on projected separation $R$, but also
on some scale $R_0$; as demonstrated in Eq.~\eqref{E:upsfromw},
$\Upsilon(R;R_0)$ is completely lacking any information from below $R_0$.  We
thus have to choose a value of $R_0$ that is appropriate for our
particular application.  We will consider several $R_0$ values in this
paper, but generally we would like this to be a few times the typical
dark matter halo virial radius (a point that we examine in more
detail in Sec.~\ref{SS:r0choice}).  As
demonstrated in detail in \cite{2010PhRvD..81f3531B}, the
advantages of such a choice are that (a) the correlation coefficient 
$r_\mathrm{cc}^{(\Upsilon)}\sim 1$ for all scales $R\ge R_0$, and (b)
 the few per cent deviations from $1$ can be calculated quite
accurately via
perturbation theory (which is only applicable in this regime outside
of halo virial radii).  
It was shown that the deviations of $r_\mathrm{cc}$ from unity can be 
described well with one free parameter related to non-linearity of the 
bias, $b_2$. In this paper, we allow the data (specifically 
galaxy auto-correlations) to determine $b_2$, which 
will in turn determine the small deviations of $r_\mathrm{cc}$ from unity.
In addition, because $\Upsilon$ is a partially
compensated statistic, it is not very susceptible to issues that can
plague the projected correlation function ($w_{\rmg\rmg}$)
such as sampling variance from large-scale modes uniformly shifting
\wgg\ up or down.


The approach described here, which entails removing
small-scale information completely, is a conservative approach that 
minimises systematic uncertainties due to all the things we do not
know on small scales (how galaxies populate dark matter halos,
baryonic effects on the matter power spectrum, etc.) at the expense of
increasing the statistical errors.  \refresp{Baryonic effects are generally
  considered to be small above scales of several \hmpc, however there
  are studies that claim that baryonic effects can change the matter
  power spectrum even by $\sim 10$ per cent at $k=1$~$h/$Mpc 
  \citep{2011MNRAS.415.3649V}, because baryons may be expelled from
  halos due to some mechanism such as AGN feedback, redistributing the
  dark matter potentially to several virial radii.  While a detailed
  study of the implications for our work would require a comparison of
  the correlation functions, we note that \refrespt{given the correspondence
  $r\sim 1/k$ (for broadband power)} it is generally the case that this $\sim 10$ per cent contamination at
  $k=1$~$h/$Mpc should correspond to $r=1$~$h^{-1}$Mpc scales, which we
  do not use in our analysis. Our minimum $r=R_0$ that is several
  times larger means that the relevant effect from that paper is the $\sim 1$
  per cent contamination that they find at $k\sim 0.3$~$h/$Mpc; this
  is comparable to our other theoretical uncertainties and well below
  our observational uncertainties, so it does 
  not have to be modeled directly. \refrespt{Alternatively, one can
    see this from the fact that the physical arguments given in that
    work suggest deviations in the power spectrum up to $\sim
    2r_\mathrm{vir}$, whereas the scales we have chosen are typically
    $>3r_\mathrm{vir}$ for the halo masses in our sample.} Future studies with increased statistical
  precision may find it necessary to model this effect on the
  correlation function directly.  It is also worth considering the
  mass-dependence of this effect, which is lower for more massive
  halos \citep{2011MNRAS.412.1965M} and could thus influence the choice of which galaxy
  samples to use for these analyses.}

\refresp{This point about baryonic
  effects is another issue for which our method should be contrasted with cosmic shear.  The problem caused by
  baryonic effects is exacerbated with shear-shear
analyses since they are not localized to a given redshift, so that a given
transverse physical scale can translate into a very large angular
scale if these galaxies are nearby.  In our case we can use the lens
galaxies with redshifts to explicitly remove scales below several
\hmpc, immunizing ourselves from this effect to a large degree.   This 
is yet another reason that the approach we advocate here can be a
powerful alternative to the shear-shear correlation functions which
have been the focus of most weak lensing cosmological analyses to date.}

Alternative approaches involving
halo occupation distribution (HOD) modeling have also been considered 
\citep{2006ApJ...652...26Y,2009MNRAS.394..929C,2011ApJ...738...45L,2012arXiv1207.0503C,2012ApJ...744..159L,2012ApJ...745...16T,2012arXiv1206.6890V}
as ways to combine the \refresp{galaxy-galaxy} lensing and clustering observations to constrain
cosmology.  Those approaches can potentially give smaller
statistical errors, since they use the small-scale lensing signals
which typically have the best $S/N$, but they are subject to 
additional systematic uncertainties both in terms of theory
interpretation and observational uncertainties that are more
pronounced on small scales (e.g., intrinsic alignments; magnification;
data processing challenges near bright lens galaxies).

To be more quantitative, our approach is that the small-scale galaxy auto-correlation 
contains no cosmological information, since there is
nothing in the distribution of galaxies within the halo that has a 
simple relation to cosmological parameters.  While small 
scale galaxy clustering can 
constrain HOD models, this by itself does not help in cosmological
constraints.  Moreover, 
it is potentially dangerous to rely on small scale
clustering to constrain cosmological models, because one can
never be sure that the HOD parametrisation is sufficiently general and
that there is no artificial breaking of degeneracies with cosmological
parameters due to insufficient generality. HOD models explored to date
\refrespt{do} not allow \refrespt{a reasonable degree of} freedom in \refrespt{how 
galaxies are} placed inside the halos.  \refrespt{For
  example, \cite{2012arXiv1207.0004M} assume the distribution of
  galaxies follows that of the dark matter, with just a 10 per cent
  uncertainty in the concentration-mass relation.  Likewise, the
  10-parameter HOD in \cite{2011ApJ...738...45L} includes no freedom
  in the radial distribution of satellite galaxies, which is assumed
  to follow that of the dark matter.  To date, no work has shown that either
  (a) cosmological information can be derived in a way that is
  completely unbiased with respect to these strong assumptions about the
  radial distribution of satellite galaxies, or (b) when one allows
  the radial distribution of satellites within halos to be free, that
  one still gets any significant cosmological information from small-scale clustering.}

Moreover, \refrespt{these HOD models} ignore issues such as assembly bias (explicit
dependence of clustering properties on assembly history, rather than
just mass alone; \citealt{2005MNRAS.363L..66G},
\citealt{2007MNRAS.377L...5G}) that can change the 
relation between small- and large-scale clustering information. 
Once we are trying to place cosmological constraints at the $\sim 5$ per cent
level, where these small details \refrespt{(such as the radial
  distribution of satellites within halos and assembly bias)} become more important, it is
more robust simply to remove the small scale clustering regime. Our approach explicitly does that.

When testing our procedure, we will apply it to a 
simulated mock sample, which we have generated using an HOD model
known to reproduce the 
galaxy two-point correlation function, but our claim is that our procedure should 
work on any sample.  \refresp{The reason for this claim is that
  despite using an HOD-based sample for the tests, the method itself
  does not assume a lack of assembly bias - in other words, the
  large-scale bias is not presumed to relate to the mean halo mass
  from the lensing measurement.  Indeed, we could carry out this
  analysis on a sample with a significant assembly bias, but that
  assembly bias would not violate our much weaker assumption, which is
that the same large-scale bias describes the weak lensing (via
$\xi_{\rmg\rmm}$) and clustering (via $\xi_{\rmm\rmm}$).  On large
enough scales, this assumption must be true.  One might worry that an
assembly bias could change the trends in $r_\mathrm{cc}$ with scale.
We see no reason {\em a priori} for this to be the case, but we
caution that our method has not been tested with samples that were
explicitly selected to include various levels of assembly bias, which
we defer to future work.}

While we believe that the galaxy auto-correlation contains 
no useful information on small scales, the galaxy-dark matter correlation does contain 
information on halo mass, which in combination with the 
galaxy auto-correlation on large scales can provide independent cosmological 
information using the method of \cite{2005PhRvD..71d3511S}. 
Our current method cannot take advantage of 
this additional information from the small scale lensing, so in this sense 
it is sub-optimal.  \refresp{But, again, using that small-scale
  lensing information would make us more obviously susceptible to errors due to
  assembly bias.}

One additional aspect to our approach that is meant to reduce
systematic uncertainties is that we do not simply use all of our lens
galaxies in one large sample to get a small statistical error.  Instead,
we have several lens samples at different redshifts. 
This way, we can check for consistency of the results with each
sample, and check for deviations from our assumptions about
$r_\mathrm{cc}$ or observational systematics that scale with 
redshift (such as our understanding of the source redshift
distribution, which is more important when the lens redshift
approaches the typical source redshifts). 


\subsection{Simulations}\label{SS:simulations}

While we argued in the previous section that our approach is, by
design, 
fairly insensitive to the details of how galaxies occupy dark matter
halos, it is nevertheless
useful to test the whole procedure on a mock data sample that is as close 
as possible to the real data. 
Here, we repeat the description of the $N$-body simulations that were used for
validation of the method in \cite{2010PhRvD..81f3531B} and that we use
for additional tests in this paper. We use the Z\"{u}rich horizon
`\texttt{zHORIZON}' simulations, a suite of forty pure
dissipationless dark matter simulations of the $\Lambda$CDM cosmology
\citep{2009MNRAS.400..851S}. Each simulation models the dark matter
density field in a box of length $L=1500$\hmpc, using
$N_\text{p}=750^3$ dark matter particles with a mass of
$M_\text{p}=5.55 \times 10^{11} \hmsun$. The cosmological parameters
for the simulations in Table \ref{tab:cosmoparam} are inspired by the
results of the WMAP cosmic microwave background experiment
\citep{2003ApJS..148..175S,2007ApJS..170..377S}. The initial conditions
were set up at redshift $z=50$ using the \texttt{2LPT} code
\citep{1998MNRAS.299.1097S}. The evolution of the $N_\text{p}$ equal mass
particles under gravity was then followed using the publicly available
$N$-body code \texttt{GADGET-2} \citep{2005MNRAS.364.1105S}.
Finally, gravitationally-bound structures were identified in each
simulation snapshot using a Friends-of-Friends \citep[FoF,
][]{1985ApJ...292..371D} algorithm with linking length of $0.2$ times
the mean inter-particle spacing. We rejected halos containing fewer
than twenty particles, and identified the potential minimum of the
particle distribution associated with the halo as the halo centre. 
In
total, we identify halos in the mass range $1.1 \times
10^{13}\hmsun\leq \mvir \leq 4 \times 10^{15} \hmsun$.

\begin{table}
        \centering
        \begin{tabular}{llllll}
        \hline
        \hline
        $\Omega_\text{m}$&$\Omega_\Lambda$&$h$&$\sigma_8$&$n_s$&$w_\text{de}$\\
        \hline
        $0.25$ &$0.75$&$0.7$&$0.8$&$1.0$&$-1$\\
        \hline
        \hline
        \end{tabular}
        \caption{Cosmological parameters adopted for the simulations: matter
density relative to the critical density, dark energy density
parameter, dimensionless Hubble parameter, matter power spectrum
normalisation, primordial power spectrum slope, and dark energy equation of state
$p=w_\text{de} \rho$.}
        \label{tab:cosmoparam}
\end{table}


We populate the halos in these simulations with galaxies using the
Halo Occupation Distribution (HOD).  This model requires us to specify
probability distributions for (a) the number of galaxies in our sample
that occupy a halo of mass\footnote{In principle the number of
  satellite galaxies could depend on other parameters such
  as formation time; however, the HOD does not include dependence on
  anything other than mass.} $M$, and (b) the radial distribution of
galaxies within halos.  The HOD can be separated into terms
representing `central' galaxies living at the centre of halos, and
`satellite' galaxies that are distributed more widely within the
halos\footnote{Due to limited resolution, we do not attempt to place
  the satellites in subhalos, but rather distribute them
  probabilistically within the host halo.}.  We assume that a halo can
only contain a satellite if it also has a central galaxy.  This
assumption may not be entirely valid for a colour-selected sample such
as LRGs, if the central galaxy is very bright but slightly too blue to
be included in the sample. This will have effects on scales below the 
virial radius: the galaxy-dark matter correlations will be reduced
on very small scales. It will also reduce galaxy clustering in cases when 
this satellite has another satellite in the same halo. We expect these 
effects to become small on scales larger than the virial radius.  

Details of the five-parameter HOD that we used, and tests of how well
it describes the sample abundance, lensing, and clustering statistics,
are given in \cite{2010PhRvD..81f3531B} and \cite{2010Natur.464..256R}.  The satellite fractions
range from $10$ per cent at the lower luminosity end, to $\sim 5$ per
cent at higher
luminosity.  These results are consistent with previous estimates of
LRG environments \citep[e.g.,][]{2009ApJ...698..143R}.

\subsection{Non-linear bias model}\label{SS:nlbiasmodel}

\begin{figure}
\includegraphics[width=\columnwidth]{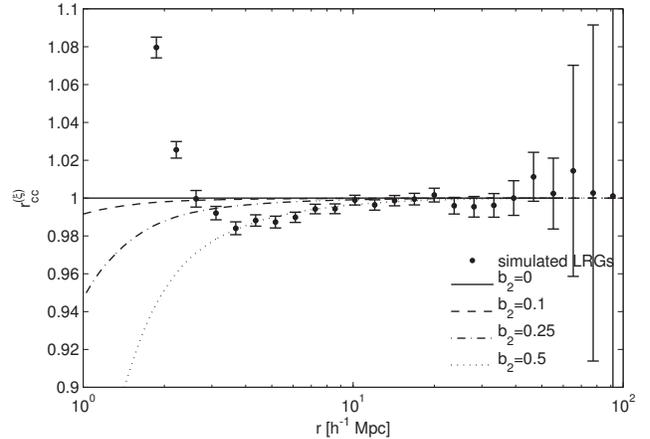}
\caption{\refresp{Correlation function cross-correlation coefficient
    between mock LRGs and dark matter (data points) in the simulations
    described in Sec.~\ref{SS:simulations}, with errors determined
    from the box to box variations in simulations.  \refresp{Lines are
      predictions from our model, with
      $\refrespm{r_\mathrm{cc}^{(\xi)}=1-(1/4)(b_2/b)^2 \xi_\text{lin}(r)}$, using a
      large-scale bias of $b=2.07$ (selected to match the observed
      large-scale $\xi_{\rmg\rmg}$ and $\xi_{\rmg\rmm}$ in the mock
      LRG sample) and} different values of $b_2$, \refrespt{with $b_2=0.5$
    providing the best fit to the simulated clustering and lensing
    observable quantities}.  The plot goes to larger $r$ compared to
    the 2d values of $R$ used in our analysis, because the measured
    observables at some $R$ depend on the 3d correlation functions to
    larger $r$.}
\label{F:rccxi}}
\end{figure}

The analysis in \cite{2010PhRvD..81f3531B} was focused on modeling 
the cross-correlation coefficient $r_\mathrm{cc}$, since this is the only quantity that is
needed to relate the measurement of galaxy auto-correlation and galaxy-dark matter 
correlation to the dark matter auto-correlation.
However, it is useful to analyse how well we model 
clustering and g-g lensing data separately with a given non-linear bias model. 
One reason to do so is that this allows us to choose different minimum
scales ($R_0$) for galaxy-galaxy lensing and galaxy clustering. We expect that lensing 
information will be quite insensitive to the details of HOD modeling: both a 
satellite and a central galaxy give approximately the same g-g lensing signal for
separations larger than the virial radius. So, we would expect the 
lensing signal to be fairly model-independent down to the virial radius. 
In contrast, the clustering signal will depend very sensitively on how satellite 
galaxies are populated within the virial radius, so the clustering signal up to at least 
 twice the virial radius will be quite model-dependent. For example, if there are no 
satellites, then the clustering signal drops to almost zero within twice the virial radius, 
while if all the halos have one central galaxy and one satellite 
radially distributed as the dark matter, then the clustering signal is similar to the 
lensing signal. 
This sensitivity to how satellite galaxies populate halos suggests that 
we should choose a larger value of $R_0$ for clustering than for lensing 
measurements. A second reason to use larger $R_0$ for clustering is that the statistical errors
are significantly smaller than for lensing, so the analysis of
clustering data is much more sensitive to inaccuracies in the
theoretical model. A third reason is that 
if we can model both of these functions with a few free parameters, 
we can use the better-measured galaxy clustering data to determine these parameters.  

To do so, we require models 
not just for $r_\mathrm{cc}$, but also for scale-dependent
bias, that are as realistic as possible. 
In order to interpret the measurements without taking ratios of noisy
quantities, we must have some well-motivated way of describing the
non-linear bias of the samples that we study.  
We consider the same local bias model of \cite{1993ApJ...413..447F} as in
\cite{2010PhRvD..81f3531B}, $\refrespm{\delta_\mathrm{h}=b\delta_\rmm+(b_2/2)\delta_\rmm^2+(b_3/3!)\delta_\rmm^3}$,
which contains a local bias relation between galaxy and dark matter
density up to third order
and combines it with standard perturbation theory (SPT), which expands the
density perturbation into a series $\delta_\rmm=\delta_\rmm^{(1)}+\delta_\rmm^{(2)}+...$, where $\delta_\rmm^{(1)}$ is the Gaussian linear theory
prediction and $\delta_\rmm^{(n)}$ is of order
$[\delta_\rmm^{(1)}]^n$, to calculate the
next-to-leading order corrections to the galaxy-galaxy and galaxy-matter
power spectra. The third order bias enters only
through a renormalisation of the leading order bias parameter, and does not have
an explicit influence on the observable correlators. At the
next-to-leading order, the corrections to the galaxy-galaxy and galaxy-matter
power spectra come from the auto-correlation of $\delta_\rmm^2$ and its cross-correlation with the
second order density perturbation $\delta_\rmm^{(2)}$. In evaluating these
terms, we can define \citep{2009PhRvD..80f3528S}
\begin{align}
A(k)=&\int \frac{\text{d}^3 q}{(2\pi)^3}F_2({\bmath q},{\bmath k}-{\bmath
q})P(q)P(|{\bmath k} -{\bmath q}|)\\
B(k)=&\int \frac{\text{d}^3 q}{(2\pi)^3}P(q)P(|{\bmath k} -{\bmath q}|),
\end{align}
where $F_2$ is the SPT mode coupling kernel (see, e.g., \citealt{2002PhR...367....1B})
\begin{align}
F_{2}({\bmath q}_1,{\bmath q}_2)=&\frac{5}{7}\alpha({\bmath q}_1,{\bmath q}_2)+\frac{2}{7}\beta({\bmath q}_1,{\bmath q}_2),
\end{align}
with
\begin{align}
\alpha({\bmath q}_1,{\bmath q}_2)=\frac{\left({\bmath q}_1+{\bmath q}_2\right)\cdot {\bmath q}_1}{
q_1^2}, && \!\!\!\beta({\bmath q}_1,{\bmath q}_2)=\frac{1}{2}\left({\bmath q}_1 +{\bmath 
q_2}\right)^2\frac{{\bmath q}_1 \cdot {\bmath q}_2}{q_1^2 q_2^2}.
\end{align}

Upon Fourier transforming, we
obtain the corresponding correlation functions, which are given by
\beq
\xi_\text{gg}(r,z)=b^2 \xi_\text{mm,NL}(r,z) +2 b\, b_2\,\xi_A(r,z) +
\frac{1}{2} b_2^2
\xi_B(r,z)\label{E:xigg}
\eeq
and
\beq
\xi_\text{gm}(r,z)=b \,\xi_\text{mm,NL}(r,z) + b_2 \xi_A(r,z).\label{E:xigm}
\eeq
$\xi_A(r)$ and $\xi_B(r)$ are the Fourier transforms of $A(k)$ and $B(k)$.
In principle, $\xi_\text{mm,NL}$ should be the correlation function
corresponding to one loop perturbation theory. Taking SPT at face value, the
Fourier transform is ill behaved and we replace it by the non-linear
correlation function measured in the $N$-body simulations. Note that
$\xi_B=\xi_\text{lin}^2$.  

As shown in \cite{2010PhRvD..81f3531B},
the above model can be used to predict the cross-correlation coefficient in the
linear and weakly non-linear regime. It predicts $r_\mathrm{cc}$ to be unity on large
scales and to drop below unity as one goes to smaller scales, with explicit
functional form given by $\refrespm{r_\mathrm{cc}^{(\xi)}=1-(1/4)(b_2/b)^2 \xi_\text{lin}(r)}$.
We know this model to be imperfect in the sense that other non-linear bias
parameters at a quadratic and cubic level may be needed to properly model the
data (\citealt{2012PhRvD..85h3509C,2012PhRvD..86h3540B}), but these higher
order parameters may not be that different in terms of their effect on the
scale dependence of the statistics we study here, so we group them into a
single parameter $b_2$ for the purpose of this paper. 

\refrespt{As seen in Fig.~\ref{F:rccxi}, this model (with parameters
  chosen to match mock LRG catalogues, and in particular, best-fitting
  $b=2.07$ and $b_2=0.5$)} describes the correlation 
coefficient down to $3h^{-1}$Mpc, below which physics from within the virial radius 
begins to affect the results.  
As argued above, we expect that these effects are more significant 
for the auto-correlation than for the cross-correlation. 
We must choose the minimum scale at which we can still adopt this
model.  Our method for doing so will be described in
Sec.~\ref{SS:r0choice}; it is based on carrying out our
analysis on simulated data, and checking that we can recover the true
cosmology in the simulations.  Before we can do so, we next describe
how we model the observable quantities in real and simulated data,
$\Upsilon_{\rmg\rmg}$ and $\Upsilon_{\rmg\rmm}$. 

\subsection{Modeling the observables}\label{SS:powspec}

Our approach is to use Eqs.~\eqref{E:xigg} and~\eqref{E:xigm} to model the
two observables.  The data are used to constrain linear $b$, quadratic
bias $b_2$, and the dark matter power spectrum times the matter
density, as in Eq.~\eqref{E:obsmm}. This is the full, non-linear
matter power spectrum, as shown in \cite{2010PhRvD..81f3531B}.  We
will use Monte Carlo Markov Chain (MCMC) methods, in which the data
are compared to the model, hence for each set of cosmological
parameters we must compute the fully non-linear dark matter power spectrum.

\subsubsection{Matter power spectrum}\label{SSS:powspec}

We obtain the estimated linear $\xi_{\rmm\rmm}(r)$ by specifying the
cosmological parameters using the {\sc camb} linear gravity solver
\citep{2000ApJ...538..473L}, which is part of the \texttt{cosmomc}
package that we use for the estimation of the cosmological parameters
\citep{2002PhRvD..66j3511L}. We increase the accuracy of the solver,
by setting \texttt{accuracy\_level=1.5}, and checked that 
increasing \texttt{accuracy\_level} does not change our results.
The correlation functions $\xi_{\rmm\rmm}(r)$ are calculated at the effective redshifts
of the three galaxy samples, given in Table~\ref{T:samples}.

\refresp{To obtain a precise prediction for the non-linear matter
  power spectrum as a function of cosmological parameters, we are
  unable to use a standardized and publicly available emulator such as
  the one presented by \cite{2010ApJ...713.1322L} for two reasons.
  First, we wish to explore variations of the Hubble parameter, $H_0$,
  which cannot be independently varied using that emulator.  Second,
  the emulator only provides predictions for the power spectrum to a
  maximum wavenumber of $k=1$~$h/$Mpc, but power at higher $k$ is
  important when computing the matter power spectrum at our minimum
  scale of $R=2h^{-1}$Mpc to the required precision.}

\refresp{Thus, given the need t}o compute the non-linear power spectrum for arbitrary cosmological
parameters ${\bmath \theta}$ that differ from our fiducial ones
(${\bmath\theta_0}$), there are several possible approaches that we could
take (given our simulations that are on a grid of cosmological
parameters).  The change in cosmological parameters affects the
non-linear power spectrum in two ways: first, via the change in the
linear power spectrum; and second, via the change in non-linearity
corrections.  Since the first effect is dominant, we account for it
accurately using analytic
calculations of the linear matter power spectrum, only interpolating on our
simulation grid to account for the second (much smaller) effect.
If we relate the non-linear and linear correlation functions via 
\beq
\xi_\mathrm{nl}(r|{\bmath\theta}) = \xi_\mathrm{lin}(r|{\bmath\theta})
\frac{\xi_\mathrm{nl}(r|{\bmath\theta})}{\xi_\mathrm{lin}(r|{\bmath\theta})}
\equiv \xi_\mathrm{lin}(r|{\bmath\theta}) \alpha(r|{\bmath\theta}), 
\eeq
then we can Taylor expand $\alpha(r)$ around our fiducial cosmological
parameters,
\beqa
\xi_\mathrm{nl}(r|{\bmath\theta}) \!\!\!\!&=& \!\!\!\!\!\xi_\mathrm{lin}(r|{\bmath\theta})
\left[ \alpha(r|\bmath\theta_0) + \sum_i
  \left . \frac{\partial\alpha(r|\bmath\theta)}{\partial\theta_i}\right|_{\bmath\theta_0}
  \!\!\Delta\theta_i\right] \notag \\
\!\!&=& \!\!\!\!\!\xi_\mathrm{lin}(r|{\bmath\theta}) \alpha(r | \bmath\theta_0)\left[1 +
  \sum_i \left .\frac{\partial\log\alpha(r|\bmath\theta)}{\partial\theta_i}\right|_{\bmath\theta_0}
  \!\!\Delta\theta_i\right]\label{der}
\eeqa
where the index $i$ runs over the parameters for which we have
$\xi_{\rmm\rmm}$ on a grid ($\sigma_8$, $n_s$, $\Omega_m$, $H_0$, and
redshift $z$).  
As an example of how this works for one parameter, changing $\sigma_8$ mostly changes
the amplitude of the correlation functions by the square of the ratio
of two values of $\sigma_8$ under consideration and this change is
propagated exactly. Only the second-order change in the shape of
the non-linear corrections is Taylor-expanded.

Our fiducial model has $\Omega_m=0.25$, $\sigma_8=0.8$, $n_s=1.0$, $h=0.7$, and
$z=0.23$;  we have 8 simulations of this cosmology. To obtain the 
derivatives of non-linear corrections with respect to cosmological parameters (needed in 
Eq.~\ref{der}), we use further models with $\Omega_m=(0.2, 0.3)$,
$\sigma_8=(0.7,0.9)$, $n_s=(0.95,1.05)$, $H_0=(65, 75)$ km s$^{-1}$
Mpc$^{-1}$, and 7 different redshift slices between $z=0$ and
$z=0.51$. 
Non-linear correlation functions for these models are obtained from
$N$-body simulations \citep{2009MNRAS.400..851S}. For each
non-fiducial model, all parameters but  one are kept at the fiducial
value. For the parameters for which we have two simulations bracketing
the fiducial value ($\Omega_m$, $\sigma_8$, $z$, $H_0$), we use different
derivatives depending on whether the
corresponding value for the target model is above or below the
fiducial value. We opted to do this rather than calculating the second
derivative to avoid numerical errors blowing up when the quadratic
term becomes dominant.
By construction, such modelling exactly reproduces the non-linear
matter correlation function for models for which we have simulations.


\subsubsection{Massive neutrinos}

We would also like to place constraints on massive neutrinos, which
requires some additional corrections to the formalism in Sec.~\ref{SSS:powspec}.  
We parametrise the neutrino mass effect as the sum of masses for the
three neutrino families,  $\sum m_\nu$, and include three different
ways that they affect the matter power spectrum. 
First, lensing is sensitive to the total gravitational potential, which includes a 
contribution from massive neutrinos. This requires us to use the Poisson 
equation to relate potential to density perturbations, the latter of
which must include 
the neutrino contribution ($\delta \rho=\rho_{\rm cdm}\delta_{\rm cdm}+\rho_{\rm b}\delta_{\rm b}+\rho_{\nu}\delta_{\nu}$, 
where cdm, b and $\nu$ subscripts denote cold dark matter, baryons and neutrinos, respectively). 
For $\sum m_{\nu}=0.15$~eV,
$f_{\nu}=\rho_{\nu}/(\rho_{\rm cdm}+\rho_{\rm b})=0.6$ per cent, 
and since neutrino perturbations go from $\delta_{\nu}=\delta_{\rm
  cdm}$ on large scales to $\delta_{\nu}=0$ on small scales, 
this effect suppresses the weak lensing power spectrum on small scales
by 1.2 per cent. 

The second effect is the usual suppression of matter fluctuations due to the fact that neutrino fluctuations are suppressed 
on small scales, which in turn leads to a suppressed growth of cold dark matter and baryon fluctuations. For 
$\sum m_{\nu}=0.15$~{\rm eV}, this effect leads to a 8 per cent suppression of the matter power spectrum. The two effects 
combined thus lead to 9.2 per cent suppression. 

The third effect is the non-linear evolution correction, which further enhances this effect. 
For $k <0.1\invhmpc$ the effect of neutrinos on the matter power spectrum in the linear regime 
can be described as a reduction of the amplitude and a red tilt (\citealt{2012MNRAS.420.2551B}). For $\sum m_{\nu}=0.15$~eV, 
this is a 4 per cent reduction in $\sigma_8$ and -0.01 reduction in $n_s$. Most of the mode coupling responsible 
for the non-linear effects comes from the long wavelength modes with $k <0.1\invhmpc$, so it is reasonable to 
assume that the non-linear effects can be described with a change of amplitude and slope, scaling linearly with 
$\sum m_{\nu}$,  
\begin{equation}
  n_{s, \rm{eff}} = n_s -0.01 \frac{\sum m_{\nu}}{0.15~\rm eV}.
\end{equation}
\refresp{The change of amplitude, $\sigma_{8, \rm{eff}}=\sigma_8-0.04[\sum m_{\nu}/(0.15~\rm eV)]$,} 
is already automatically included since we compute $\sigma_8$ for a given cosmological model 
using its power spectrum. 
The spectral index seen by the non-linear correction is not
the actual one, but the effective one given by
the above equation. This is justified by noting that the most
important quantities that determine the shape of the non-linear
correction are the amplitude and slope of the power spectrum at a
relevant pivot scale ($k \sim 0.1\invhmpc$ in our case). Since the
change in amplitude is already included in the change of $\sigma_8$,
we just approximate the change in the slope of the linear power
spectrum at the pivot point for $m_\nu$ as the change in the
spectral index.

To test this procedure, we compare the resulting non-linear to linear
power spectrum correction for massive neutrinos to the full
simulations presented in \cite{2012MNRAS.420.2551B}.  For example, for the total
neutrino mass of $\sum m_{\nu}=0.15$~eV, we find that the reduced linear
amplitude of the power spectrum at the pivot point is 8 per cent,
corresponding to a 4 per cent change in $\sigma_8$.  This 
leads to a further non-linear suppression of power, up to 3 per cent at $k \sim 1\invhmpc$. 
In addition, the
effective slope is reduced by 0.01 at the pivot point. This means that
for massive neutrinos there is more power on large scales than in the zero
mass case, relative to the pivot point.  As a result, there is more
mode-mode coupling which increases the small scale non-linear power,
countering the effect from the reduced linear amplitude. The net
effect is that the non-linear correction peaks at $k \sim 1\invhmpc$,
but this quickly reverses sign above $2$\invhmpc. The
overall effect 
is in a good
agreement with the results of the full simulations of massive
neutrinos in \cite{2012MNRAS.420.2551B}. This suggests that we can
parametrise the 
non-linear effect of massive neutrinos simply with the change in the
effective amplitude and slope of the linear power spectrum.

To summarise the neutrino mass effects: at $k \sim 0.5\invhmpc$, which is the peak of 
the contribution to $\Upsilon$ at $R=5\hmpc$ and $R_0=3\hmpc$ (Fig.~2 of \citealt{2010PhRvD..81f3531B}) and where 
we expect to have the most stringent constraints from our data set,  
we expect about 10 per cent suppression of the power for $\sum
m_{\nu}=0.15$~eV, relative to the zero mass
case. 

\subsubsection{Cosmology corrections}\label{SSS:coscorr}

When estimating the signal from the data, we assume a
cosmological model in order to convert angular distances
$\Delta\theta$, shears $\gamma_t$, and redshift-space separations
$\Delta z$ to transverse separation $R$, lensing surface density
contrast 
$\Delta\Sigma$, and line-of-sight separation $\Pi$.  Thus, for the
model predictions for any other cosmology than the fiducial cosmology,
we should in principle include a factor in both the transverse
separation and the amplitude of the measured signals to account for
the fact that the wrong cosmology was used to do these conversions
from observed to physical separations. However, for the highest
redshift 
sample (for which this is most important), the size of the corrections
is typically $\lesssim 1$ per cent for the range of allowed
cosmological models. The correction is even smaller for the other
samples, therefore it is well within the statistical error for this
analysis, and we do not include it.

\subsubsection{Combining the model ingredients}\label{SSS:combmodel}

Finally, we need to combine these model ingredients to obtain
$\xi_{\rmg\rmm}(r)$ and $\xi_{\rmg\rmg}(r)$.  We do so by using the
non-linear matter power spectrum for a given cosmology from
Sec.~\ref{SSS:powspec} along with Eqs.~\ref{E:xigg} and~\ref{E:xigm}.

The results are then
integrated to obtain the projected statistics that we use in reality.
For the lensing signals, we integrate the correlation function along
the line-of-sight to $\pm 140$\hmpc, consistent with the fact that the
lensing window is extremely broad.  We do not include that window
directly, but as shown in \cite{2010PhRvD..81f3531B} fig. 9, its
effects are very small on the scales we use for science, and can be
corrected for in a single factor that includes the clustering
line-of-sight integration length, redshift-space distortions (RSD), and the lensing
window. Given that this correction factor is $\sim 3$ per cent at
60\hmpc, much smaller than the observational errors, and $\sim 1$ per
cent below 30\hmpc, we neglect this correction\footnote{Technically,
  we have only done this test for the LRG sample.  However, for the
  higher redshift and mass sample, the galaxy bias is higher and therefore RSD are
  even less important.  For the lower redshift and mass sample, while the galaxy bias is lower
  and RSD are more important, we will see that the observational
  errors are also larger than for LRG.}.  For clustering, we integrate along the
line-of-sight to $\pm 60$\hmpc, consistent with the observational
measurements.  Finally, for both the lensing and clustering we use the
projected surface densities to obtain $\Upsilon$.

Because we have some uncertainty in the calibration of the lensing
signal due to several systematic errors 
 (Sec.~\ref{SS:calibbias}), 
we include a
nuisance calibration bias parameter for the g-g lensing, which is assumed
to have a mean zero and a Gaussian width of $(4, 4$, and $5)$ per cent
for the 3 samples.  The calibration bias is assumed to be the same at
all radii and for all samples - i.e., if the calibration bias is 4 per
cent for Main-L5 then it is 4 and 5 per cent for LRG and LRG-highz.
This treatment is appropriate since the lensing calibration biases
originate from the same sources for each sample, and improper
estimation and removal of those biases would affect all samples nearly equally.

\subsection{Choice of $R_0$}\label{SS:r0choice}

\cite{2010PhRvD..81f3531B} considered relatively large values of $R_0$
such as $3$ and $5$\hmpc.  Use of a large value of $R_0$ is
advantageous from the perspective of systematic error, because it
means that we are less sensitive to several effects that tend to be
worse at small scales: cross-correlation coefficient deviations from $1$, deviations
from our model for non-linear bias, and observational issues such as
intrinsic alignments.  

However, use of larger $R_0$ will
necessarily remove more of the measured signal, resulting in a noisier
measurement.  We therefore revisit the choice of $R_0$ in order to
achieve a fair compromise between statistical and systematic error.
Moreover, unlike in \cite{2010PhRvD..81f3531B}, we permit different
$R_0$ for the galaxy-galaxy lensing and the galaxy clustering, which
is possible in the case that we explicitly model the signals (i.e.,
using different $R_0$ means that results cannot be obtained by taking
ratios of the two signals).  
The galaxy clustering signal is more sensitive than the
galaxy-galaxy lensing to the
fidelity of the non-linear bias model 
(because it has higher signal-to-noise), so it
might require a higher $R_0$ to avoid systematic errors.  

\begin{figure}
\includegraphics[width=\columnwidth]{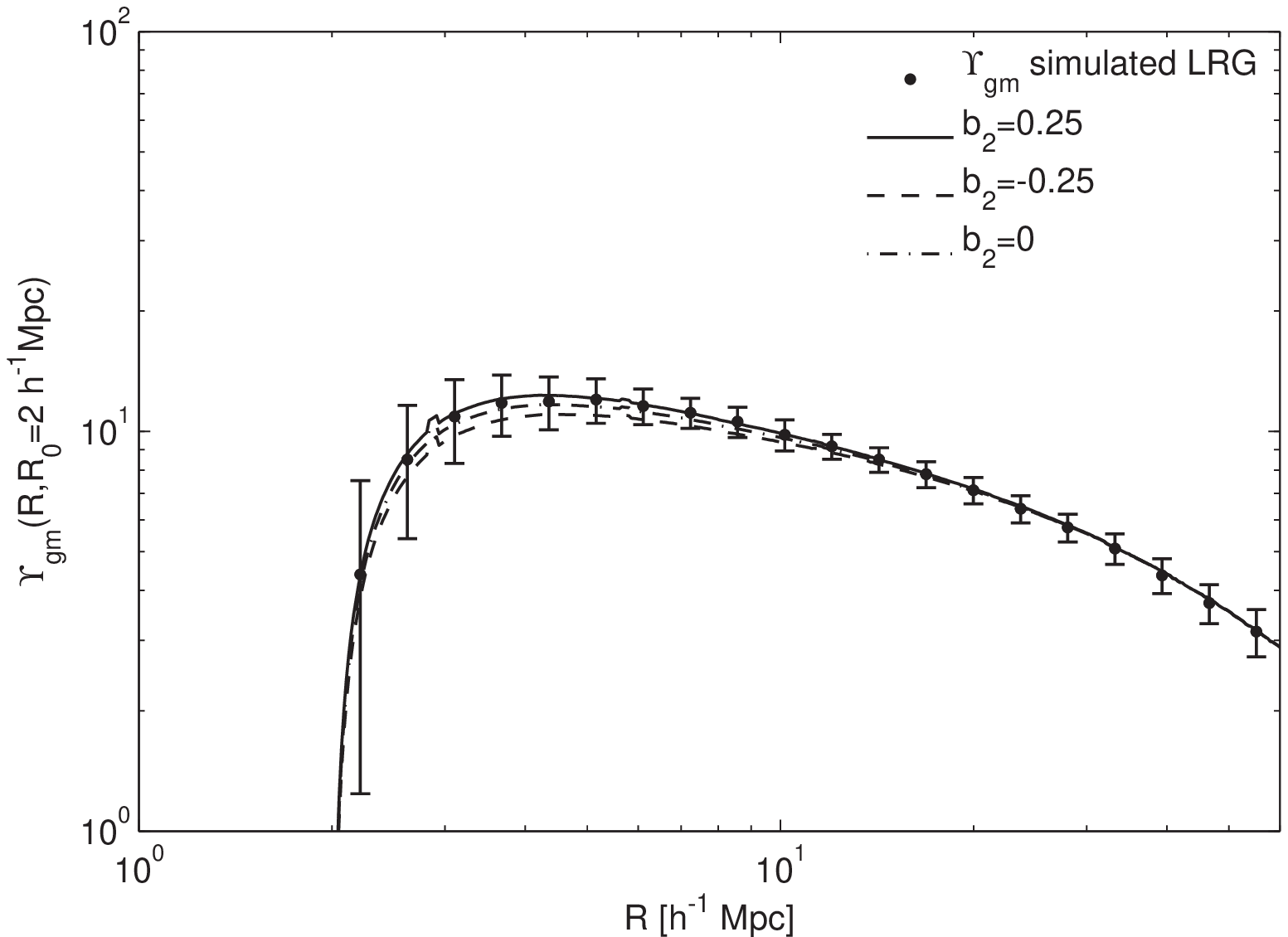}
\includegraphics[width=\columnwidth]{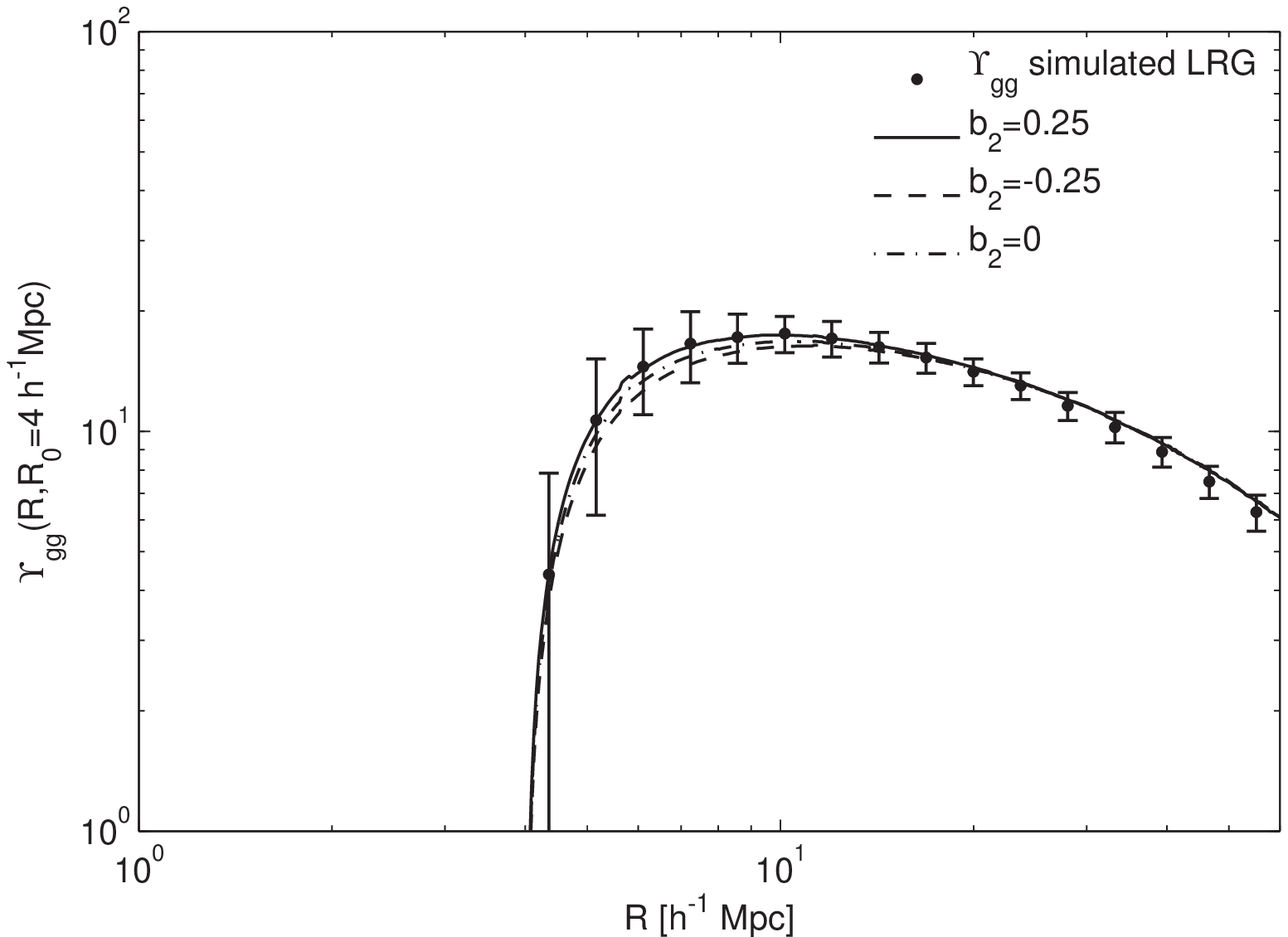}
\caption{\refresp{\emph{Top: } $\Upsilon_\text{gm}$ for mock LRGs and the model predictions.
\emph{Bottom: }$\Upsilon_\text{gg}$ for mock LRGs and the model predictions.}\label{F:nlbiasmodel}}
\end{figure}

We choose values of $R_0$ for the two measurements based on modeling
the simulated LRG sample using the same machinery that we use to model
the data (but without adding lensing shape noise, so that deviations
from the real cosmology in the simulations are due to a real
analysis bias).  
We decided to use $R_0=2$ and 4\hmpc\ for
galaxy-galaxy lensing and galaxy clustering, respectively, since 
we find this choice still gives a small systematic error compared to 
the statistical error, as shown in Appendix~\ref{SS:paramconstraints-sims}. The corresponding plots of $\Upsilon$ are 
shown in Fig.~\ref{F:nlbiasmodel}, 
\refrespt{and we see that the simulated data and the best model 
agree reasonably well both for galaxy-galaxy lensing and galaxy clustering, 
with $b_2=0.25$ working acceptably. Note that $b_2=0.5$
  best describes the cross-correlation coefficient for $\xi$, as seen
  in Fig.~\ref{F:rccxi}.  In fact for the simulated data, values
  in the range from $b_2=0.25$ and $0.5$ are able to describe the data
within the limits of our errorbars.}  \refresp{A study carried out
  independently from this work \citep{2012MNRAS.426..566C} has also studied the
  value of $R_0$ for which we can safely achieve $r_\mathrm{cc}\sim 1$
at all $R>R_0$, in the context of a more general HOD model.  They
identify $3$\hmpc\ as a reasonable choice of $R_0$ (assuming the same
$R_0$ is used for both measurements), nicely consistent
with our findings.}

\subsection{Parameter values and constraints}

Our
convention is to quote parameter values based on the median of the
posterior distribution after marginalisation over all other
parameters.  \refrespt{Except where explicitly stated otherwise, we
  include a prior that is a function of $\sigma_8$, 
based on our calibration on simulations in
Appendix~\ref{SS:paramconstraints-sims}.}  The quoted
error-bars come from using the PDF to determine the 68, 95, and 99.7
per cent confidence intervals.

\section{Data}\label{S:data}

Here we describe the data used in this paper, all of
which comes from the Sloan Digital Sky Survey (SDSS).   The SDSS \citep{2000AJ....120.1579Y} imaged roughly $\pi$ steradians
of the sky, and followed up approximately one million of the detected
objects spectroscopically \citep{2001AJ....122.2267E,
  2002AJ....123.2945R,2002AJ....124.1810S}. The imaging was carried
out by drift-scanning the sky in photometric conditions
\citep{2001AJ....122.2129H, 2004AN....325..583I}, in five bands
($ugriz$) \citep{1996AJ....111.1748F, 2002AJ....123.2121S} using a
specially-designed wide-field camera
\citep{1998AJ....116.3040G}. These imaging data were used to create
the cluster and source catalogues that we use in this paper.  All of
the data were processed by completely automated pipelines that detect
and measure photometric properties of objects, and astrometrically
calibrate the data \citep{2001ASPC..238..269L,
  2003AJ....125.1559P,2006AN....327..821T}. The SDSS-I/II imaging
surveys were completed with a seventh data release
\citep{2009ApJS..182..543A}, though this work will rely as well on an
improved data reduction pipeline that was part of the eighth data
release, from SDSS-III \citep{2011ApJS..193...29A}; and an improved
photometric calibration \citep[`ubercalibration',][]{2008ApJ...674.1217P}.

Below we describe the samples that are used as
lenses and as sources. 

\subsection{Main sample lenses}\label{SS:main}

The first lens sample that we use for this work is the flux-limited Main
galaxy sample \citep{2002AJ....124.1810S} from SDSS DR7.  The nominal
flux limit is $r<17.77$, when defined using Petrosian
magnitudes\footnote{All magnitudes quoted in this paper are corrected
  for Galactic extinction using the dust maps from \cite{1998ApJ...500..525S} and the
  extinction-to-reddening ratios from \cite{2002AJ....123..485S}.}
(based on a modification of \citealt{1976ApJ...209L...1P} described in
\citealt{2001AJ....121.2358B} and \citealt{2001AJ....122.1104Y}). In
reality, the actual flux limit varies slightly across the survey
area.  We use the Main sample selection from the NYU Value-Added
Galaxy Catalog \citep[VAGC,][]{2005AJ....129.2562B}, which includes 7966 deg$^2$ of
spectroscopic coverage (though we will employ further area cuts,
described below).  

We select
our sample using the `LSS sample' DR7-2 in the VAGC, which carefully
tracks the spectroscopic flux limit and completeness across the sky.  The particular
LSS samples that we use are `dr72full8' through `dr72full10', where `full' samples have the
following properties:
\begin{itemize}
\item They use all galaxies from $r>10$ to the position-dependent
  flux limit.
\item They use areas with any level of completeness (even $<0.5$,
  which occurs very rarely).
\item Galaxies that did not get a spectrum due to fibre-collisions are
  assigned a redshift using the nearest-neighbour method.
\end{itemize}
The `8', `9', and `10' subsamples have the following properties, some of which will be
subject to more cuts described below:
\begin{itemize}
\item Redshift $0.001<z<0.4$
\item The $k$-corrections are to $z=0.1$ ({\sc kcorrect v4\_1\_4}; \citealt{2007AJ....133..734B}).
\item The distance modulus $\mu=5\log{[D_L/(10\mathrm{pc})]}$ is determined using $\Omega_m=0.3$,
  $\Omega_\Lambda=0.7$.  This is formally inconsistent with the
  numbers used in the rest of this paper.  However, this is not a significant issue here where we
  simply seek a reasonably volume-limited and consistent sample
  (particularly given the weak dependence of the distance modulus on
  cosmology for these redshifts).
\item The luminosity evolution is assumed to have the form
\beq\label{E:lumev}
M(z) = M(z=0.1) - 2[1 - (z-0.1)](z-0.1),
\eeq
\refresp{which is chosen to match the number counts of SDSS
  spectroscopic
  galaxies\footnote{\texttt{http://sdss.physics.nyu.edu/vagc/lss.html}}.
  Given the redshift limits of our sample, this correction is
  constrained to lie within the range $[-0.17, 0.10]$.}
\item The absolute magnitude is defined, in terms of the
  Petrosian magnitude $r$ and galaxy redshift $z$, correcting for
  luminosity evolution, as
\beq\label{E:absmagdef}
M_r = r - [\mu + k_{0.1}(z) + M(z) - M(z=0.1)].
\eeq
Given that the luminosity evolution is such that galaxies were
brighter in the past, the definition here removes that
trend, connecting galaxies at one redshift to those that were
suitably brighter at earlier times according to the
empirically-determined evolution law in Eq.~\eqref{E:lumev}.
\item The absolute magnitude is then required to be in the range
  $[-22, -21]$, $[-21, -20]$, and $[-20, -19]$ for the three samples, respectively.
\end{itemize}

The effective area of the LSS sample is 7279 deg$^2$.  We then imposed
some additional cuts on the area, removing regions without any source
galaxies in the 
background or within a Tycho bright star map
\citep{2000A&A...355L..27H}.  These cuts reduce 
the effective area to 7131 deg$^2$.

 We wish to
avoid overlap with the LRG lens samples described in the upcoming
subsection, so that the cross-covariance between the signals with
different lens samples can be assumed to be zero.  Thus, we first
require $0.02<z<0.155$ (where the lower redshift cut removes galaxies
for which it would be computationally prohibitive to measure
correlations to 70\hmpc, and the upper redshift cut removes overlaps
with LRGs).   We then defined the three LSS
samples 
using the notation from \cite{2006MNRAS.368..715M}: L3 with $-19>M_r\ge-20$, L4 with
$-20>M_r\ge-21$, L5 with $-21>M_r\ge-22$.  We do not define any
brighter samples because their low abundance 
means that there are very few galaxies in those samples after the $z<0.155$ cut.  

In practise, carrying out the analysis with all three samples
requires caution: if the redshift ranges overlap (as
naturally occurs for a flux-limited sample), then 
for scales above $R\sim 8h^{-1}$Mpc, we find that the clustering and lensing signals
exhibit high cross-correlations between the samples -- typically
$80$ per cent -- because they trace similar large-scale structures.  When
limiting to volume-limited samples that do not overlap, the
statistical power of L3 and L4 becomes relatively low on cosmological scales.
In addition,
our non-linear bias model in Sec.~\ref{S:theory} was only tested on
simulations with relatively high-mass halos ($\mvir\gtrsim 1\times 10^{13}\hmsun$).
Given the typical halo masses for L3 and L4 in \cite{2006MNRAS.368..715M}, we
conclude that the optimal way of including the Main sample in this
analysis is to use
L5 only.  This sample (referred to as `Main-L5' in the rest of the
paper) includes 69~150 galaxies, and is 
volume-limited for the redshift range that we use.

For computation of cosmological observables, we require a set of
random points that are distributed in the same way as the lens
sample.  We therefore use the sets of random points distributed for
this sample in the VAGC LSS sample; this includes a proper redshift
distribution that depends on the position-dependent flux limit at each
point.  For weighting, we use the inverse of the `sector
completeness' defining the redshift success rate.  The sector
completeness for this sample has a
median value of 0.972; 95 per cent of the galaxies have
completeness above 0.924.

\begin{figure}
\begin{center}
\includegraphics[width=0.8\columnwidth]{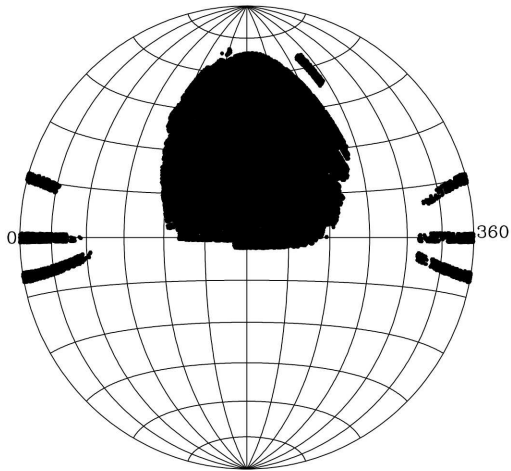}
\caption{Area coverage of the lens samples used in this paper.\label{F:area}}
\end{center}
\end{figure}

The area coverage of the lens sample (7131 deg$^2$, or
$f_\mathrm{sky}=0.17$) is shown in Fig.~\ref{F:area}.  This coverage is strictly a subsample
of the source catalogue from \cite{2012MNRAS.425.2610R} that we describe in Sec.~\ref{SS:source}. 

\begin{figure}
\includegraphics[width=\columnwidth]{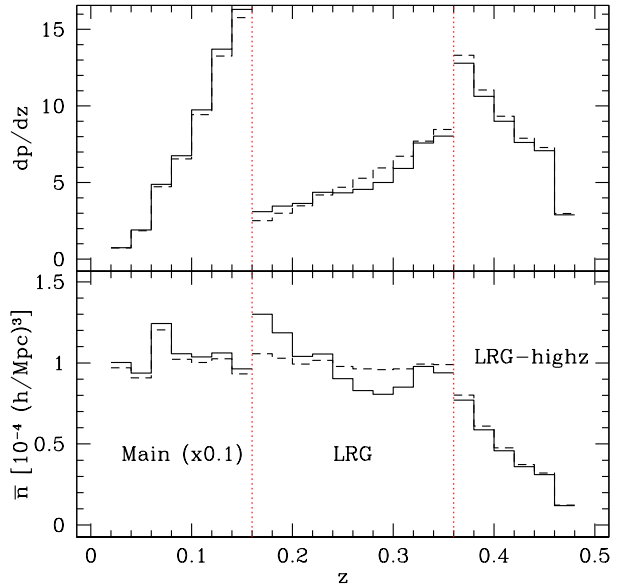}
\caption{{\em Top:} Redshift distribution $\rmd p/\rmd z$ for the
  three lens samples used in this work, as labeled on the plot. The
  solid and dashed lines show the unweighted and weighted histograms,
  respectively.  {\em
    Bottom:} Comoving number density $\bar{n}$, in units of
  $10^{-4}(h/\text{Mpc})^3$ (multiplied by $0.1$ for Main-L5 for
  easier viewing). The dotted vertical lines show the
  delineation between the different lens samples.\label{F:lensz}}
\end{figure}

The redshift distribution $\rmd p/\rmd z$ and the comoving number
density $\bar{n}$ are shown for the Main L5 sample, and for the other
samples described in subsequent subsections, in Fig.~\ref{F:lensz}.
For the Main-L5 sample, the comoving number density is $\sim 10^{-3}(h/\text{Mpc})^3$.  

The properties of this
sample (and those introduced in subsequent sections) are summarised in
Table~\ref{T:samples}. For the lensing, the effective
redshift $z_\mathrm{eff,gm}$ is determined not just by the lens redshift distribution, but
also by geometric factors related to the relation between lens and
source redshifts that come into the weighting scheme we use for
estimating the signals (Sec.~\ref{S:observations})\refresp{:} 
\beq\label{E:zeffgm}
z_\mathrm{eff,gm} = \frac{\sum_{ls} w_{ls} z_l}{\sum_{ls} w_{ls}}
\eeq

\begin{table*}
\begin{tabular}{cccccc}
\hline\hline
Sample & $N_\mathrm{gal}$ & $z$ range & $\langle z\rangle$ &
$z_\mathrm{eff,gm}$ & $\bar{n}$ \\
name & & & & & $[(h/\mathrm{Mpc})^3]$ \\ 
\hline
Main-L5 & 69~150 & $0.02\le z<0.155$ & $0.115$ & $0.109$ & $10^{-3}$ \\
LRG & 62~081 & $0.16 \le z < 0.36$ & $0.278$ & $0.257$ & $10^{-4}$ \\
LRG-highz & 35~088 & $0.36\le z <0.47$ & $0.407$ & $0.398$ & $[8 - 6((z-0.36)/0.11)]\times 10^{-5}$ \\
\hline
\end{tabular}
\caption{Properties of the lens samples described in
  Sec.~\ref{S:data}.  For each sample, we show the redshift range,
  the mean redshifts, the effective redshift for the lensing 
  as \refresp{in Eq.~\ref{E:zeffgm}}, and the number density.\label{T:samples}}
\end{table*}

\subsection{LRG sample lenses}\label{SS:lrg}

We also define two lens samples consisting of Luminous Red Galaxies,
or LRGs \citep{2001AJ....122.2267E}.  These galaxies have been used 
for numerous cosmology analyses with SDSS, most notably the detection
of Baryon Acoustic Oscillations (BAO), which is enabled 
 by the high galaxy bias of $\sim 2$ and the large volume probed
by this sample, $\sim 0.65 (h^{-1}\text{Gpc})^3$.

For selection of Luminous Red Galaxies, we follow the
methodology\footnote{\texttt{http://cosmo.nyu.edu/\mytilde
  eak306/SDSS-LRG.html}} of \cite{2010ApJ...710.1444K}, which also
starts from the NYU VAGC LSS sample described in the previous
subsection.  In this case, only regions with completeness $\ge 0.6$
are included; this definition is inconsistent with that used
for Main-L5, but in practise, the discrepancy only affects 13 deg$^2$, or
0.2 
per cent of the area.  Our selection is otherwise identical to that
from \cite{2010ApJ...710.1444K}, with the exception of area cuts to
restrict to the footprint of the source catalogue, eliminating 8 per
cent of the LRGs.

Rest-frame absolute magnitudes in the $g$ band are calculated starting from the
$r$ band extinction-corrected apparent Petrosian magnitude.  The
distance modulus assumes $\Omega_m=0.25$, $\Omega_\Lambda=0.75$.
$k$-corrections and evolution corrections from
\cite{2001AJ....122.2267E} are used to convert $r$ to $M_g$.  

\cite{2010ApJ...710.1444K} have a well-defined procedure for
calculating weights, completeness factors, dealing with fibre
collisions, and distributing random points.  In brief, they begin with
a calculation of sector completeness \refresp{to account for all
  sources of incompleteness except for fibre collisions (i.e., this
  calculation 
accounts for galaxies {\em that were allocated fibres} and did not get
a spectrum)}.  This completeness is used when distributing random points
in the survey area; in any given region, they are diluted according to
that region's sector completeness.  To deal with the $\sim 2$ per cent
of LRG targets that were not allocated fibres due to fibre collisions,
a special weight is assigned; e.g., in a group of 3 LRGs of which only
2 were allocated a fibre, those two would each get a weight of $1.5$.
The random points -- of which there are fifteen times as many as real
points -- are assigned a random redshift drawn from the $p(z)$
of the real LRGs.

We define two redshift samples, which we call `LRG' ($0.16\le z<0.36$)
and `LRG-highz' ($0.36\le z<0.47$).  In both cases, the absolute
magnitude limits are $-23.2<M_g<-21.2$; the former is approximately
volume-limited, whereas the latter is flux-limited but relatively
narrow\footnote{\refresp{One might legitimately wonder whether the method
    described in Sec.~\ref{S:theory} can be applied to a
    flux-limited sample, in which the sample properties clearly evolve
    with redshift.  However, as emphasized there, all we
    are assuming is that the large-scale bias describing the galaxy
    auto-correlation is the same as that describing the galaxy-mass
    cross-correlation, and the stochasticity is near one on the scales
    we use.  Using the notation from Sec.~\ref{S:theory} it is
    possible to show that our method should be broadly applicable for
    galaxy populations with mixes of properties, provided that the
    above assumptions are true.  In contrast, methods that use the
    small-scale lensing and/or clustering signals have additional
    assumptions that {\em would} be violated at some level in a
    flux-limited sample, because the small- and large-scale lensing
    signals scale with $M^{2/3}$ and $b$, respectively, so the
    effective mean halo mass and bias inferred from small- and
    large-scale lensing signals would not in general lie on the
    cosmological halo mass versus bias relation.}} (see
Fig.~\ref{F:lensz}).  In the first case, we \refresp{adopt a radial
  weighting scheme that reduces the impact of large-scale structure
  fluctuations on the redshift histogram.  This scheme is taken
  directly from \cite{2010ApJ...710.1444K} appendix A2, is
  optimized for BAO studies, and does not significantly change the
  results, but we use it directly in order to enable a
  comparison between our results and other large-scale structure
  measurements with LRGs.  In short, they bin the redshift histogram
  into bins of width $\Delta z=0.015$, and define a smooth $n(z)$ by
  doing a spline fit to that histogram.  Then the radial weight is
  defined as $1/(1+n(z)P_{\rm fid})$ where $P_{\rm fid}=4\times
  10^{4}(h/\mathrm{Mpc})^3$.}  Thus, for the LRG
sample, the weights used for real points are \beq\label{E:wlrg}
w_\mathrm{LRG} = \left[\frac{\text{fibre collision
      weight}}{\text{completeness}}\right]\left[\frac{1}{1+n(z)P_{\rm fid}}\right]
\eeq and for random points, the same but without any fibre collision
weights\footnote{When normalising ratios of real versus random points,
  we use weights rather than absolute numbers of galaxies, and must
  watch out for the fact that if $N_\mathrm{real}=N_\mathrm{rand}$,
  $\sum w_\mathrm{real}\ne \sum w_\mathrm{rand}$, because of the fibre
  collision weighting on the real points.}.  For the LRG-highz sample,
we use \beq\label{E:wlrg-highz} w_\mathrm{LRG-highz} =
\frac{\text{fibre collision weight}}{\text{completeness}}.  \eeq
\refresp{In this case, since the $n(z)$ is a stronger function of
  redshift, it is not clear that it makes sense to include it in the
  weighting, and particularly not in a lensing analysis where the
  source density is dropping rapidly with redshift.}

Once we include the
  redshift-dependent weighting, the LRG sample has a comoving number
  density that is nearly
  constant at $10^{-4}(h/\text{Mpc})^3$, a factor of ten smaller than
  for Main-L5.  The LRG-highz sample has a
  comoving number density that drops with redshift because the sample
  is 
  flux-limited.  More details of these samples are
  shown in Table~\ref{T:samples}.


\subsection{Sources}\label{SS:source}

The catalogue of source galaxies with measured shapes used in this
paper is described in \cite{2012MNRAS.425.2610R}, hereafter R12, which
uses the re-Gaussianization method \citep{2003MNRAS.343..459H} of
correcting for the effects of the point-spread function (PSF) on the
observed galaxy shapes. The treatment of systematic errors is updated
and improved compared to the previous SDSS source catalogue using this
PSF-correction method \citep{2005MNRAS.361.1287M}, in part using tests
of simulated SDSS images using real galaxies from COSMOS and real SDSS
PSFs \citep{2012MNRAS.420.1518M}.  To estimate source redshifts, we
use photometric redshifts (\photoz) based on the five-band photometry from the Zurich Extragalactic
Bayesian Redshift Analyzer \citep[ZEBRA,][]{2006MNRAS.372..565F}, which were
characterised by \cite{2012MNRAS.420.3240N}, hereafter
N12. \refresp{In particular, we use the maximum-likelihood mode for
  ZEBRA, and choose the best-fitting photo-$z$ after marginalizing
  over the SED template.}

The catalogue production procedure was described in detail in R12, so
we describe it only briefly here.  Galaxies were selected in a 9243
deg$^2$ region, with an average number density of $1.2$ arcmin$^{-2}$.
The selection was based on cuts on the imaging quality, data reduction
quality, galactic extinction $A_r<2$ defined using the dust maps from
\cite{1998ApJ...500..525S} and the extinction-to-reddening ratios from
\cite{2002AJ....123..485S}, apparent magnitude (extinction-corrected
$r<21.8$), \photoz\ and template used to estimate the \photoz, and galaxy size compared to the
PSF.  The apparent magnitude cut used model
magnitudes\footnote{\texttt{http://www.sdss3.org/dr8/algorithms/\\magnitudes.php\#mag\_model}}, which
are defined by fitting the galaxy profile in the $r$ band to a
S\'ersic profile with $n=1$ (exponential) and $n=4$ (de Vaucouleurs),
choosing the better of the two models, and then using that same
rescaled profile to get magnitudes in all the bands.   For comparing
the galaxy size to that of the PSF, we use the resolution factor $R_2$
which is defined using the trace of the moment matrix of the PSF
$T_\mathrm{P}$ and of the observed (PSF-convolved) galaxy image
$T_\mathrm{I}$ as
\beq
R_2 = 1 - \frac{T_\mathrm{P}}{T_\mathrm{I}}.
\eeq
We require $R_2>1/3$ in both $r$ and $i$ bands.

The software pipeline used to create this catalogue obtains galaxy images in the $r$ 
and $i$ filters from the SDSS `atlas images' 
\citep{2002AJ....123..485S}.  The basic principle of shear measurement 
using these images is to fit a Gaussian profile with elliptical
isophotes 
to the image, and define the components of the ellipticity
\beq
(e_+,e_\times) = \frac{1-(b/a)^2}{1+(b/a)^2}(\cos 2\phi, \sin 2\phi),
\label{eq:e}
\eeq
where $b/a$ is the axis ratio and $\phi$ is the position angle of the 
major axis.  The ellipticity is then an estimator for the shear,
\beq
(\gamma_+,\gamma_\times) = \frac{1}{2\cal R}
\langle(e_+,e_\times)\rangle,
\eeq
where ${\cal R}\approx 0.87$ is called the `shear responsivity' and 
represents the response of the ellipticity (Eq.~\ref{eq:e}) to a small 
shear \citep{1995ApJ...449..460K, 2002AJ....123..583B}; ${\cal R} =
1-e_\mathrm{rms}^2$.  In the course of the re-Gaussianization
PSF-correction method, corrections are applied to account for
non-Gaussianity of both the PSF and the galaxy surface brightness
profiles \citep{2003MNRAS.343..459H}.

\begin{figure}
\begin{center}
\includegraphics[width=0.8\columnwidth,angle=0]{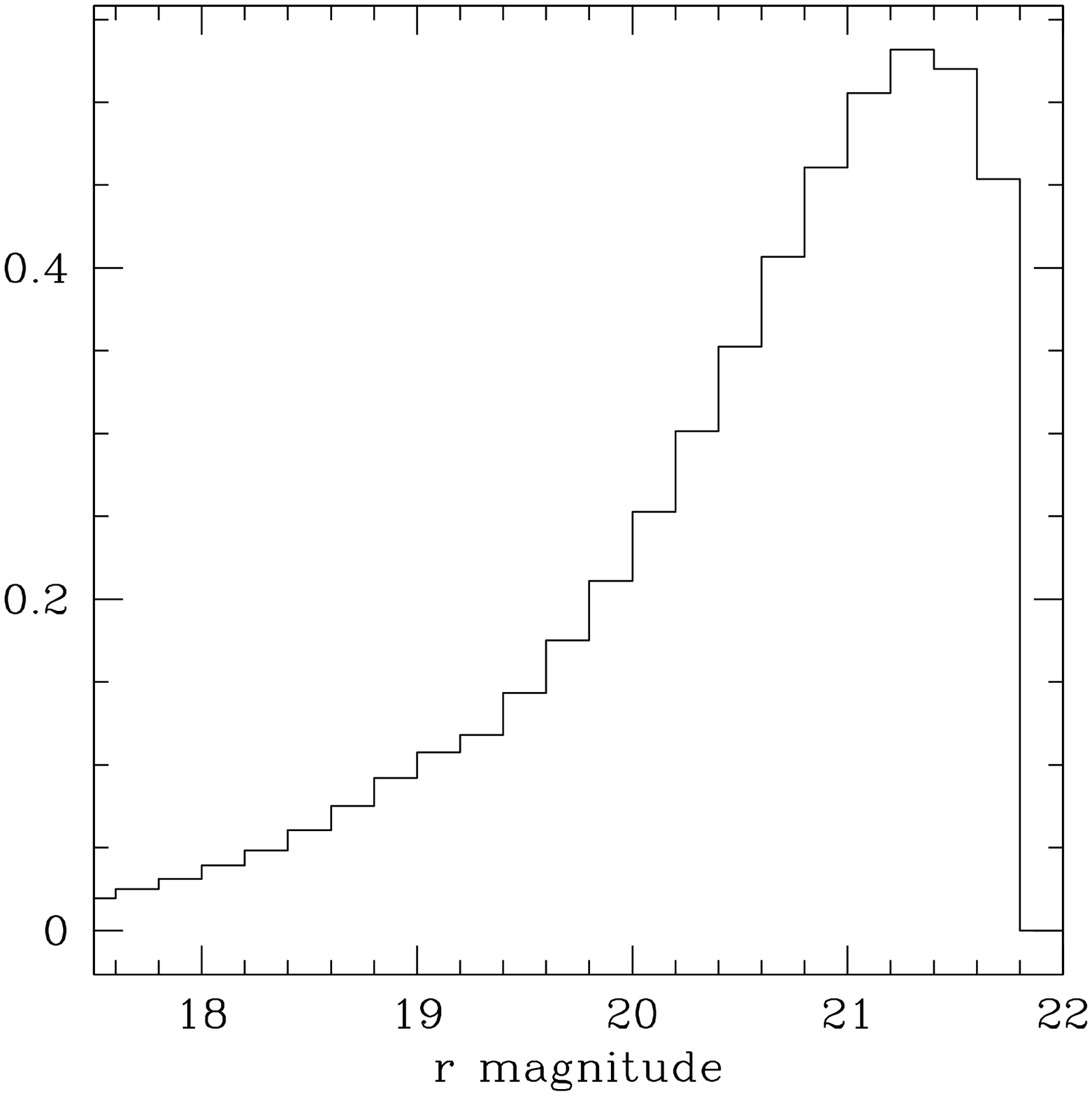}
\includegraphics[width=0.8\columnwidth,angle=0]{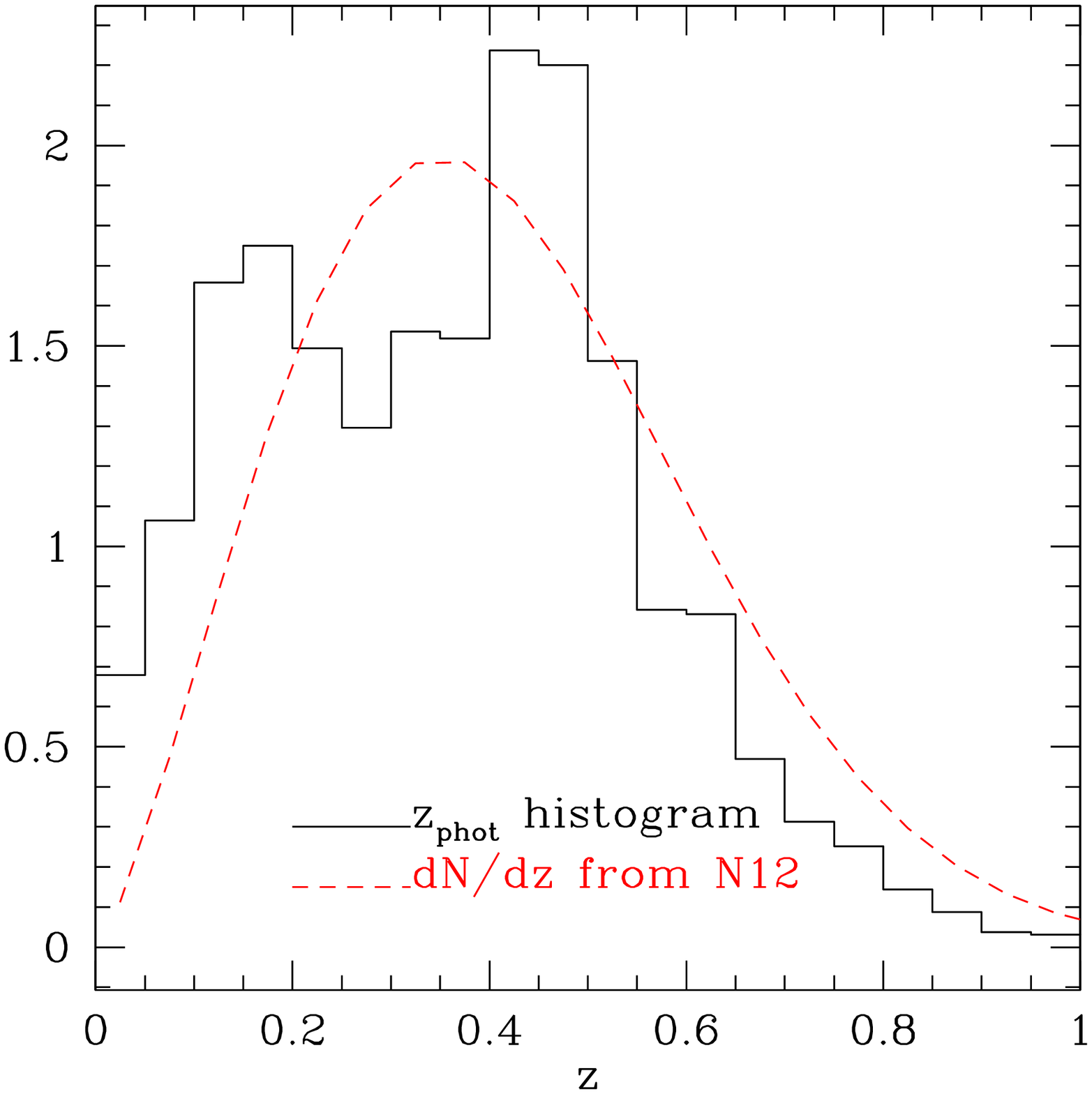}
\caption{\label{F:srchist} Histograms of
  source galaxy properties, derived from a random subsample of 5 per
  cent of the catalogue after imposing all cuts
  ($\sim 2\times 10^{6}$ galaxies).  {\em Top:} Histogram of $r$-band
  extinction corrected model magnitude. 
{\em Bottom:} \photoz\ histogram, and the inferred true
  $\rmd N/\rmd z$ from N12.}
\end{center}
\end{figure}

For this work, we do not use the entire source catalogue, only the
portion overlapping the aforementioned lens samples and around the
edges.  Fig.~\ref{F:srchist} shows histograms of the source galaxy
$r$-band apparent magnitude 
and \photoz. 

\section{Observational method}\label{S:observations}

In this section, we describe how we use the galaxy catalogues from
Sec.~\ref{S:data} to measure our two observables, the galaxy-galaxy
lensing and the galaxy clustering.

\subsection{Galaxy-galaxy lensing}\label{SS:obs-gglensing}

Here we describe the computation of the lensing signal.  For this
computation, we rely on the lens catalogues in Sections~\ref{SS:main}
and~\ref{SS:lrg}, and the catalogues of random lenses with the
corresponding area coverage and redshift distributions.  First, pairs
of lenses and sources that are physically close on the sky and satisfy
$z_s > z_l$ (using \photoz\ for sources) are identified.  Here,
``physically close'' is determined using the comoving transverse
separation at the lens redshift; we require $0.02<R<32.9$\hmpc\ for
Main-L5, and $0.02<R<73.1$\hmpc\ for the two LRG samples.  These
ranges are split into 37 or 41 logarithmic radial bins \refresp{with $\Delta (\ln{R})=0.2$.}

Lens-source
pairs are assigned weights according to the error on the source shape
measurement via
\beq
w_{ls} = \frac{1}{\Sigma_c^{2}(\sigma_s^2 + \sigma_{SN}^2)}
\eeq
\refresp{where $\sigma_s^2$ is the estimated shape measurement error due to
pixel noise (validated in R12 by comparing measured shapes in repeat
observations),} and $\sigma_{SN}^2$, the intrinsic shape noise, was found in R12 to be $0.365$, independent of magnitude.  The factor of
$\Sigma_c^{-2}$ means that we weight by inverse variance of the expected lensing
signal, thus downweighting pairs that are close in redshift because
the lensing geometry is suboptimal.

Once we have computed these weights, the lensing signal in each annular
bin can be computed via
a summation over lens-source pairs ``ls'' and random lens-source pairs
``rs'':
\beq
\ds(R) = \frac{\sum_{ls} w_{ls} e_t^{(ls)} \Sigma_{c,ls}}{2 {\cal
    R}\sum_{rs} w_{rs}} 
\eeq
where the factor of $2{\cal R}$ arises due to our definition of
ellipticity \refresp{and is needed to convert tangential ellipticity
  $e_t$ to shear $\gamma_t$}.  Note that this is equivalent to the procedure in
previous works such as \cite{2005MNRAS.361.1287M} of using $\sum_{ls} w_{ls}$ in the
denominator and then multiplying the result by the boost
factor, 
\beq\label{E:boostdef}
B(R) = \frac{\sum{w_{ls}}}{\sum{w_{rs}}}.
\eeq
The division by $\sum w_{rs}$ is necessary to account for the fact
that some of our `sources' are physically associated with the lens,
and therefore not lensed by it\footnote{This correction is formally
  correct in the limit that other effects that modulate the source
  number density, such as magnification or difficulty in detecting
  sources due to software or light from the lens, are negligible.
  This condition is satisfied on all scales used in this work.}.

Due to the observational strategy in SDSS, there is a tendency for the
PSF to align coherently along the scan direction.  This tendency gives
rise to 
a so-called `systematic shear' if the PSF correction is not
perfectly efficient at removing the PSF ellipticity from the galaxy
shapes, which turns out to be the case at a low level for our
PSF-correction method.  If a lens has a uniform distribution of
sources around it, the contribution of the coherently-aligned
systematic shear to the average tangential shear is zero.  Thus, the systematic shear can contribute to the
lensing signal primarily due to the inclusion of lenses near survey boundaries,
since they lack a symmetric distribution of sources.  To remove
this systematic shear, we can simply subtract the lensing signal
\dsrand\ measured around random points, which will capture the
geometry-dependent effect of the systematic shear.  As noted by
\cite{2005MNRAS.361.1287M}, this correction may be imperfect in the case
that the lens density fluctuates due to some effect that also modulates
the systematic shear, if this modulation is not included in the
random point distribution. We will return to this issue in
Sec.~\ref{SSS:largescale}.

To compute $\Upsilon$ using Eq.~\eqref{E:upsilon} on our noisy, binned
data, we use the following procedure.  We first 
determine $\ds(R_0)$; a discussion of how this procedure can affect
results is in \cite{2010MNRAS.405.2078M}.  For this project, as we
will show in Sec.~\ref{SSS:calcupsgm}, we are helped by the fact that around
$R_0$ there is a range of scales on which \dsr\ is well-approximated
by a power law.  This situation is different from that of
\cite{2010MNRAS.405.2078M}, which used  galaxy clusters with
$R_0$ well within the cluster virial radius, so that the scaling of
\dsr\ was  inconsistent with a power law.
Thus, in this paper (unlike previous work) we can fit $\ds$ to a power law over a fixed range
of scales, which will minimise the noise in estimation of $\ds(R_0)$
that gets propagated into $\Upsilon$.  
We present details and tests of this procedure in
Sec.~\ref{SSS:calcupsgm}. 
After estimating $\ds(R_0)$, we compute $\Upsilon$ in each radial bin
using Eq.~\eqref{E:upsilon} directly.

For science, we only use scales above 2\hmpc\ (4\hmpc) for the lensing (clustering), which results in the
inclusion of 18 radial bins for LRGs and 14 for Main sample lenses (14
and 10). These bin counts take into account the fact that we exclude
the nominal first radial bin above 4\hmpc\ for the clustering, because
while the bin center is above $R_0$, the lower edge is below $R_0$,
and we do not attempt to model $\Upsilon$ in this rapidly-varying
regime.  The
maximum scales of $\sim 30$ and $\sim 70$\hmpc\ for Main-L5 and the
LRG samples, respectively, were chosen based on considerations related
to 
systematic error, which will be described in Appendix~\ref{SSS:largescale}.  In
brief, our finding is that there seem to be
fluctuations of lens number densities that correlate with systematic
errors in the shear for larger scales and lead to a situation where
our systematic uncertainty exceeds the statistical error on the signal.

To determine errors on the lensing signal \ds\ or derived quantities
like $\Upsilon$, we divide the
survey area into 100 equal-area\footnote{For an arbitrary survey
  geometry, it is difficult to achieve equal-area and contiguous
  regions.  We have opted for equal-area regions, roughly 10 per cent
  of which are not contiguous.} jackknife subregions, each of size
$\sim 71$ deg$^2$ or typical length scale $\sim 8.4$ degrees.  The same division of
the area into regions will be used when computing the galaxy
clustering signal, so that we can also estimate the covariance between
the two.  This number of regions was motivated by a desire to balance
two competing effects.  First, we require that the number of regions
be significantly larger than the number of radial bins (18), to reduce
the noise in the covariance matrix \citep{2004MNRAS.353..529H}.  Second, we require that the
region size be larger than the maximum scale used for science.  For
the Main  sample, 30\hmpc\ (comoving) at $z=0.11$ corresponds to 5.3 degrees; for
LRGs, 70\hmpc\ at $z=0.27$ corresponds to 5.2 
degrees.  Thus, our typical region size is 60 per cent larger than the
maximum angular scale used for science.  

 When computing the
covariance for derived quantities such as $\Upsilon$, we estimate
$\Upsilon$ for each jackknife sample to get the covariance
matrix, rather than using the covariance matrix for \ds\ and
propagating errors.  

It is well known that jackknife covariance matrices cannot be used to
get cosmological constraints without some correction due to the finite
level of noise
\citep[e.g.,][]{2004MNRAS.353..529H,2007A&A...464..399H}; this is a
consequence of the fact that the inverse of a noisy, unbiased
estimator of the covariance matrix is {\em not} an unbiased estimator
of the inverse covariance matrix.  We handle this issue by modeling
the covariance matrix to eliminate noise.  (While this might seem
to eliminate the need to make many jackknife regions to reduce noise, as
we have already 
done, we still need the covariance matrix to be reasonably
well-determined in order to easily model it empirically.)  Details of
this approach will be described in Sections~\ref{SS:reslensingerr}
and~\ref{SS:resclusterr}.  However, we note that our results are
insensitive to whether we use the noisy jackknife covariances with a
correction factor \citep{2007A&A...464..399H} after inverting to
obtain the inverse covariance, or whether we use the covariance
matrices that we have modeled to reduce the noise.  This finding
suggests that our results are not significantly impacted by systematics related to
our handling of covariance matrices.

\subsection{Lensing systematic errors}\label{SS:lensingsys}

A thorough treatment of systematic errors with this source catalogue
is in R12.  Here we include only a brief
summary of the issues, along with the impact for this
work. 

\subsubsection{Calibration biases}\label{SS:calibbias}

In \cite{2012MNRAS.425.2610R} we considered several different types of
systematic errors for which we applied corrections and estimated a
total error budget.  In this work we consider the same set of
systematic errors, with the only change being
that the lens samples are at different redshifts, thus changing the
values of many of the systematic errors and their uncertainties.

To summarise briefly, our approach to estimating the systematic error
budget is to consider a full list of systematic errors that affect the
lensing signal calibration.  We correct for our best estimate of
any biases, and assign systematic errors using the following
prescription: for those types of biases that are inherently connected,
we assume that systematic uncertainties add linearly (e.g., two sources of
1 per cent-level uncertainty become a combined 2
per cent uncertainty); for those that are independent, we add them in
quadrature (i.e. in the previous example, the combined uncertainty
would be $\sqrt{2}$ per cent).  

There are three calibration biases related to shear estimation that we
consider to be inherently connected: errors in the correction for PSF
dilution due to the PSF correction method not being perfect; noise
rectification bias; and selection biases (due to our resolution cut
favouring galaxies that are aligned with the shear).  In R12 we
described tests using realistic galaxy simulations
\citep{2012MNRAS.420.1518M} to constrain these three uncertainties
together, which yielded a combined 3.5 per cent uncertainty largely
independent of the lens redshift.

The other calibration biases that we consider to be independent are
the impact of \photoz\ error (as discussed thoroughly in N12); stellar
contamination, which we constrain using space-based data; PSF model
uncertainty; and shear responsivity errors due to incorrect estimation
of the RMS galaxy ellipticity.  Of these, the first is the dominant
one (1, 2, and 3 per cent uncertainty for Main-L5, LRG, and LRG-highz
respectively -- because as shown in N12, the systematic uncertainty is
larger for higher redshift samples, where cosmic variance in the
calibration samples is more important).   Both stellar
contamination and PSF model uncertainties are $\lesssim 0.5$ per cent.
Shear responsivity uncertainty is 1 per cent for all samples.  Thus,
the three shear biases listed previously are the dominant uncertainty
for all samples; when we add up the independent effects in quadrature,
we obtain a 4, 5, and 5 per cent systematic uncertainty for Main-L5,
LRG, and LRG-highz, respectively.

For the purpose of simplifying the modeling, we assume that this final
calibration uncertainty has a Gaussian error distribution, which may
not be quite correct in detail.  Moreover, since the errors were
assessed in the same way for each lens sample, we assume that they are
100 per cent correlated -- i.e. if the calibration is really 4 per
cent too high for Main-L5, then it is 5 per cent too high for LRG and
LRG-highz.  We include this calibration uncertainty in the modeling of
the lensing signal. 

To test our understanding of the calibration biases, we present
several ratio tests \citep{2005MNRAS.361.1287M}, i.e. comparisons of
the signal computed using the same lens samples, but with different
subsamples of the source catalogue.  After we correct for our
understanding of the calibration biases, we should find that the
ratios of these signals are consistent with $1$ within the
errors\footnote{For the combinations of lens and source redshifts used
  here, the predicted differences in those ratios due to reasonable
  variations on our adopted cosmological model are at the 0.1 per cent
  level, well within the errors.}.

\subsubsection{Scale-dependent systematics}

\cite{2010MNRAS.405.2078M} includes a list of scale-dependent
systematic errors that complicate the inference of cluster masses from
the cluster lensing signal.  Fortunately, many such errors are
sufficiently small for galaxy-scale lenses and/or on the $>2$\hmpc\ scales that we use for science
that we can ignore them.  
The scale-dependent systematic errors that we do consider are
intrinsic alignments of galaxy shapes
\citep[e.g.,][]{2004PhRvD..70f3526H}, given that we know some of our
`sources' are really physically associated with the lens and therefore
may tend to point towards the lens.  In principle, this effect can be
quite important if we have no way of removing galaxies that are
physically associated with lenses from our source sample;
fortunately, our \photoz\ are sufficiently good that we are fairly
successful at doing so.  In Sec.~\refresp{\ref{SSS:ia}, we} estimate
the importance of this effect based on the fraction of
physically-associated galaxies as a function of scale (see also \citealt{2012JCAP...05..041B}).

The other main scale-dependent systematic error is the `systematic
shear' described in Sec.~\ref{SS:obs-gglensing}.  While we can use
the procedure described there to correct for it, we also must test the
validity of that correction procedure, which we will do once we
present the results in Sec.~\ref{SSS:largescale}.  Moreover, the systematic
shear is the main factor that determines the maximum scale that we
use; our maximum scales of $\sim 30$ (Main-L5) and $\sim 70$\hmpc\
(LRG, LRG-highz) are motivated by a desire to avoid a situation where
the correction for systematic shear is comparable in size to the real
lensing shear.

\subsection{Galaxy clustering}\label{SS:method-clustering}

We compute the galaxy clustering signals using the same logarithmic
binning size and maximum $R$ as for the lensing, but with a minimum
$R=0.3$\hmpc\ (which provides some measurements below
$R_0$ for estimating $w_{\rmg\rmg}(R_0=4)$).  

The estimation of clustering signals for the lens samples relies on
SDSSpix\footnote{\texttt{http://dls.physics.ucdavis.edu/\mytilde scranton/SDSSPix/}} software to rapidly identify galaxy pairs within the
required separation on the sky.  To compute the galaxy
auto-correlation $w_{\rmg\rmg}(R)$, we begin by computing the 3D galaxy
correlation function $\xi_{\rmg\rmg}$ on a grid of values in $(R,\Pi)$
where $\Pi$ is the comoving line-of-sight separation with respect to
the mean position of the galaxies in the pair.  Our estimator
for the correlation function is a generalisation of that from \cite{1993ApJ...412...64L}, 
\beq
\xi_{\rmg\rmg}(R,\Pi) = \frac{DD-2DR+RR}{RR},
\eeq
using sums of products of weights rather than numbers
of pairs of data-data, data-random, and random-random
pairs\footnote{To reduce the noise, we have many
  more random points than real points.  Here, all numbers such as $DR$
and $RR$ (or their generalisation in terms of  pairs of products of
weights) are properly normalised to account
for this fact.}.  Here, the weights for a given pair come from the
product of the weight for each galaxy in the pair, where the weight
per galaxy is initially defined as in Sec.~\ref{S:data} (e.g.,
Eq.~\ref{E:wlrg}).  For the LRG and LRG-highz samples, there is an
additional factor in the weight, to account for the fact that the
g-g lensing and galaxy clustering measurements would have different
effective weights since the g-g lensing automatically includes a
lensing weight factor that depends on the redshift distribution of the
source galaxies.  This lensing weight is a decreasing function of
redshift, and we include it in the clustering analysis so that the two
measurements will not have different effective  amplitudes
(or even shapes, since the full scale-dependent matter clustering and
non-linear bias can evolve with redshift).  \refresp{We define this weight by taking a grid of lens 
  redshifts starting at our minimum lens redshift and having $\Delta z =
  0.01$, and for each lens redshift on the grid, we 
  use our source sample to identify lens-source pairs in the
  full range of $R$ used for this analysis, with $w(z) = \sum_{ls}
  w_{ls}$.  The weight therefore includes the photometric redshift
  distribution of the sources, and all appropriate weight factors.}  In practise, it turns out
that this weight is not important for Main-L5 because those galaxies
are well below the bulk of the source redshift distribution and
because the Main-L5 redshift distribution is rather narrow.  For LRGs,
it is more important, changing the effective redshift of the
clustering measurement by $\Delta z=0.03$. 

The projected correlation function $w_{\rmg\rmg}(R)$ is formally defined as
\beq
w_{\rmg\rmg}(R) = \int \xi_{\rmg\rmg}(r=\sqrt{R^2+\Pi^2}) \,\rmd\Pi
\eeq
integrated along the entire line-of-sight.  In practise, we compute
$w_{\rmg\rmg}(R)$  via a limited summation,
\beq\label{E:estwgg}
w_{\rmg\rmg}(R) = \Delta\Pi \sum_i \xi_{\rmg\rmg}(R,\Pi_i),
\eeq
in $40$ bins in $\Pi$ that are linearly spaced with $\Delta\Pi=3$\hmpc,
spanning a range $-\pimax\le\Pi\le\pimax$, for $\pimax=60$\hmpc\
(we consider the impact of this choice of \pimax\ in Sec.~\ref{SS:resclustsys}).

To estimate $\upsgg(R)$, we use Eq.~\eqref{E:upsfromw},
replacing $\Sigma$ with $\wgg$,
which requires an estimate of $\wgg(R_0)$.  As for the lensing signal, we
identify a range of scales for which the signal is
approximately a power law, and fit \wgg\ to a power-law on those
scales to estimate $\wgg(R_0)$.  Tests of this determination of
$\wgg(R_0)$ are presented in Sec.~\ref{SSS:calcupsgg}.  We then employ Eq.~\eqref{E:upsfromw} to
get $\upsgg$; the first term, an integral over all scales
above $R_0$, is done via summation.  For each bin, we effectively
assume a constant $w$ within the bin, and we carefully account for
partial bins that fall in the $R$ range of interest. 

As for the lensing signal, we use the division of the lens samples
into equal-area jackknife regions to compute covariance matrices, 
and we present
the results of the jackknife covariances and modeling them to reduce
the noise in Sec.~\ref{SS:resclusterr}.

\subsection{Galaxy clustering systematic errors}

There are several possible systematic errors in the calculation of
the galaxy clustering statistics.

One issue is the handling of the integral in Eq.~\eqref{E:estwgg},
with the finite line-of-sight cutoff.  Naively, we can account for
this in our modeling of the clustering signal by integrating the
theoretical 3D correlation function to the same line-of-sight cutoff.
However, as illustrated by \cite{2007MNRAS.376.1702P}, this approach
is uncertain at the level of tens of per cent on large scales 
by redshift-space distortions (RSD), and a linear treatment \citep{1987MNRAS.227....1K}
is not likely to be adequate on these scales \citep[e.g.,][]{2011MNRAS.417.1913R}.  Fortunately,
as illustrated by \cite{2010PhRvD..81f3531B}, the uncertainty induced by the
redshift-space distortions is far less important for the observable
that we use for science, $\upsgg$.  As shown in figure 8 (right panel)
in that paper, the impact of RSD and the finite line-of-sight cutoff
is reduced from $\sim 30$ per cent bias on $\wgg$ to
$\sim 5$ per cent bias on $\upsgg$ at the maximum scale that we use for science, and is very
small below $30$\hmpc.  We could apply a correction for this small, residual
systematic error, but as described in Sec.~\ref{SSS:combmodel}, a
combined correction for this effect and others (e.g., the lensing
window, which goes in the opposite direction) is so small as to be
negligible for our analysis.

Because $\upsgg$ is a partially compensated statistic, it is also
less sensitive to large-scale density fluctuations that can shift
$\wgg$ up and down.  We will demonstrate the effect of this on the
cosmic variance component of the errors in Sec.~\ref{SS:resclusterr}.  The same argument is
true for the integral constraint, which will lead to a constant offset
in $\wgg$ which goes away when computing $\upsgg$.

There are also a question of how sample definition choices affect the
measured statistics.  \cite{2010ApJ...710.1444K} consider several such effects for $\wgg$,
including the method of distributing the random points in redshift,
and 
the way of handling fibre collisions in the weighting.  Their
conclusion that these issues are only important at the few per cent
level at most is also applicable to our results, and thus this systematic error is
subdominant to the statistical and systematic uncertainties in the
lensing signal.

%

\section{Results of lensing measurements}\label{S:results-lensing}

In this section we present the galaxy-galaxy lensing measurements
(Sec.~\ref{SS:reslensingobs}), the error estimates
(Sec.~\ref{SS:reslensingerr}).  Tests for systematic errors
are in Appendix~\ref{SS:reslensingsys}.

\subsection{Observations}\label{SS:reslensingobs}

\begin{figure}
\includegraphics[width=\columnwidth]{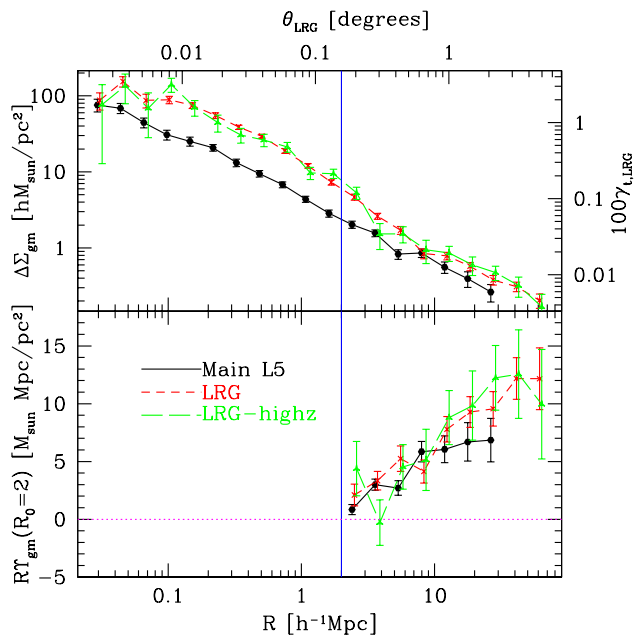}
\caption{{\em Top:} Observed lensing signal $\ds_{\rmg\rmm}(R)$ for all three
  lens samples.  The vertical line indicates the minimum scale used
  for cosmological constraints.  The axes on the top and right
  indicate the angular scale $\theta$ and the tangential shear
  $100\gamma_t$ for the LRG sample.  {\em Bottom:} $R\Upsilon_{\rmg\rmm}(R;R_0=2)$
  for all three samples as labeled on the plot.  \label{F:showlensing}}
\end{figure}

The lensing signals for all three samples are shown in
Fig.~\ref{F:showlensing}.  This figure shows the observable quantity,
$\ds$, and also the quantity used for cosmological constraints,
$\upsgm(R_0=2h^{-1}\text{Mpc})$, plotted as $R\Upsilon$ for easier
viewing on a linear scale. Clearly the $S/N$ of the
observable is quite high -- typically $\sim 25$ averaged over all scales, using
\beq\label{E:sn}
\frac{S}{N} = \left({\bmath x}^T {\bf C}^{-1} {\bmath x}\right)^{1/2},
\eeq
where ${\bmath x}$ is the vector of $\ds$ values in each radial bin
and ${\bf C}$ is their covariance matrix, to account for correlations
between bins. The shear is well-constrained down to a level
of $\sim 5\times 10^{-5}$ (at $\sim 5$ degree angular separations), and the results are statistically consistent with
previously-published ones for LRGs with $R<4$\hmpc\ \citep{2006MNRAS.372..758M}.
Comparison with previous Main-L5 observations is complicated by our
imposition of a redshift cut $z<0.155$, and this is the first such
galaxy-galaxy lensing observation for LRG-highz.

$\upsgm$ is a lower $S/N$ quantity, with
total average $S/N$ of 11, 14, and 8 for Main-L5, LRG, and LRG-highz, averaged over the 
range of scales ($2<R<70$\hmpc) shown on the plot.

A detailed discussion of tests for systematic errors in the lensing signal is in
Appendix~\ref{SS:reslensingsys}.


\subsection{Covariance matrices}\label{SS:reslensingerr}

Here, we present the error estimates for the \upsgm\ results shown above.  As stated in Sec.~\ref{SS:obs-gglensing},
the noisiness of the jackknife covariance matrices requires some
correction in order to get cosmological parameter estimates.  Rather
than modifying the procedure for using them to get confidence
intervals, as in \cite{2004MNRAS.353..529H} or
\cite{2007A&A...464..399H}, we instead model the matrices to make 
noiseless versions.

The process begins by modeling the diagonal
terms of the covariance matrix as a function of $R$.  We refer to the
covariance matrix for \upsgm\ as ${\bf C}_{\rmgm}^{(\Upsilon)}$ with
elements corresponding to radial bins $i$ and $j$ of ${\bf C}_{\rmgm}^{(\Upsilon)}(R_i,R_j)$.  Our
  smooth model is
\beq\label{E:upsgmdiag}
{\bf C}_{\rmgm}^{(\Upsilon)}(R_i,R_i) = A R^{-2} [1 + (R/R_t)^2],
\eeq
with an amplitude $A$ and a turnover radius $R_t$.
This two-parameter model is motivated as follows: on all scales, we
expect sampling variance to be minimal because of the large area 
and the compensated
nature of $\Upsilon$, so shape noise should be the dominant source
of error.  The shape noise variance scales like $R^{-2}$ for
logarithmically spaced annular bins, and we fit for the amplitude $A$
of this term.  However, as shown in 
\cite{2009PhRvD..80l3527J}, above some radius the shape noise fails to
decrease as rapidly with $R$, in the regime
where $R$ is significantly larger than the typical separation between
lenses.  In that case, the lens-source pairs in the annular bin
include many of the same sources around nearby lenses, so the shape
noise does not decrease by adding more lenses.  The term in brackets
in Eq.~\eqref{E:upsgmdiag} represents this flattening of
the errors with scale. (There will also be a corresponding increase in bin-to-bin
correlations on those scales, as will shortly be apparent.)

\begin{figure}
\includegraphics[width=0.935\columnwidth]{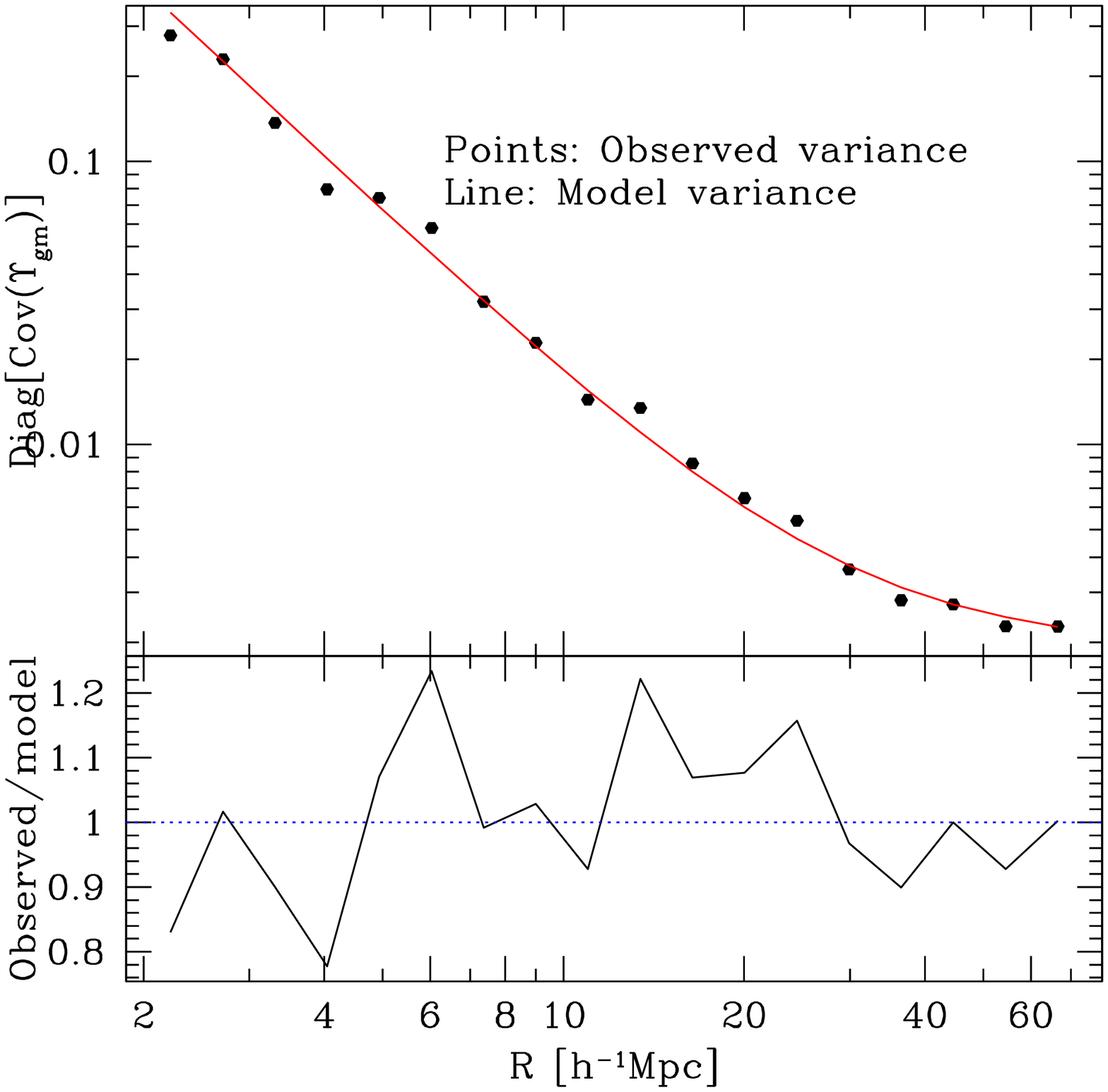}
\includegraphics[width=0.935\columnwidth]{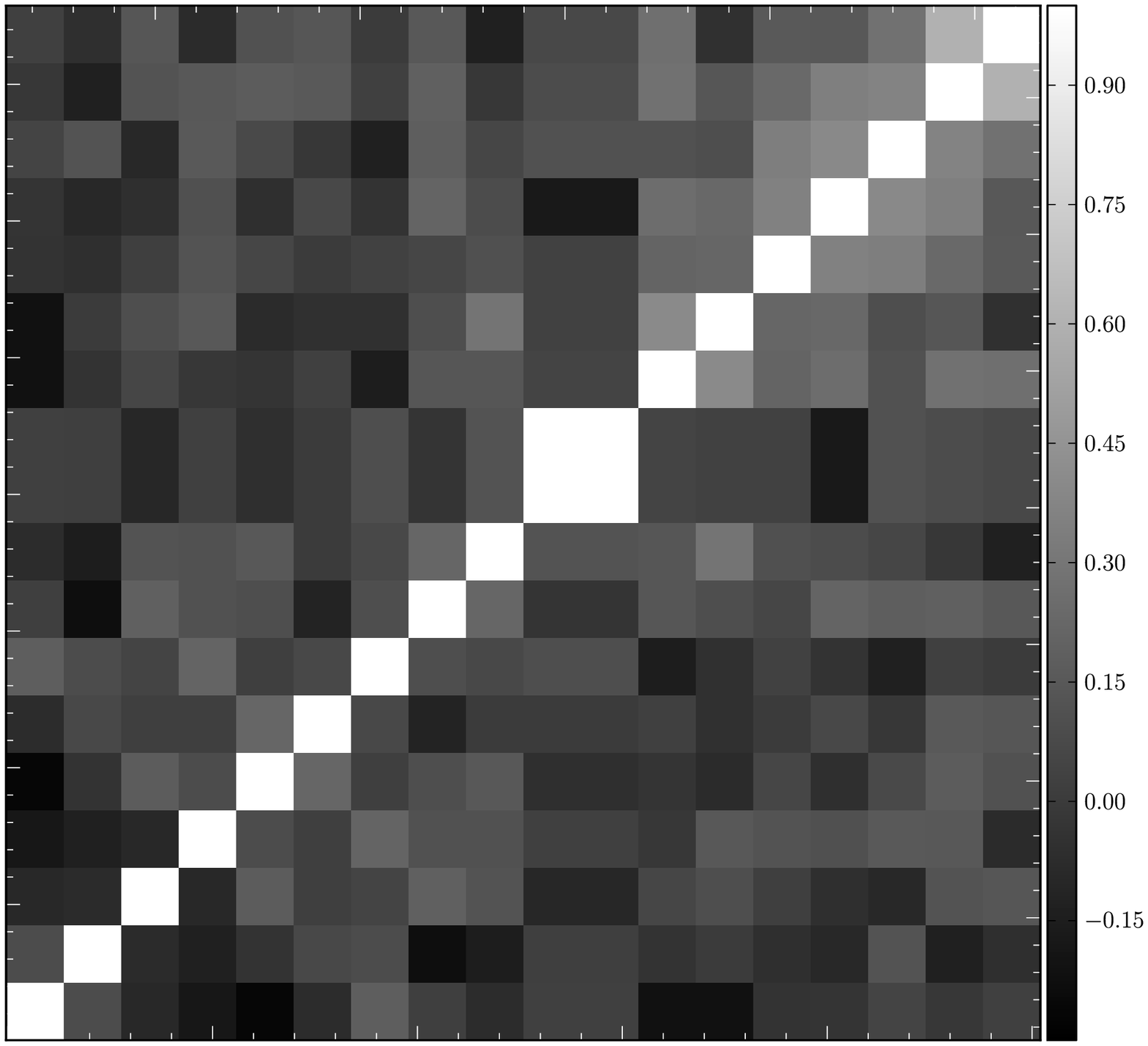}
\includegraphics[width=0.935\columnwidth]{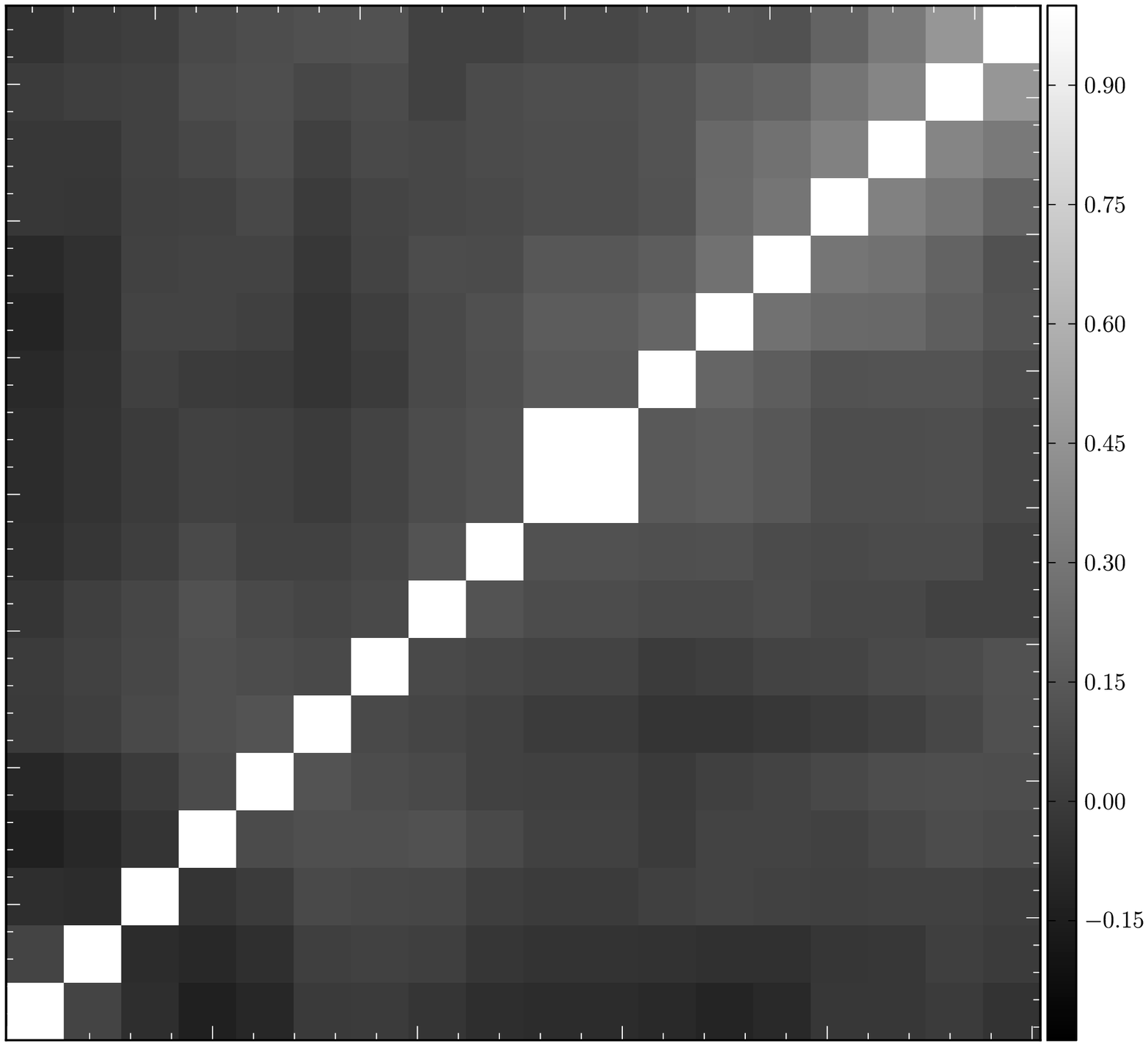}
\caption{{\em Top:} A comparison of the jackknife covariance matrix
  diagonal terms for \upsgm\ for the LRG lens sample with the model covariance matrix terms as a function
  of transverse separation $R$.  {\em Middle:} Jackknife correlation matrix for
  \upsgm, again for LRG lenses.  {\em Bottom:} Smoothed correlation matrix for \upsgm.\label{F:covgm}}
\end{figure}

Our approach is to  model the diagonal elements of the covariance
matrix by directly fitting for $(A,
R_t)$ for each lens sample using
$\chi^2$ minimisation, doing an unweighted fit for $\log{
  C_{\rmgm}^{(\Upsilon)}(R_i,R_i)}$ as a function of $\log R$.  The
scale on which the term in brackets in Eq.~\eqref{E:upsgmdiag} becomes
important is $R_t = (8, 31, 41)$\hmpc\ for the three samples.  This
trend of $R_t$ is unsurprising given the trends in lens number
density for the three samples.  In
the top panel of Fig.~\ref{F:covgm}, we show a comparison between the
jackknife covariance diagonal terms and those from the model, for the
LRG sample.  As shown, the \refresp{RMS} level of fluctuations of the jackknife 
variances compared to those in the model is 12 per
cent.

Next, we model the off-diagonal terms, which are also somewhat noisy.
Off-diagonal terms can arise due to (a) cosmic variance (not very
significant for this sample), (b) correlated shape noise due to the
large $R$ compared to the separation between lenses, and (c) the fact
that $\Upsilon(R)$ at some radius depends on the $\ds(R_0)$, which
tends to anti-correlate bins at $R\sim R_0$ with each other.  Since there
are several sources of correlations, they are not as simple to
model analytically.  Thus, we take a non-parametric approach, by
generating the correlation matrix, i.e. ${\bf C}_\mathrm{corr}$, defined by
\beq
{\bf C}_{\mathrm{corr},i,j} = \frac{{\bf
    C}_{\rmgm}^{(\Upsilon)}(R_i,R_j)}{\sqrt{{\bf
    C}_{\rmgm}^{(\Upsilon)}(R_i,R_i) {\bf
    C}_{\rmgm}^{(\Upsilon)}(R_j,R_j) }}
\eeq
We then apply a boxcar smoothing algorithm with a length of 3 bins in
radius to this matrix, to reduce the noise.  
The middle
and bottom panels of Fig.~\ref{F:covgm} show the unsmoothed and
smoothed correlation matrix for the LRG sample.  As shown, the
smoothing has not resulted in any significant modification of the
apparent real trends, but has eliminated the majority of the noise.

\section{Results of clustering measurements}\label{S:results-clustering}

In this section we present the galaxy clustering measurements
(Sec.~\ref{SS:resclustobs}), the error estimates
(Sec.~\ref{SS:resclusterr}), and cross-covariance with the lensing results
(Sec.~\ref{SS:rescross}).  Tests for systematic errors in the
clustering measurements are in Appendix~\ref{SS:resclustsys}.

\subsection{Observations}\label{SS:resclustobs}

\begin{figure}
\includegraphics[width=\columnwidth]{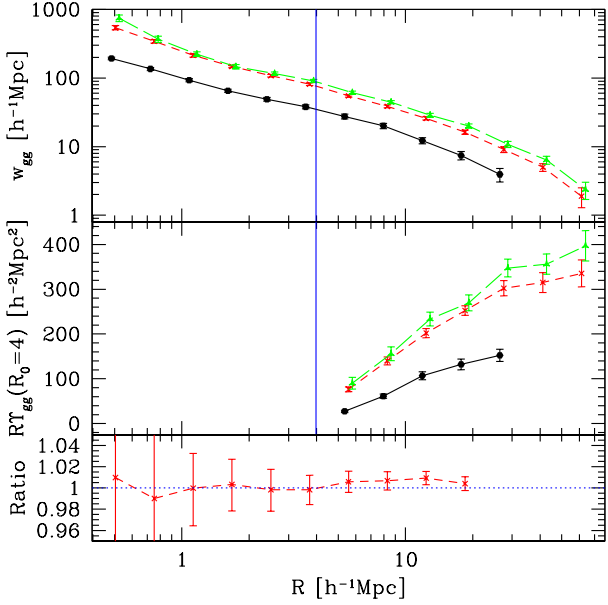}
\caption{{\em Top:} Observed clustering signal $w_{\rmg\rmg}(R)$ for
  all three lens samples, including lens-weighting factors for LRG and
  LRG-highz.  The vertical line indicates the minimum
  scale used for cosmological constraints.  Line colours and types
  indicate the sample, using the same scheme as
  Fig.~\ref{F:showlensing}. {\em Middle:}
  $R\Upsilon_{\rmg\rmg}(R;R_0=4)$ for all three samples as labeled on
  the plot.  {\em Bottom}: For the LRG sample, the ratio of \wgg\ from
  \protect\cite{2005ApJ...621...22Z} to that from our work without lens-weighting, after
  accounting for the different radial binning. \label{F:showclust}}
\end{figure}

The clustering signals for all three samples are shown in
Fig.~\ref{F:showclust}.  This figure shows the observable quantity,
$w_{\rmg\rmg}$, and also the quantity used for cosmological constraints,
$\Upsilon_{\rmg\rmg}(R_0=4h^{-1}\text{Mpc})$ (plotted as $R\Upsilon$ for easier
viewing on a linear scale). Clearly the $S/N$ of the
observable is quite high, significantly more so than for the lensing
observable.  $\Upsilon$ gives a 
total average $S/N$ of 19, 33, and 32 (Main, LRG, LRG-highz) when
averaged over scales $R>4$\hmpc\ using Eq.~\eqref{E:sn}.  

As discussed in Sec.~\ref{SS:method-clustering}, these results include
a redshift-dependent weighting factor so that the effective redshift
will be the same as for the galaxy-galaxy lensing measurement.  Thus,
they cannot be directly compared with previous measurements of LRG and
LRG-highz sample clustering without checking the effect of this
weighting.  We find that for the LRG (LRG-highz) sample, inclusion of the lens-weighting
factor has lowered the amplitude of $w_{\rmg\rmg}$ by 4 (2) per
cent\footnote{The expected sign and magnitude of the effect is not
  completely clear; for a passively evolving population the sign
  should in fact be the opposite.  However, we have split the LRG
  samples into redshift slices and confirmed that within the redshift
  range of the LRG sample, the amplitude of $w_{\rmg\rmg}$ evolves by
  as much as 10 per cent, with lower amplitude at lower redshift,
  consistent with our findings with lens-weighting included.} and increased the errors by $\sim 10$ per cent.  Thus, if we had not included it, then with
the same lensing signal but a higher clustering signal we would have inferred a lower
$\sigma_8$ by 2 (1) per cent when analysing these samples, which is of similar
order as other calibration factors we have considered and therefore
validates our choice to include this redshift-weighting properly.  

The bottom
panel of Fig.~\ref{F:showclust} shows the ratio of the measured signal for the LRG sample from
\cite{2005ApJ...621...22Z} to our measurement {\em without the lens-weighting}, with the differences
being due to our use of nearly twice as much area (7131 versus 3836
degrees$^2$) and the different line-of-sight integration lengths ($\pm
60$
versus $\pm 80$ \hmpc).  Even with these differences, the results
agree at the 1 per cent level\refresp{; the results agree to well
  within the naively propagated errors because the results are
  actually correlated to some extent, and this agreement shows that we
  have not done anything substantively different from the perspective
  of systematic errors}. 
Additional tests for systematic errors in the
clustering measurements are in Appendix~\ref{SS:resclustsys}.
 
\subsection{Covariance matrix}\label{SS:resclusterr}

Here, we present the error estimates for the quantities shown above,
in particular, the $\upsgg$ results.  We follow the approach from
Sec.~\ref{SS:reslensingerr} of modeling the covariance matrices to
reduce the noise, again showing results for the LRG sample
as an example.  

The process begins by modeling the diagonal
terms of the covariance matrix as a function of $R$.  We refer to the
covariance matrix for \upsgg\ as ${\bf C}_{\rmgg}^{(\Upsilon)}$ with
elements corresponding to radial bins $i$ and $j$ of ${\bf C}_{\rmgg}^{(\Upsilon)}(R_i,R_j)$.  Our
  smooth four-parameter model is
\beq\label{E:upsggdiag}
{\bf C}_{\rmgg}^{(\Upsilon)}(R_i,R_i) = A_1 R^{\alpha_1} + A_2 R^{\alpha_2},
\eeq
a sum of a shallower and a steeper power-law, which dominate on larger
and smaller scales respectively.  The choice of this functional form
comes from the fact that for \upsgg\ there is an additional noise component on
small scales due to propagated uncertainty in $\wgg(R_0)$.

\begin{figure}
\includegraphics[width=0.935\columnwidth]{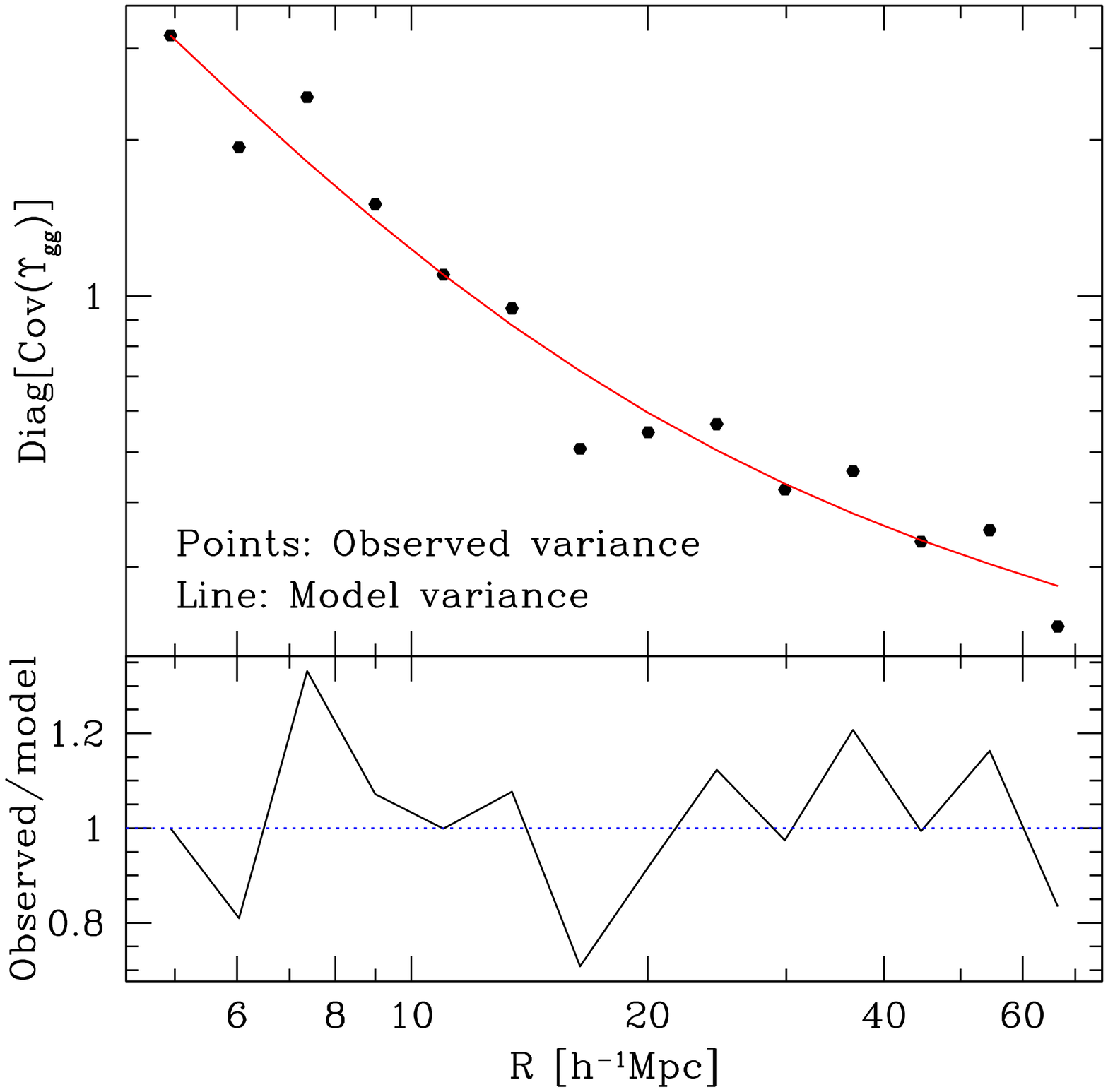}
\includegraphics[width=0.935\columnwidth]{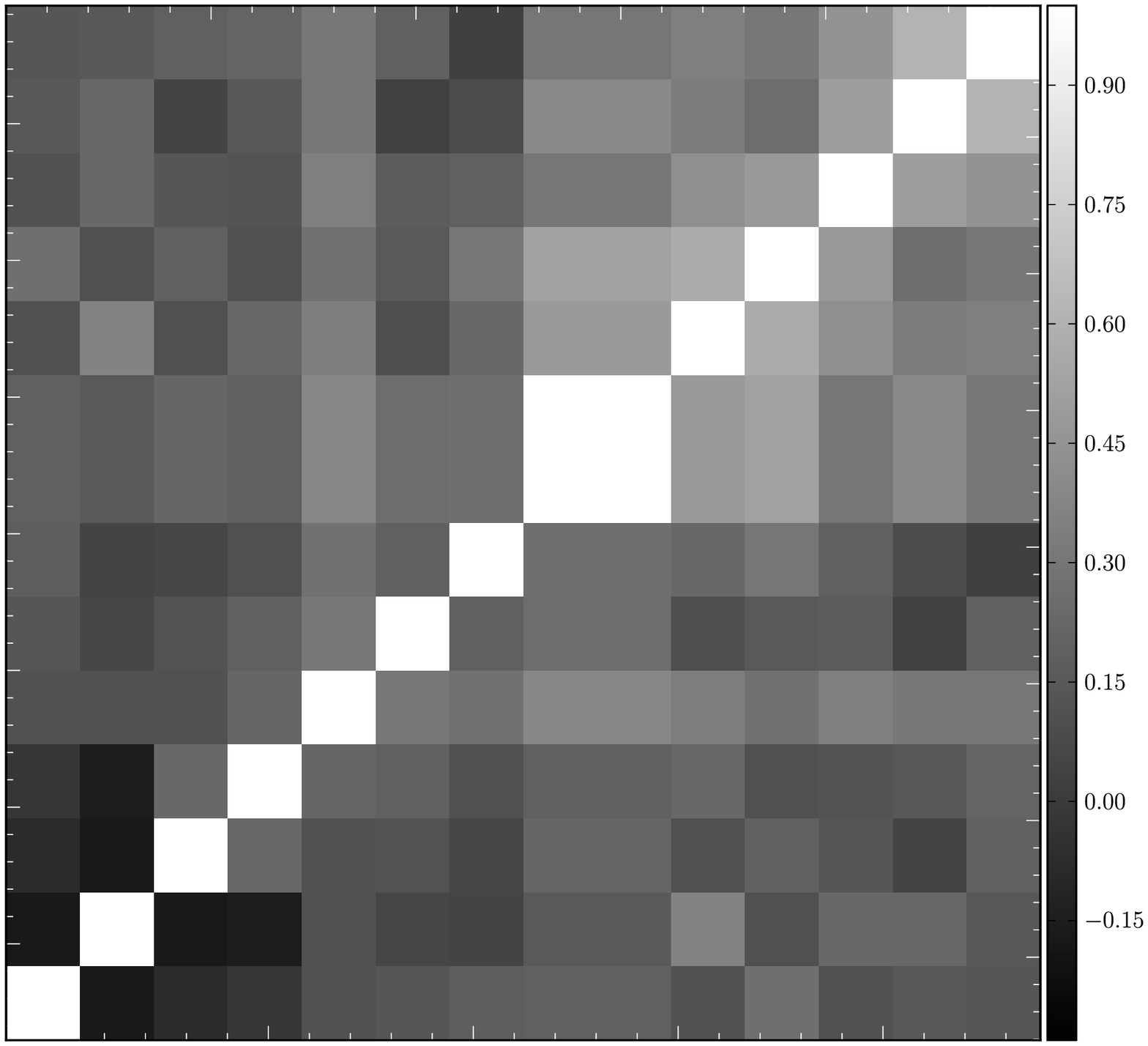}
\includegraphics[width=0.935\columnwidth]{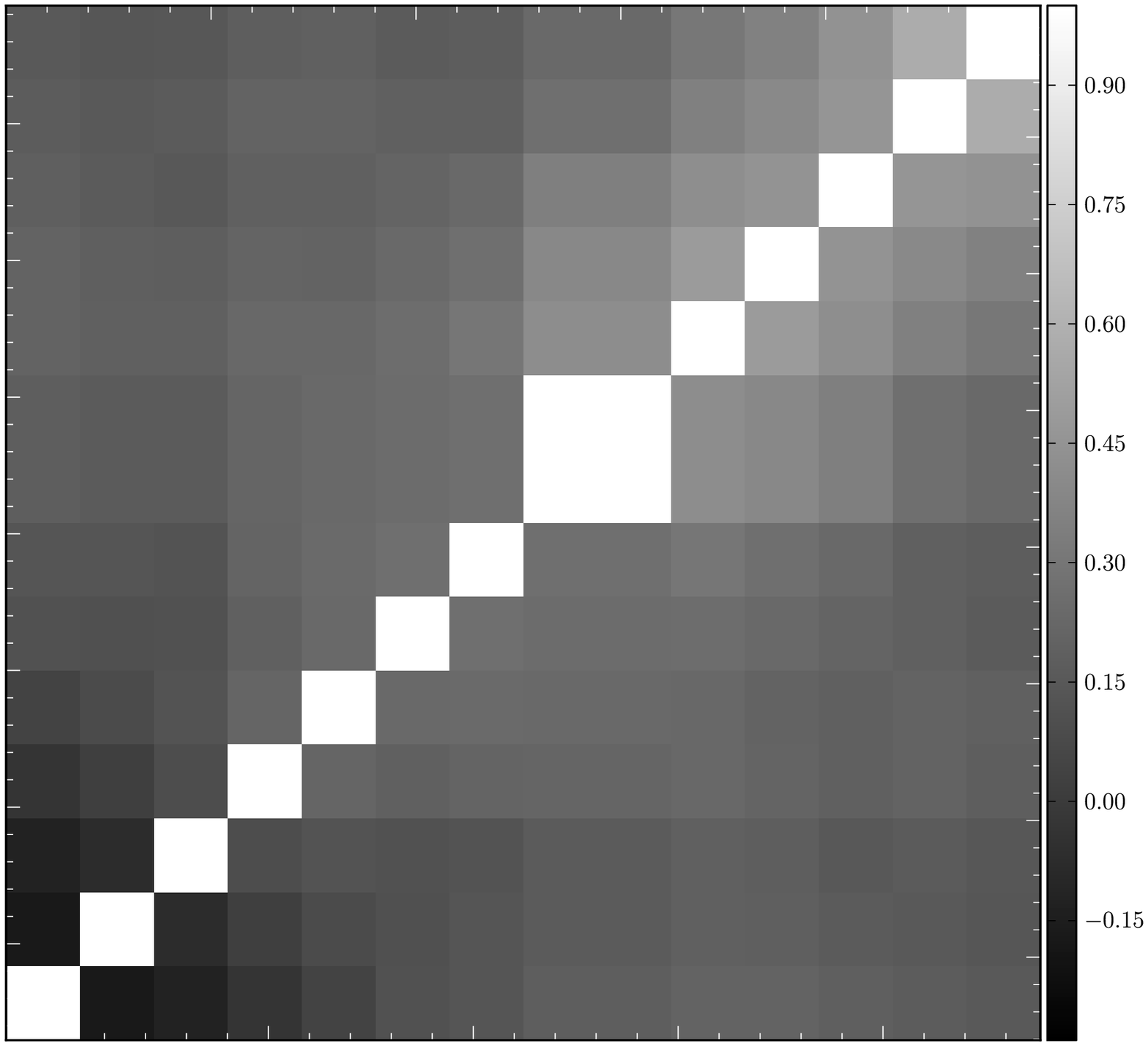}
\caption{{\em Top:} A comparison of the jackknife covariance matrix
  diagonal terms for \upsgg\ for the LRG sample, with the model covariance matrix terms as a function
  of transverse separation $R$.  {\em Middle:} Jackknife correlation matrix for
  \upsgg\ for the LRG lens sample.  {\em Bottom:} Smoothed correlation matrix for \upsgg.\label{F:covgg}}
\end{figure}

We model this term by directly fitting for
$(A_1, \alpha_1, A_2, \alpha_2)$ for each lens sample using
$\chi^2$ minimisation, doing an unweighted fit for $\log{\bf
  C}_{\rmgg}^{(\Upsilon)}(R_i,R_i)$ as a function of $\log R_i$.  In
the top panel of Fig.~\ref{F:covgg}, we show a comparison between the
jackknife covariance diagonal terms and those from the model, for the
LRG sample.  As shown, the level of fluctuations of the observed
variances compared to those in the model is 17 per
cent (but there is a fairly extreme outlier; excluding that
one noticeably decreases the estimated scatter).  Smoothing the
diagonal terms to reduce the influence of this outlier is
important; too low a variance would result in that bin
unfairly dominating the fits.

Next, we model the off-diagonal terms, which are also somewhat noisy.
Off-diagonal terms can arise due to cosmic variance (not very
significant for this compensated statistic \upsgg), but also a small
contribution from the fact
that $\upsgg(R)$ at some radius depends on $\wgg(R_0)$, which
tends to anti-correlate bins at $R\sim R_0$ with each other.  Since there
are several sources of these correlations, they are not as simple to
model analytically.  Thus, we take a non-parametric approach, by
generating the correlation matrix, i.e. ${\bf C}_\mathrm{corr}$, defined by
\beq
{\bf C}_{\mathrm{corr},i,j} = \frac{{\bf
    C}_{\rmgg}^{(\Upsilon)}(R_i,R_j)}{\sqrt{{\bf
    C}_{\rmgg}^{(\Upsilon)}(R_i,R_i) {\bf
    C}_{\rmgg}^{(\Upsilon)}(R_j,R_j) }}
\eeq
We then apply a boxcar smoothing algorithm with a length of 3 bins in
radius to this matrix, to reduce the noise.  
The middle
and bottom panels of Fig.~\ref{F:covgg} show the unsmoothed and
smoothed correlation matrix for the LRG sample.  As shown, the
smoothing has not resulted in any significant modification of the
apparent real trends, but has eliminated the majority of the noise.

\begin{figure}
\includegraphics[width=\columnwidth]{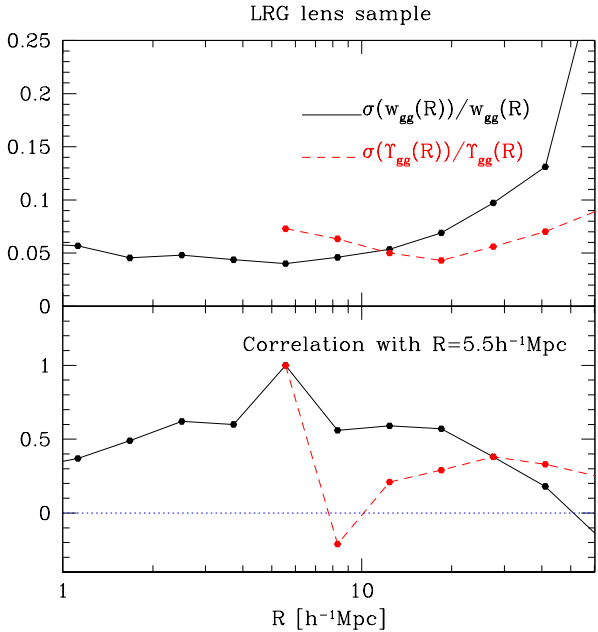}
\caption{Error comparison for \wgg\ (black solid lines) and \upsgg\
  (red dashed lines) for the LRG lens
  sample.  {\em Top:} Standard deviation of these statistics (i.e.,
  the square root of the diagonal covariance matrix elements) as a
  function of $R$, divided by the
  signal itself.  {\em Bottom:} Correlation of the data at each bin
  in radius with the bin above $R=R_0$\hmpc. \label{F:upswcov}}
\end{figure}

We also compare the covariance matrices for \wgg\ and \upsgg, to check
that they behave in the way that we expect with respect to reduced
cosmic variance in the latter due to its compensated nature
(Sec.~\ref{S:theory}).   Fig.~\ref{F:upswcov} illustrates this
difference, using the LRG sample as an example.  Here we have rebinned
the data and covariances by a factor of two, because we do not have a
smoothed version of the covariance matrix for \wgg.  As shown, the
relative error (top panel) on \upsgg\ is larger near $R_0$, because
\upsgg\ is defined to be near zero there, but on large scales, \wgg\
has a larger fractional error due to cosmic variance.  We also study
the correlation properties of these statistics.  As shown (bottom panel), \wgg\
exhibits larger correlations between nearby bins.  This is
a consequence of cosmic variance: large-scale modes can coherently
shift \wgg\ up or down, resulting in bin-to-bin correlations.  In
contrast, \upsgg\ shows less significant correlation patterns.

We also compare these error properties versus the expected ones from
the simulations.  For the simulations, the volume has been divided
into $1.5\times 1.5\times 0.3=0.67 (\mathrm{Gpc}/h)^3$ sub-volumes,
almost exactly the size of our observed volume but with slightly
different geometry.  There are $40$ 
such sub-volumes, which we use to estimate covariance
matrices by comparing the signals computed in each one.  This test
serves as a check on the jackknife method that we carry out on the
real data.  We find that the scaling of
$\sigma(\wgg)/\wgg$ in the simulations and data is extremely similar
for all scales that we use.  For \upsgg\, the same is generally true,
though the errors near $R_0$ seem to be $\sim 20$ per cent larger in the simulations than
in the data, but the opposite is true at higher $R$, with the two in agreement 
around 8\hmpc.  This may be attributed to the method of determining
$\wgg(R_0)$, which differs slightly between the data analysis and
simulations.  Nonetheless, the comparison between errors in the
simulations and in the real data is very similar, and should alleviate
any concerns about the jackknife errorbars used in practise. 
We have also confirmed that the correlation properties for \wgg\ and
\upsgg\ in the simulations are consistent with those we observe in the
real data (bottom of Fig.~\ref{F:upswcov}).

\subsection{Cross-covariance with lensing}\label{SS:rescross}

Using the jackknife resampling, we are able to compute the
cross-covariance between the lensing and clustering observations.  Our
findings are that on all scales, these correlations are consistent
with zero, except possibly on the largest scales where they 
reach 10--15 per cent. This result is unsurprising given that the
dominant source of error in our observations is the lensing shape
noise.  So, for the purpose of cosmological parameter constraints, we
set the cross-terms between lensing and clustering in the covariance
matrix to precisely zero.

For future work with surveys that are not as limited by shape noise,
these covariances will be important to model accurately. 

%

\section{Cosmological parameter constraints}\label{S:paramconstraints}

In this section, we present cosmological parameter constraints derived
from the data that were shown in the previous section.

\subsection{Constraints with these data alone}\label{SS:paramconstraints-alone}

\begin{figure*}
\includegraphics[width=0.935\columnwidth]{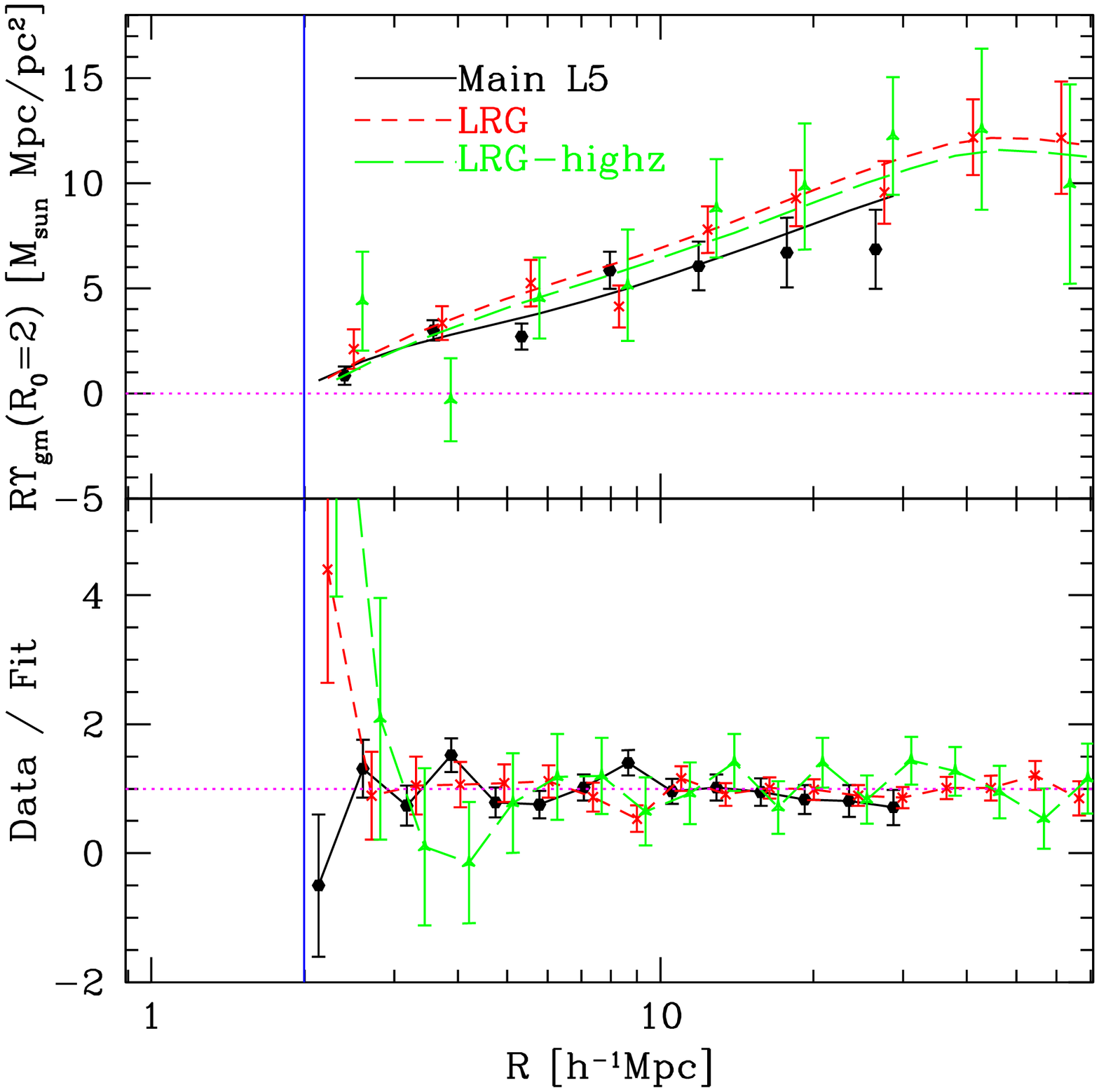}
\includegraphics[width=0.935\columnwidth]{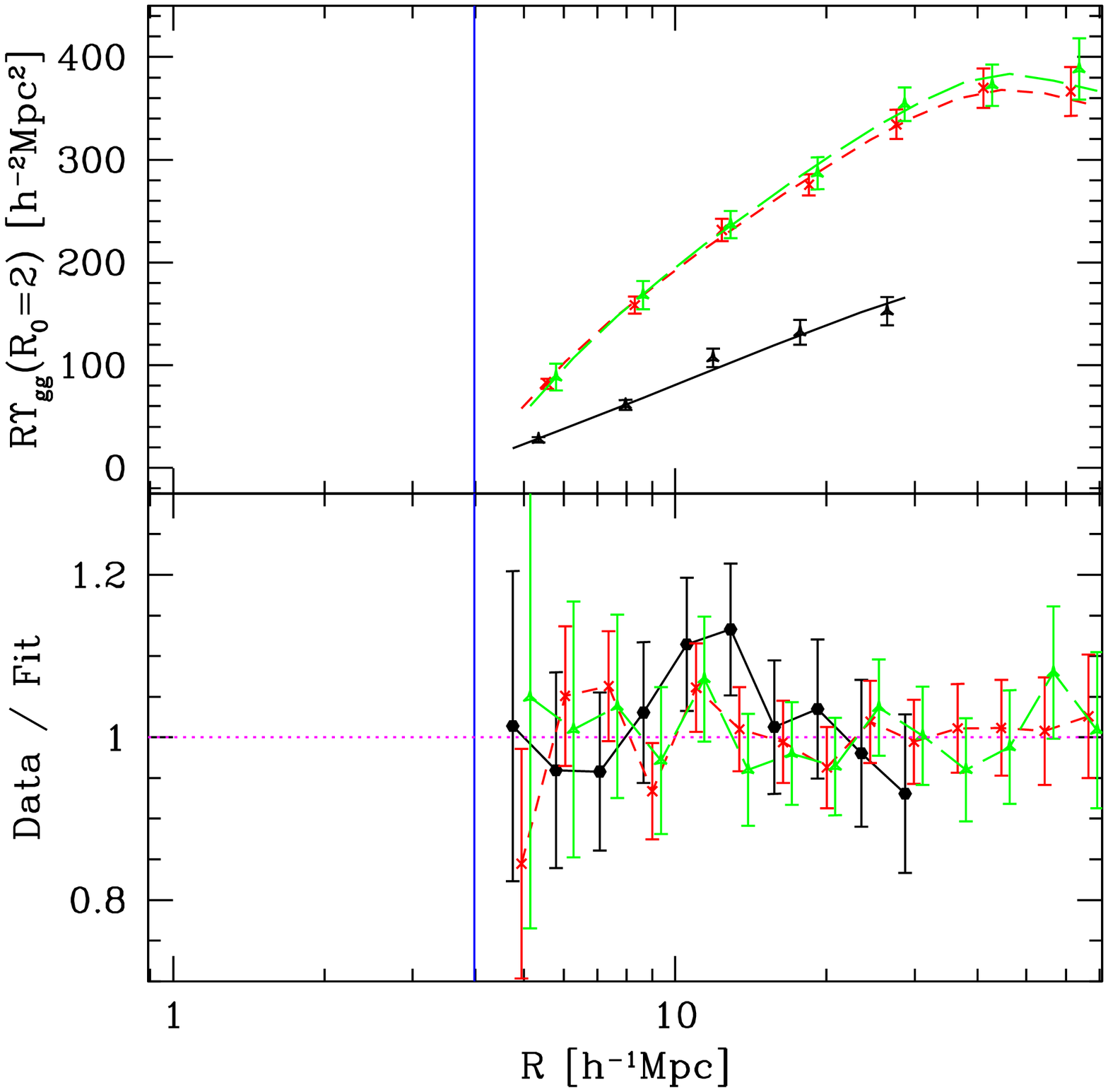}
\caption{{\em Top left:} Observed $R \upsgm$ with best-fit signals
  from fits 1--3 in Table~\ref{T:fits}; for these fits, the data were
  fit separately for each sample.  Data have been rebinned for easier
  viewing.  {\em Bottom left:} Ratio of observed to best-fit signal
  from the top panel, using the original narrower binning as for the
  actual fit. {\em Right:} Same as left, but for clustering $R\upsgg$.\label{F:bestfitsig}}
\end{figure*}
\begin{figure*}
\includegraphics[width=1.6\columnwidth]{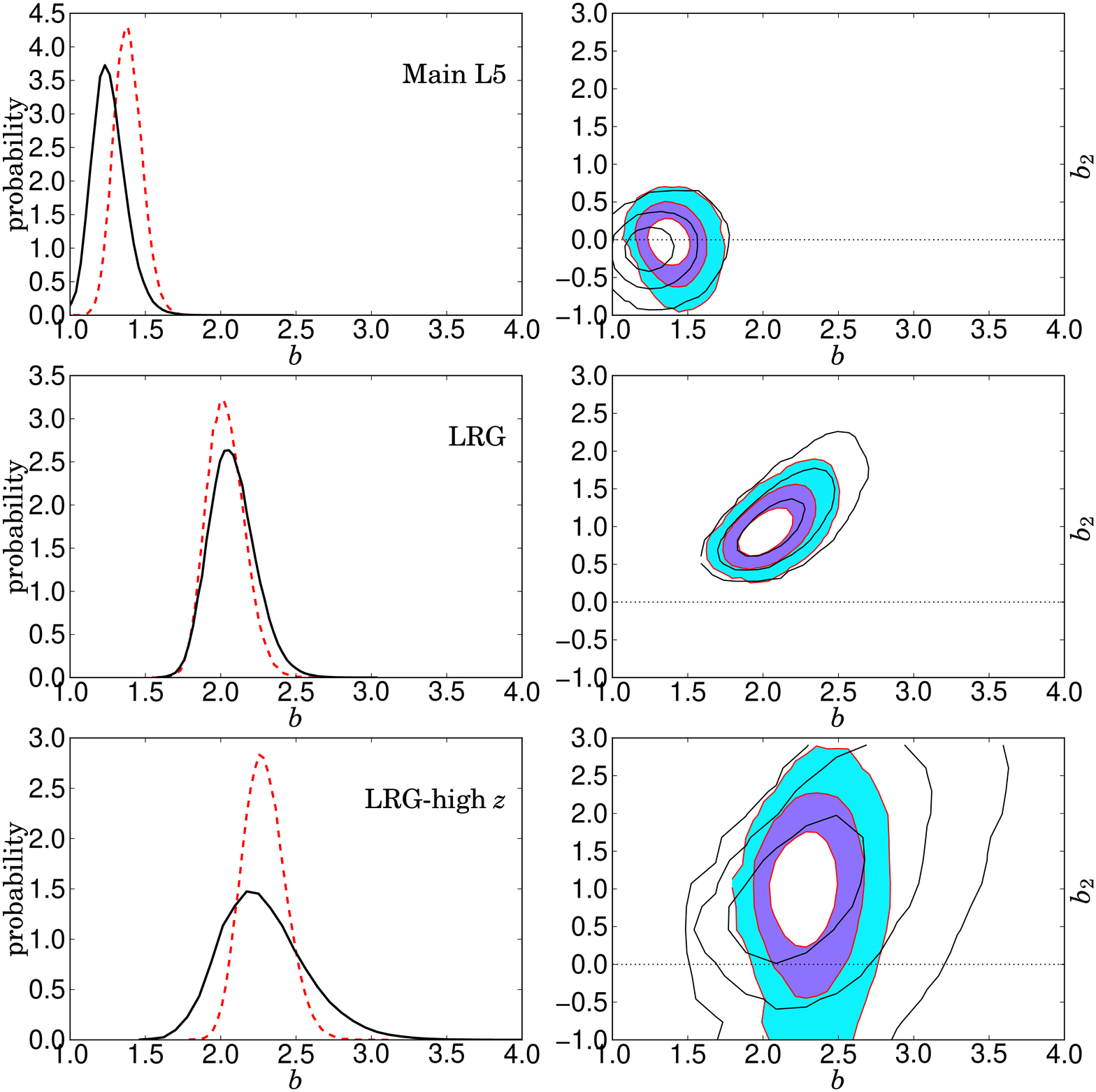}
\caption{Best-fitting galaxy bias parameters for Main-L5 (top), LRG
  (middle), and LRG-highz (bottom).  {\em Left:} The posterior
  probability distribution for
  the large-scale galaxy bias $b$, marginalised over other parameters
  including the non-linear
  bias parameter $b_2$.  The solid black lines show the results for fits
  1--3, fitting for each sample separately; the dashed red lines are the
  results from fit 4, jointly fitting all samples. {\em Right:}
  Contour plots for the  large-scale bias $b$ versus
  non-linear bias parameter $b_2$ 
  ($1$, $2$, and $3\sigma$).  The black line contours show the results
  from fits 1--3, fitting for each sample separately; the coloured
  contours show the results from fit 4, jointly fitting all samples.
 \label{F:bplots}}
\end{figure*}
We use the standard \texttt{cosmomc} \citep{2002PhRvD..66j3511L} package
to make statistical inferences from our data. The package has been
extended with a module that models our signals as described above and
provides the likelihood that can be turned into posterior probability
either alone or in conjunction with another dataset.  We use the
standard \texttt{cosmomc} parametrisation of cosmology with flat priors on the
following parameters $\omega_b=\Omega_b h^2$ (baryon density),
$\omega_{\rm cdm}$ (cold dark matter density), $\theta$ (angular
diameter distance of the surface of last scattering, a proxy for
Hubble's constant), $\tau$ (optical depth to the surface of last
scattering), $w$ (dark energy equation of state), $n_s$ (spectral
index of primordial fluctuations), $A$ (amplitude of the primordial
fluctuations; however, we quote our results in terms of the more
commonly-used parameter $\sigma_8$). \refrespt{On top of these priors,
  we include a prior that is flat in $\sigma_8^5$,
based on our calibration on simulations in
Appendix~\ref{SS:paramconstraints-sims}.  W}e vary parameters specific to our dataset:
two bias parameters for each dataset and a common calibration
parameter $c$ \refrespt{as discussed in Sec.~\ref{SSS:combmodel} and~\ref{SS:calibbias}}.

We start by performing the analysis using \refresp{SDSS lensing and
  clustering data without any external priors from e.g. CMB data; this
  analysis is carried out} both in
individual redshift bins and combined. For these tests, we fix the
cosmological parameters 
to $\Omega_m=0.25$, $n_s=0.96$, and $h=0.7$,  only varying $\sigma_8$, the bias
parameters, and the lensing calibration parameter.
 We do this fit for each sample separately, and for all
three together (Table~\ref{T:fits}, fits 1--4).
  
\begin{table*}
\begin{tabular}{ccccccccccccc}
\hline\hline
$\!\!\!\!$Fit & sample & $\sigma_8$ & $\Omega_m$ & $b$  & $b_2$ 
& $b$ & $b_2$ & $b$  & $b_2$$\!\!\! $&
$\chi^2$ & $\nu$ & $\!\!\!\!\!\!\!p(>\chi^2)$\\
 & & & & (L5) & (L5) & (LRG) &
 (LRG) & $\!\!\!\!\!$(LRG-highz)$\!\!\!\!\!$ &
 $\!\!\!\!\!\!\!$(LRG-highz)$\!\!\!\!\!$ & & & \\
\hline
$\!\!\!$1 & L5 & $0.89^{+0.07}_{-0.08}$ & \textbf{0.25} & $1.25\pm 0.05$ & $\!\!\!-0.12\pm 0.18\!\!\!$ & - & - & - &
- & $29.8$ & $19$ & $0.06$ \\
$\!\!\!$2 & LRG & $0.79\pm 0.06$ & \textbf{0.25} & - & - & $2.07\pm 0.05$ & $\!\!\!0.98^{+0.28}_{-0.24}\!\!\!$ & - &
- & $19.3$ & $27$ & $0.86$ \\
$\!\!\!$3 & $\!\!\!\!\!\!\!\!\!$LRG-highz$\!\!\!\!\!\!\!$ & $0.81\pm 0.10$ & \textbf{0.25} & - & - & - & -
& $2.26\pm 0.06$ & $\!\!\!0.94^{+0.66}_{-0.54}\!\!\!$ & $20.5$ & $27$ & $0.81$ \\
$\!\!\!$4 & All & $0.80\pm 0.05$ & \textbf{0.25} & $1.38\pm 0.05$ &
$\!\!\!-0.02\pm 0.20\!\!\!$ & $2.03\pm 0.05$ & $\!\!\!0.94^{+0.24}_{-0.20}\!\!\!$ & $2.28\pm 0.06$ &
$\!\!\!1.00^{+0.46}_{-0.50}\!\!\!$ & $71.3$ & $77$ & $0.66$ \\
$\!\!\!$5 & All & $0.76\pm 0.08$ &
$\!\!\!\!\!\!0.269^{+0.038}_{-0.034}\!\!\!\!$ & $1.46\pm 0.06$ & $\!\!\!-0.06\pm
0.26\!\!\!$ & $2.15\pm 0.07$ & $\!\!\!0.96^{+0.36}_{-0.26}\!\!\!$ & $2.44\pm 0.11$ & $\!\!\!0.84^{+0.58}_{-0.64}\!\!\!$ & $71.6$ & $76$ & $0.63$ \\
\hline
\end{tabular}
\caption{Fits for cosmological parameters described in
  Sec.~\ref{SS:paramconstraints-alone}.  In each case, parameters that are
  fixed to a single value are in bold; those that are fit are shown with 68 per cent confidence limits after marginalising over
  all other fitted parameters. \label{T:fits}}
\end{table*}

With these fits, we can do a basic sanity test for consistency between
samples.  In Fig.~\ref{F:bestfitsig} we show the data with the best-fitting
signals.  The model appears to fit the data quite well, without any
signs of systematic tension, consistent with the reasonable $\chi^2$
and $p$-values in Table~\ref{T:fits}. 

\begin{figure}
\includegraphics[width=\columnwidth]{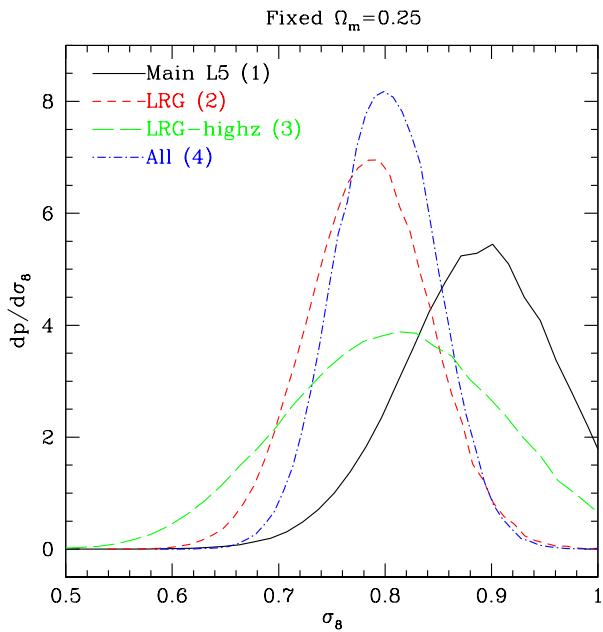}
\caption{Posterior distribution for $\sigma_8$ marginalised over all other fit
  parameters, for fits 1--4 ($\Omega_m$ fixed).\label{F:ps8}}
\end{figure}

Fig.~\ref{F:bplots} shows the posterior distributions for the bias
parameters for each sample; we show the results from
fits 1--3 (fitting for the samples separately) and 4 (fitting all
together).  Fig.~\ref{F:ps8} shows the 1D posterior distributions for $\sigma_8$
for each sample separately and the combined result (fits 1--4), in all cases
marginalised over the bias parameters.  The corresponding limits on
$\sigma_8$ when using all samples together (as in Table~\ref{T:fits},
fit 4) are $0.80\pm 0.05$ (68 per cent confidence level,
stat. $+$ sys.).  
Figures~\ref{F:bestfitsig}, \ref{F:bplots}, and~\ref{F:ps8} demonstrate
clearly that there is no statistically significant tension between
cosmological parameter constraints 
from the different samples. 

The results in Fig.~\ref{F:bplots} suggest that for LRG and LRG-highz, we
detect non-linear bias at the $>3\sigma$ and $\sim 1.5\sigma$ levels,
respectively (the results in Table~\ref{T:fits} are deceptive; they
suggest less significant detections, because the errorbars are quite
non-Gaussian).  The non-detection for L5 does not mean we find that the
bias is linear on all scales, only above $\sim 4$\hmpc.  Quantitatively, we
find a $\Delta\chi^2$ for the best-fit model with $b_2$ free versus
that with $b_2=0$ (linear bias only) of $(0.2,32,4.5)$ for Main-L5, LRG,
and LRG-highz, respectively.  That number for Main-L5 confirms the
non-detection of non-linear bias, but the $\Delta\chi^2$ for LRG
suggests a $>3\sigma$ detection of non-linear bias, with a marginal detection
for LRG-highz.  As expected, the best-fit model with
$b_2=0$ has higher $\sigma_8=0.85$, $0.83$ for LRG and
LRG-highz, to accommodate the increased clustering below 10\hmpc\  
due to non-linear bias. \refresp{Note that the hierarchy of bias
  values for these samples roughly mirrors the trends previously
  detected for galaxy bias as a function of luminosity determined in
  SDSS using relative bias measurements, for example by \cite{2008MNRAS.385.1635S}.  Furthermore,
the LRG linear and quadratic biases are consistent with values that
were previously 
measured using a combination of the two- and three-point correlation
functions \citep{2011ApJ...737...97M}.}  \refrespt{While the value of
$b_2$ for the LRG sample is roughly $2\sigma$ above the value that
describes the simulated sample in Fig~\ref{F:rccxi}, this may simply
reflect slightly stronger clustering on small scales, which can result
from even slightly larger satellite fractions.}

As described in Sec.~\ref{SSS:combmodel}, we have included an
arbitrary galaxy-galaxy lensing signal calibration uncertainty with a
Gaussian standard deviation of $(0.04, 0.04, 0.05)$ for the three
samples, assuming the calibration is perfectly correlated between
them.  Given this prior, we find that the best fit is obtained with a
calibration that is $0.1\sigma$ from the expected value (this
decrease in the signal is illustrated in Fig.~\ref{F:bestfitsig} via
an increase of the theoretical model).   If
we instead fix the lensing calibration without allowing any freedom,
the best-fit $\sigma_8$ changes by an amount that is well below our
quoted error, and the errors become smaller by 20 per cent.
These findings suggest that systematic uncertainty due to uncertain lensing
calibration is not completely negligible, but does not dominate our
error budget.

We have also carried out these fits with the off-diagonal elements of
the covariance matrix set to zero for both lensing and clustering for
all samples, in order to test how sensitive we are to the treatment of
off-diagonal elements.  We find that the best-fitting $\sigma_8$
changes by $0.01$ ($0.2\sigma$), and the size of the error regions
becomes smaller by 15 per cent.  Thus, our results are relatively
stable to inclusion of correlations between radial bins.

\refrespt{As an empirical test of the $\sigma_8^5$ prior that was
  justified using simulated data in
  Appendix~\ref{SS:paramconstraints-sims}, we confirm that when we
  remove the $\sigma_8^5$ prior for fit 4, $\sigma_8$ for the most likely point in
  the full fit parameter space does not correspond to that of the median of
  the posterior distribution after marginalizing over nuisance
  parameters.  The sign of the difference is the same of that in the
  simulated data in the Appendix; the magnitude of the difference is
  slightly larger than that in the simulations, but the effects are
  consistent in the simulations and real data once we take into
  account the far larger noise in the real data.  Thus, there is no
  indication from the data that the assumptions behind our
  $\sigma_8^5$ prior are incorrect, so we use this prior for all fits
  that include the SDSS lensing and clustering data.}

\begin{figure}
\includegraphics[width=\columnwidth]{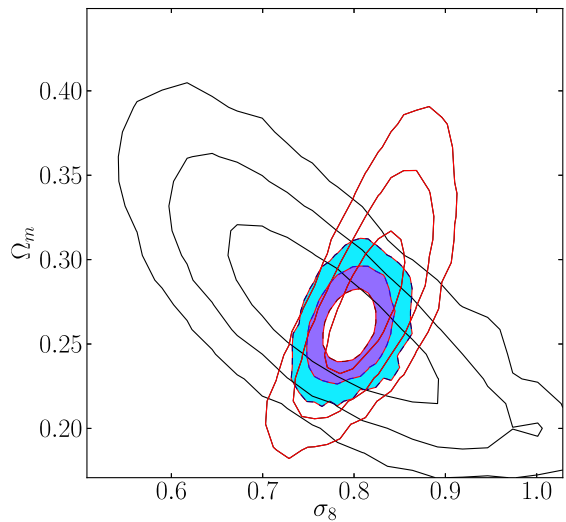}
\caption{Open black lines show the contours in the $\sigma_8$
  vs. $\Omega_m$ plane (fit 5) for our dataset, marginalising over all linear and non-linear
  bias parameters and lensing calibration.  The contours that are
  shown are $1$, $2$, and $3\sigma$.  The nearly orthogonal open red 
  contours are for WMAP7 (also fitting for $n_s$, $H_0$, and other
  parameters in flat $\Lambda$CDM as in
  Sec.~\ref{SS:paramconstraints-comb}), and the filled contours are
  for WMAP7 combined with our data.\label{F:oms8}}
\end{figure}

Next, we have allowed the matter density $\Omega_m$ to vary (fit 5 in
Table~\ref{T:fits}, using all three samples together).  In this case, there is a classic degeneracy
between $\sigma_8$ and $\Omega_m$ in the lensing data, with higher
$\sigma_8$ requiring a lower $\Omega_m$ to fit the data.  Because of
this degeneracy, the allowed range of $\sigma_8$ becomes broader,
$\sigma_8=0.76\pm 0.08$ when marginalised over $\Omega_m$,
the bias parameters, and the lensing calibration 
(still a 10 per cent constraint on $\sigma_8$ even with this
degeneracy).  When marginalising over $\sigma_8$, we can constrain
$\Omega_m=0.269^{+0.038}_{-0.034}$. The resulting
2D contour plot for these two parameters is
shown in Fig.~\ref{F:oms8}.  The best-constrained parameter
combination, which is a better illustration of our overall $S/N$, is $\sigma_8
  (\Omega_m/0.25)^{0.57}=0.80\pm 0.05$ ($1\sigma$, stat. $+$ sys.),
  representing a 6 per cent uncertainty in the amplitude of matter
  fluctuations.  We do not show the best-fitting signal for
this case, because it differs very little from that shown in
Fig.~\ref{F:bestfitsig}.  

\subsection{Combination with other data}\label{SS:paramconstraints-comb}

We combine these data with WMAP7 CMB results
\citep{2011ApJS..192...18K}, anticipating significant benefit from a
tight prior on the amplitude of matter fluctuations at early times.
For the first combined fit, we vary $(A,\omega_{dm}, \omega_b, \theta,
\tau, n_s)$ (see definitions at start of
Sec.~\ref{SS:paramconstraints-alone}), the six bias parameters for the
three galaxy samples, and the lensing calibration with its $(4, 5, 5)$
per cent Gaussian prior -- with fixed $w_\mathrm{de}=-1$ and
$\Omega_k=0$. We include lensing of the CMB and marginalise over
an SZ template as in \cite{2011ApJS..192...18K}.
 All results in this section that use our data include our
 $\sigma_8^5$ prior from Appendix~\ref{SS:paramconstraints-sims}, but that
 prior was not included for the analyses that use WMAP7 data alone. 

\begin{figure}
\begin{center}
\includegraphics[width=0.8\columnwidth]{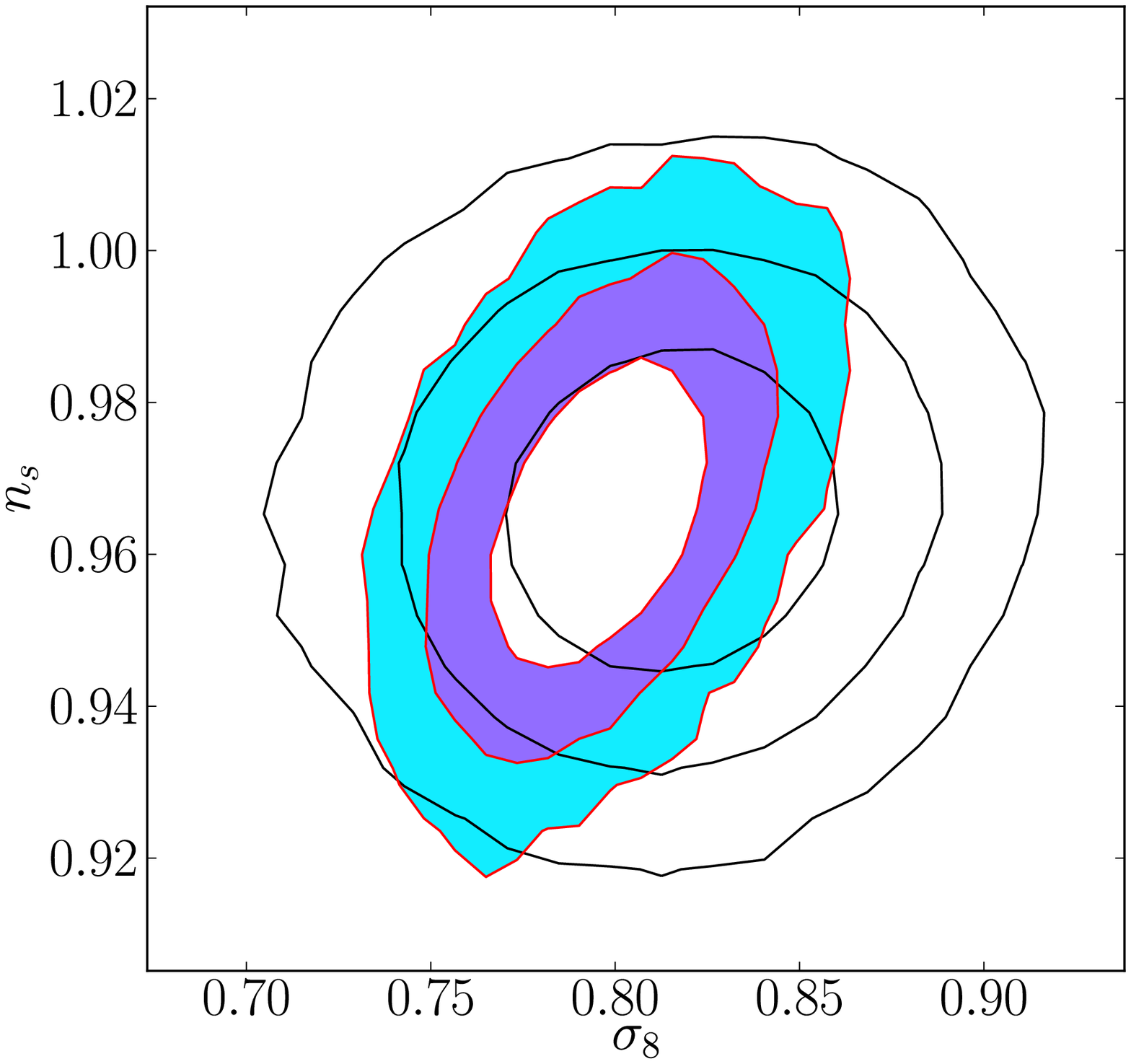}
\includegraphics[width=0.8\columnwidth]{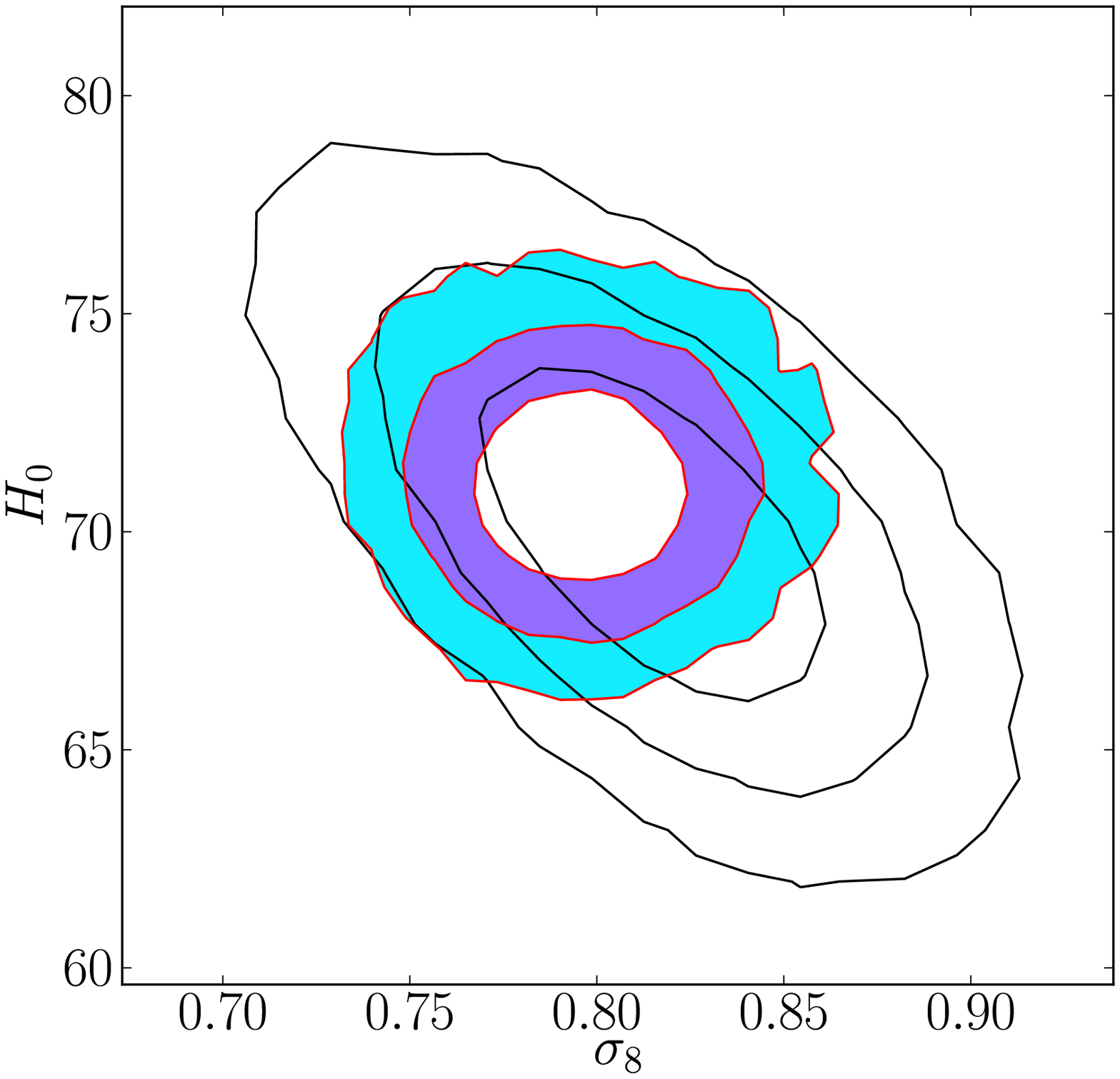}
\end{center}
\caption{Contour plots of 2D probability distributions for fits using the WMAP7 data
  along with our new results for galaxy-galaxy lensing and galaxy
  clustering in SDSS, assuming flat $\Lambda$CDM.  In all cases, we have marginalised over the
  nuisance parameters (bias and calibration) for our data, and over
  any cosmological parameters not shown on the plot.  The black lines
  show $1$, $2$, and $3\sigma$ contours for WMAP7 alone.  The contours
  shown in colour are WMAP7 and our data together.  {\em Top:} $\sigma_8$ vs. $n_s$; {\em bottom:}
  $\sigma_8$ vs. $H_0$.\label{F:wmap-simple}}
\end{figure}

\begin{figure}
\begin{center}
\includegraphics[width=\columnwidth]{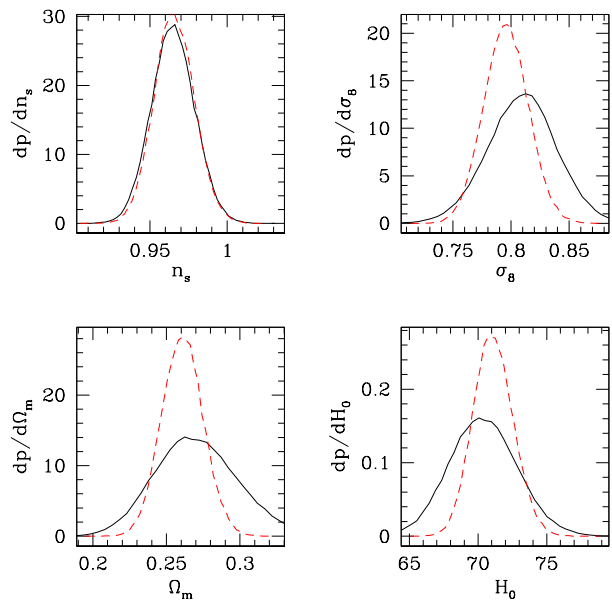}
\end{center}
\caption{1D probability distributions for fits using the WMAP7 data
  along with our new results for galaxy-galaxy lensing and galaxy
  clustering in SDSS, assuming flat $\Lambda$CDM as in Fig.~\ref{F:wmap-simple}.  In all cases, we have marginalised over the
  nuisance parameters (bias and calibration) for our data, and over
  any cosmological parameters not shown on the plot.  The solid black lines
  are for WMAP7 alone; red dashed lines are for WMAP7 and our data together.\label{F:wmap-simple-1d}}
\end{figure}

Fig.~\ref{F:wmap-simple} shows  2D parameter contours with
WMAP7 alone and combined with these data (the $\Omega_m$
vs. $\sigma_8$ contour was already shown in Fig.~\ref{F:oms8}).  We can see that since the degeneracy
direction for $\sigma_8$ vs. $\Omega_m$ for the CMB data is orthogonal
to that for lensing data (Fig.~\ref{F:oms8}), adding our data to
the WMAP7 data roughly halves the size of the allowed regions in
parameter space.  However, the top panel of Fig.~\ref{F:wmap-simple} shows that the constraints
on $n_s$ are, unsurprisingly, set exclusively by the WMAP7 data.  The
bottom panel shows that we also provide additional constraining power
on $H_0$, through the stronger constraint on $\Omega_m$ (and
the fact that CMB constrains a different parameter combination,
$\Omega_m h^2$).  The resulting 1D probability distributions for $(n_s, \sigma_8,
\Omega_m, H_0)$, marginalised over other parameters, are shown in Fig.~\ref{F:wmap-simple-1d}. 

Table~\ref{T:fits-wmap} gives best-fitting parameters and their 68 per cent
confidence intervals from these fits, for WMAP7 alone and for WMAP7 plus these
data (fits 6 and 7).  As shown in Figs.~\ref{F:wmap-simple} and~\ref{F:wmap-simple-1d}, the parameters for which constraints
improve significantly by combination of these datasets are $\sigma_8$,
$\Omega_m$, and $H_0$. In these combined fits, there is little tension
between the datasets, and the best-fitting galaxy bias and lensing
calibration bias parameters are largely unchanged from their values
when fitting only to the SDSS data.   As in the
previous section, we can identify the best-fitting combination of
$\sigma_8$ and $\Omega_m$, which has changed from $\sigma_8
  (\Omega_m/0.25)^{0.57}=0.80\pm 0.05$ (our data alone) to $\sigma_8
  (\Omega_m/0.25)^{-0.13} = 0.79 \pm 0.02$ (WMAP7 $+$ our data)

\begin{table*}
\begin{tabular}{ccccccccc}
\hline\hline
Fit & Data & $\sigma_8$ & $\Omega_m$ & $n_s$ & $H_0$ & $w_\mathrm{de}$ & $\Omega_k$ & $\sum m_\nu$ \\
 & & & & & km s$^{-1}$ Mpc$^{-1}$ & & & eV \\
\hline
6 & WMAP7 & $0.810\pm 0.029$ & $0.270^{+0.030}_{-0.027}$ & $0.965\pm
0.014$ & $70.4\pm 2.5$ & \textbf{-1} & \textbf{0} & \textbf{0} \\ 
7 & WMAP7$+$our data & $0.796\pm 0.019$ & $0.261\pm 0.014$ &
$0.966\pm 0.013$ & $71.1\pm 1.5$ & \textbf{-1} & \textbf{0} &
\textbf{0} \\ 
8 & WMAP7 & $0.83^{+0.10}_{-0.11}$ & $0.26^{+0.10}_{-0.07}$ &
$0.969\pm 0.014$ & $72\pm 11$ & $-1.05^{+0.33}_{-0.30}$ &
\textbf{0} & \textbf{0} \\
9 & WMAP7$+$our data & $0.82\pm 0.08$ & $0.25^{+0.04}_{-0.03}$ &
$0.968\pm 0.014$ & $73\pm 6$ & $-1.07\pm 0.20$ & \textbf{0} &
\textbf{0} \\
10 & WMAP7 & $0.77^{+0.09}_{-0.07}$ & $0.42^{+0.21}_{-0.17}$ & $0.964\pm
0.015$ & $57^{+16}_{-10}$ & $-0.94^{+0.30}_{-0.34}$ &
$\!\!-0.03^{+0.03}_{-0.05}\!\!$ & \textbf{0} \\
11 & WMAP7$+$our data & $0.81^{+0.07}_{-0.08}$ &
$0.25^{+0.04}_{-0.03}$ & $0.969\pm 0.014$ & $72^{+6}_{-5}$ &
$-1.05^{+0.22}_{-0.24}$ & $\!\!0.00\pm 0.01\!\!$ & \textbf{0} \\
12 & WMAP7 & $0.72^{+0.07}_{-0.08}$ & $0.32^{+0.07}_{-0.05}$ &
$0.961\pm 0.016$ & $66\pm 4$ & \textbf{-1} & \textbf{0} & $<1.1$ \\
13 & WMAP7$+$our data & $0.76^{+0.04}_{-0.05}$ & $0.28^{+0.03}_{-0.02}$ &
$0.968\pm 0.013$ & $69\pm 2$ & \textbf{-1} & \textbf{0} & $<0.56$ \\
\hline
\end{tabular}
\caption{Fits for cosmological parameters using WMAP7 and our data as described in
  Sec.~\ref{SS:paramconstraints-comb}.  In each case, parameters that are
  fixed to a single value are in bold; those that are fit are shown with 68 per cent confidence limits after marginalising over
  all other fitted parameters. The exception to this convention is the sum of neutrino
  masses, for which the 95 per cent upper limit is shown.\label{T:fits-wmap}}
\end{table*}

\begin{figure}
\begin{center}
\includegraphics[width=\columnwidth]{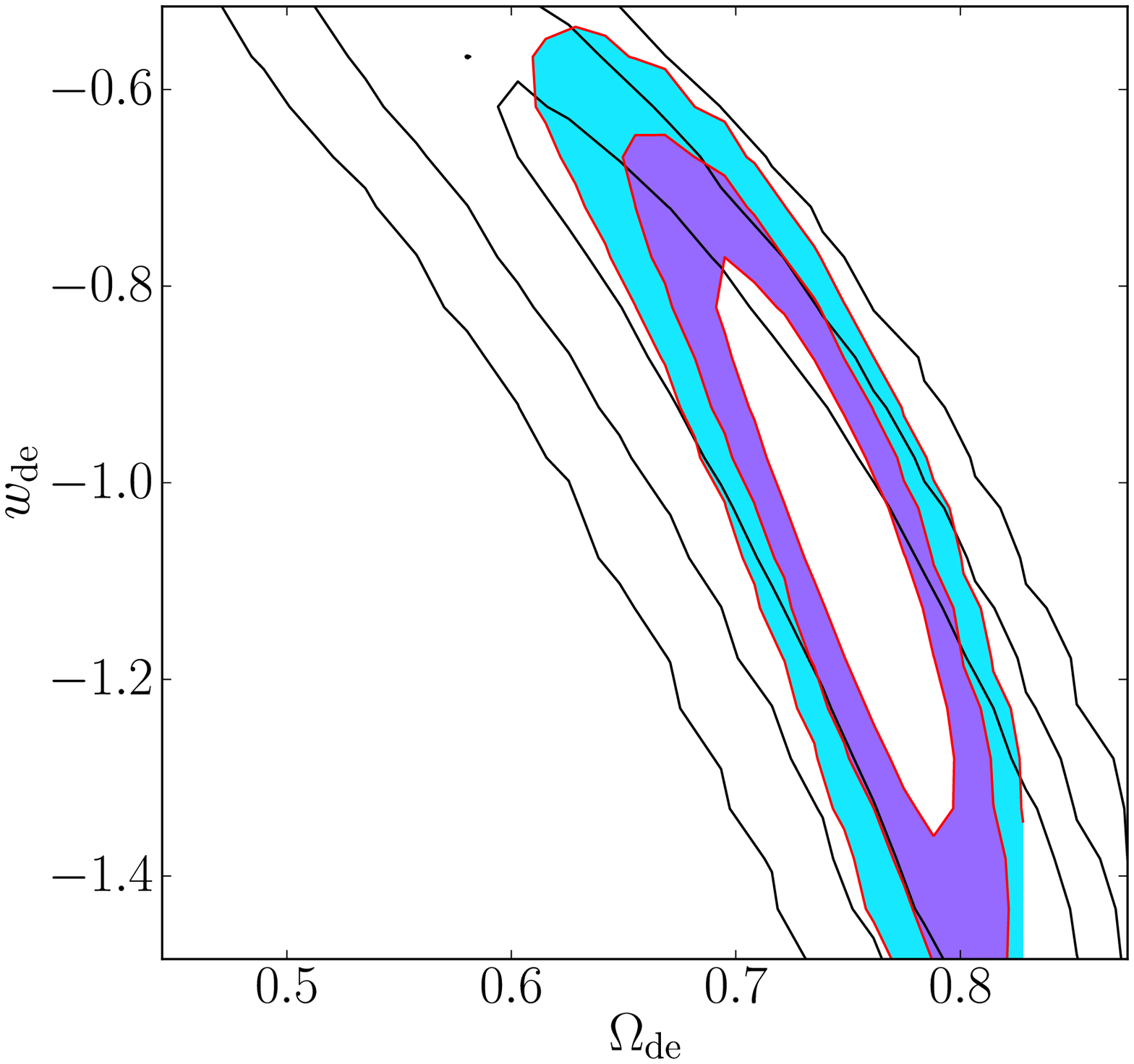}
\end{center}
\caption{2D contour plot for flat $w$CDM fits with the equation of state of
  dark energy allowed to vary from our fiducial value of $-1$ (but
  assumed to be constant in time).  Line and contour styles are as in
  Fig.~\ref{F:wmap-simple}.\label{F:wmap-freew}}
\end{figure}

\begin{figure}
\begin{center}
\includegraphics[width=\columnwidth]{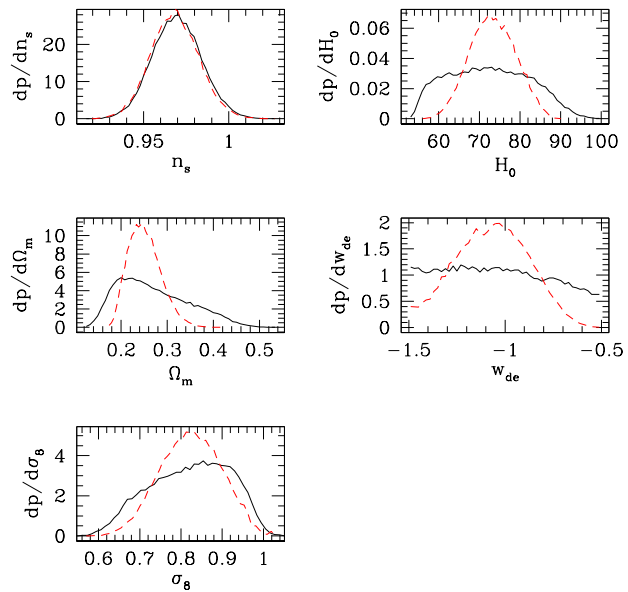}
\end{center}
\caption{1D probability distributions for flat $w$CDM fits with the equation of state of
  dark energy allowed to vary from our fiducial value of $-1$ (but
  assumed to be constant in time), as in Fig.~\ref{F:wmap-freew}.  Line styles are as in
  Fig.~\ref{F:wmap-simple-1d}.\label{F:wmap-freew-1d}}
\end{figure}

Next, we allow the equation of state of
dark energy ($w_\mathrm{de}$) to vary from a cosmological constant.
Figs.~\ref{F:wmap-freew} and~\ref{F:wmap-freew-1d} show
2D contours and 1D parameter distributions, respectively, and Table~\ref{T:fits-wmap} gives best-fitting parameter
constraints, again with WMAP7 alone (fit 8) and with our data included
(fit 9).  As in the $\Lambda$CDM case, our data do not provide
significant additional constraining power on $n_s$, but it does
improve the constraints on all the other parameters, most dramatically
on $w_\mathrm{de}$ (because adding our data provides a constraint on the amplitude
of matter fluctuations at times well after dark energy has become
important).

\begin{figure*}
\begin{center}
\includegraphics[width=\columnwidth]{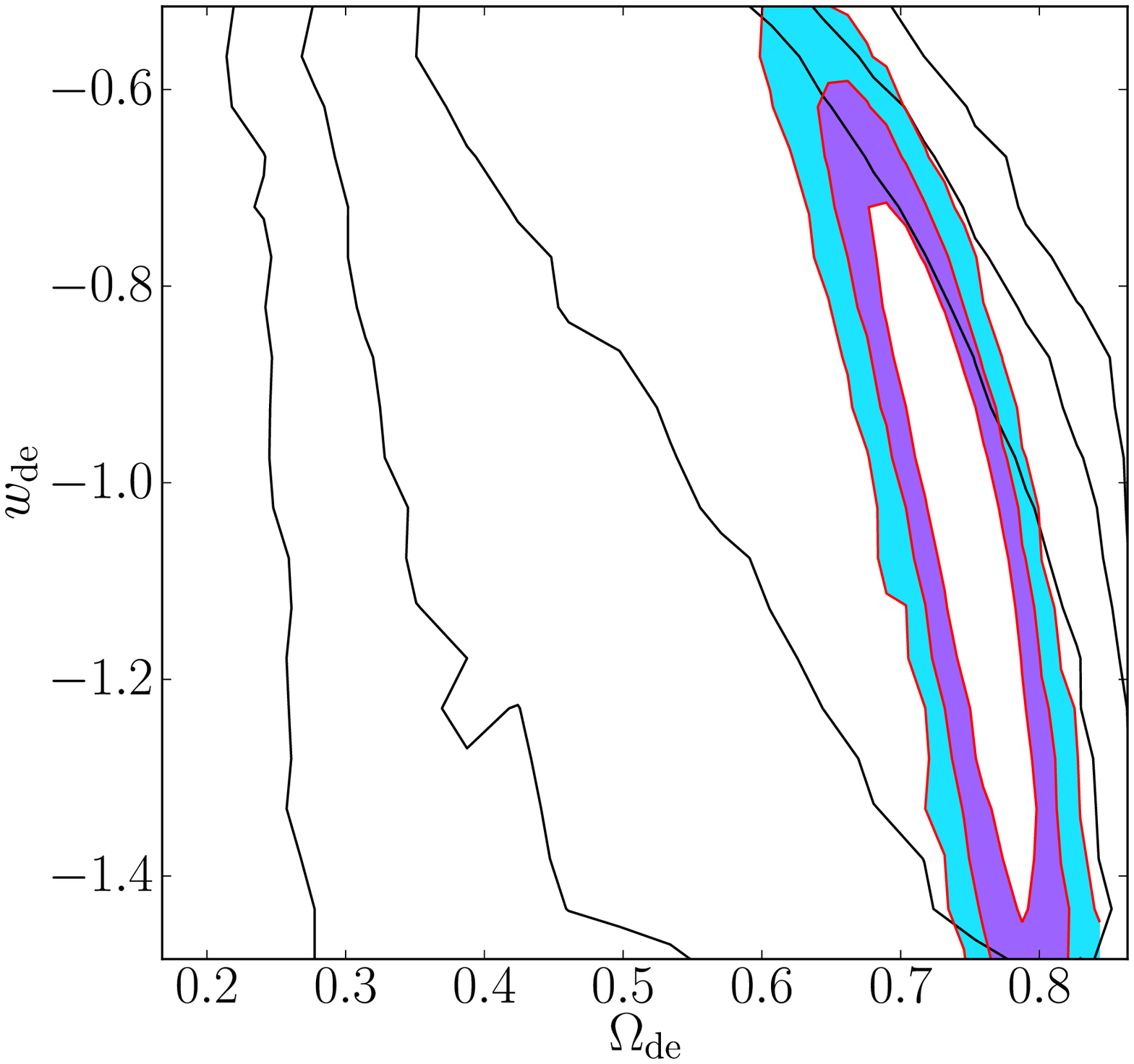}
\includegraphics[width=\columnwidth]{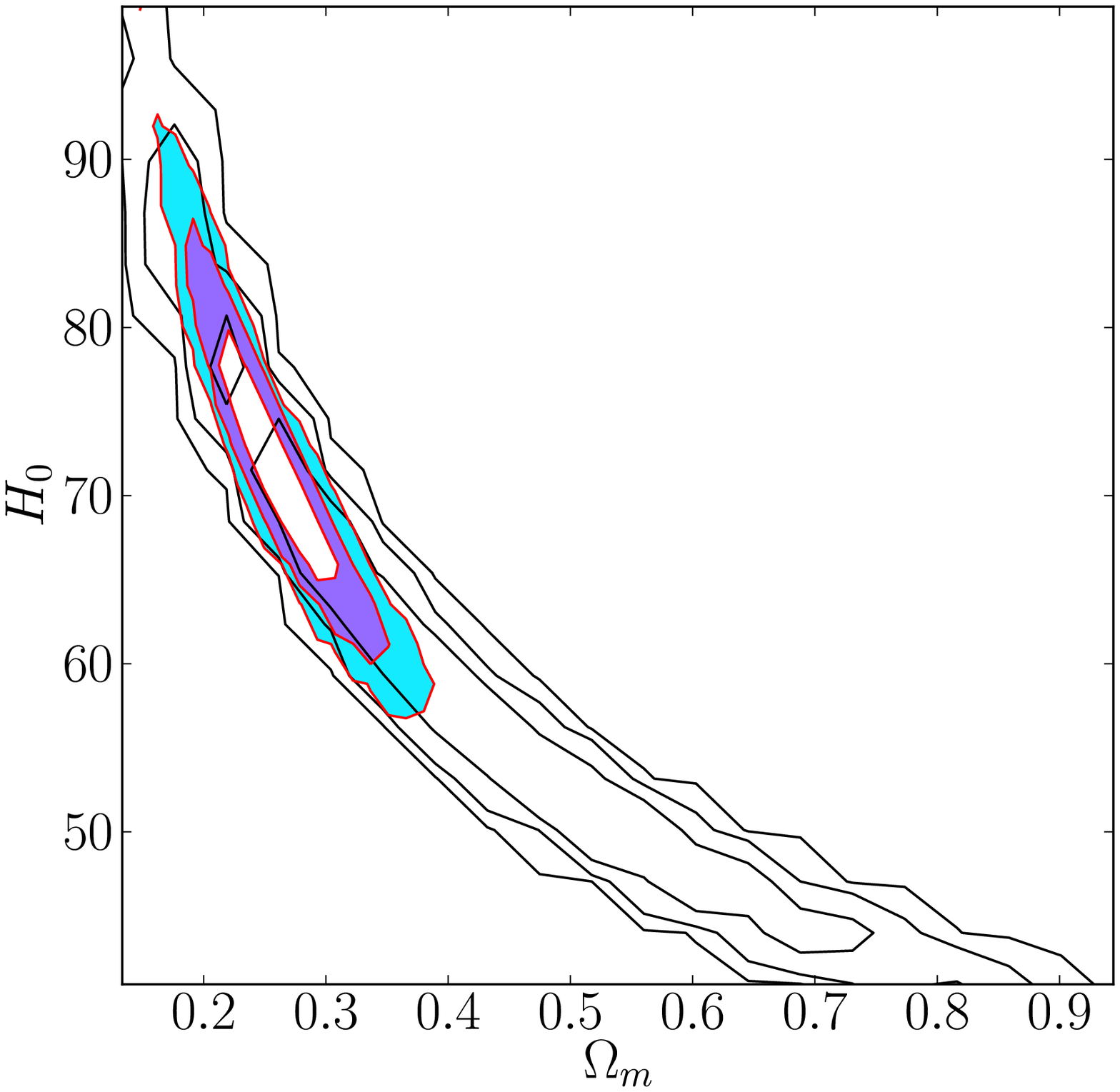}
\includegraphics[width=\columnwidth]{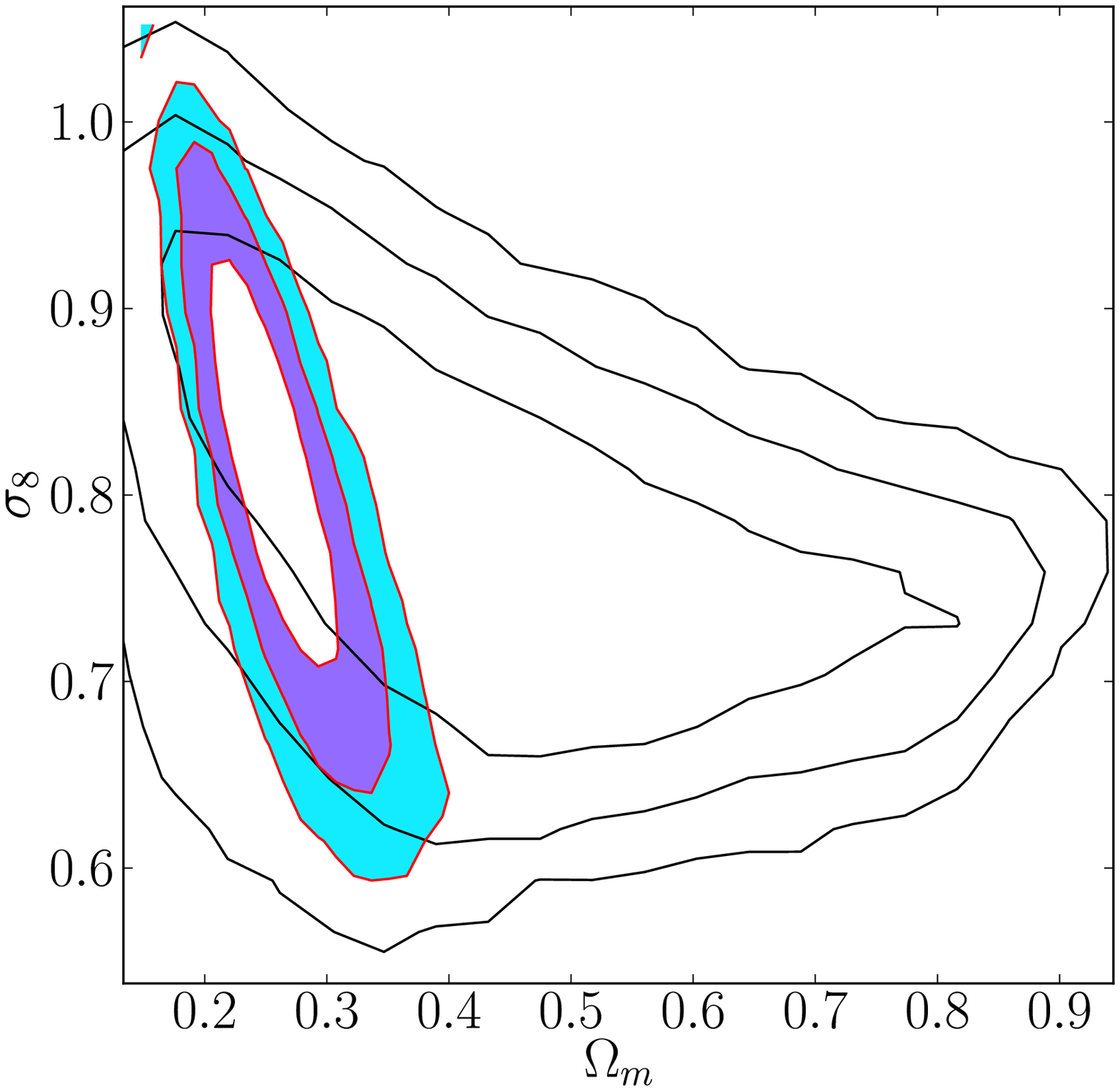}
\includegraphics[width=\columnwidth]{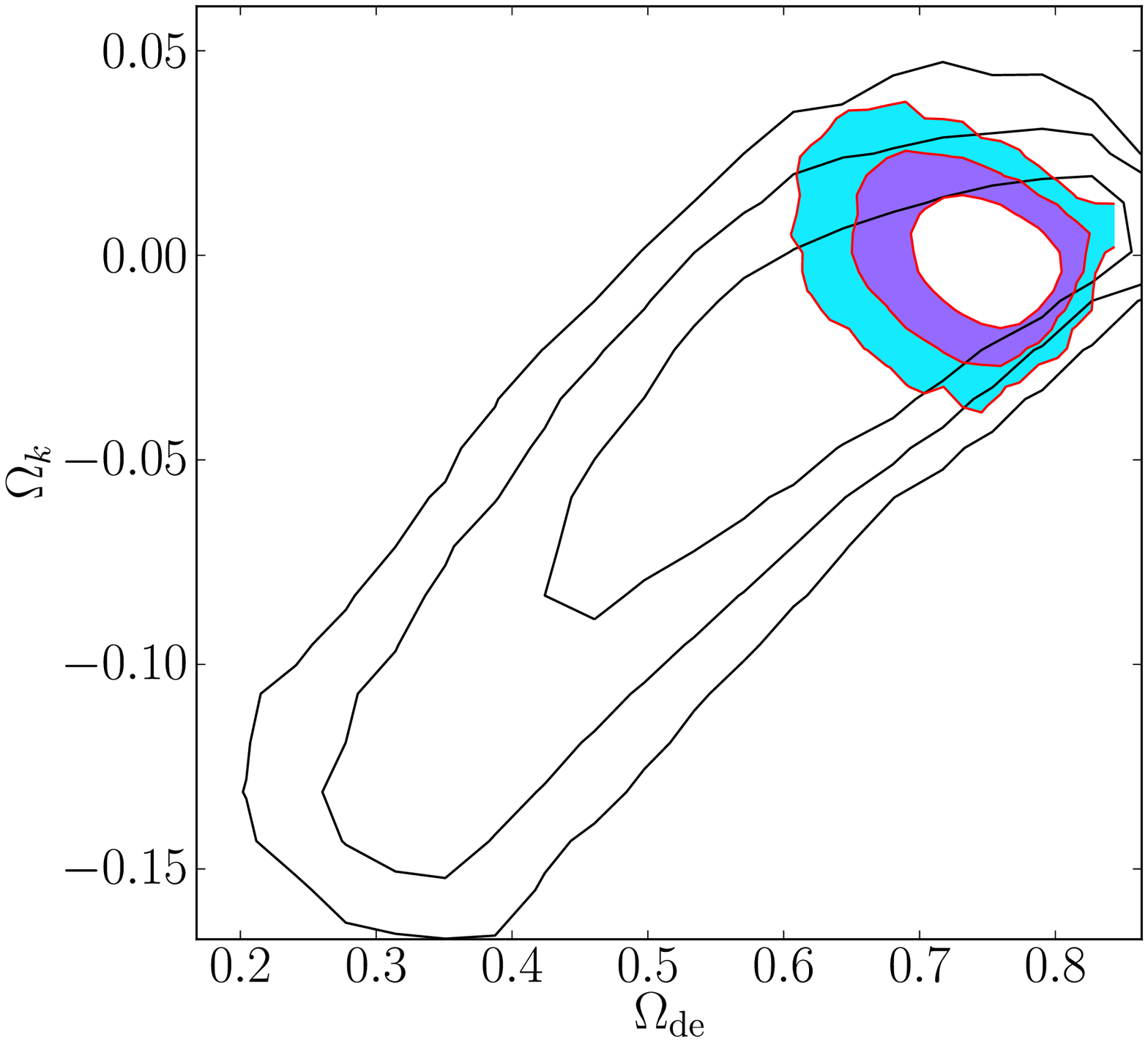}
\end{center}
\caption{2D contour plot for $w$CDM fits without the
  assumption of flatness.  Line and contour styles are as in
  Fig.~\ref{F:wmap-simple}. \label{F:wmap-nonflat}}
\end{figure*}

\begin{figure}
\begin{center}
\includegraphics[width=\columnwidth]{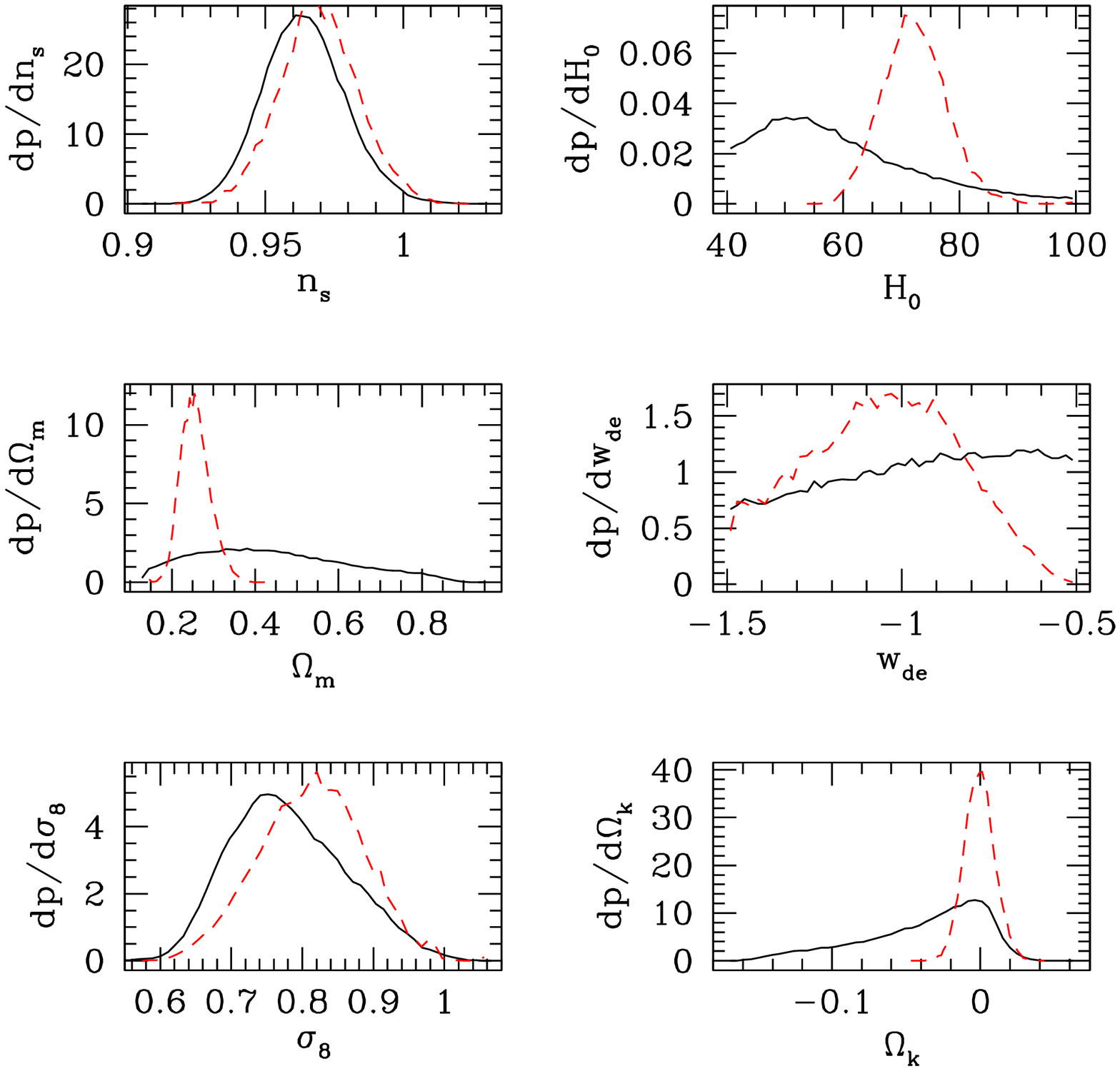}
\end{center}
\caption{1D probability distributions for $w$CDM fits without the assumption of flatness, as in Fig.~\ref{F:wmap-nonflat}.  Line styles are as in
  Fig.~\ref{F:wmap-simple-1d}.\label{F:wmap-nonflat-1d}}
\end{figure}

Next, we relax the assumption of flatness (while still allowing the equation of state of
dark energy to vary from a cosmological constant).
Figs.~\ref{F:wmap-nonflat} and~\ref{F:wmap-nonflat-1d} show
2D and 1D constraints, respectively, and Table~\ref{T:fits-wmap} gives best-fitting parameter
constraints, again with WMAP7 alone (fit 10) and with our data included
(fit 11).  We can see that with these relaxed assumptions about $w_\mathrm{de}$
and $\Omega_k$, the posterior probability distributions for $\Omega_m$, $H_0$, and $w_\mathrm{de}$ are
very broad; our data play a crucial role in reducing the allowed
region of parameter space for all parameters except $n_s$.

\begin{figure}
\begin{center}
\includegraphics[width=\columnwidth]{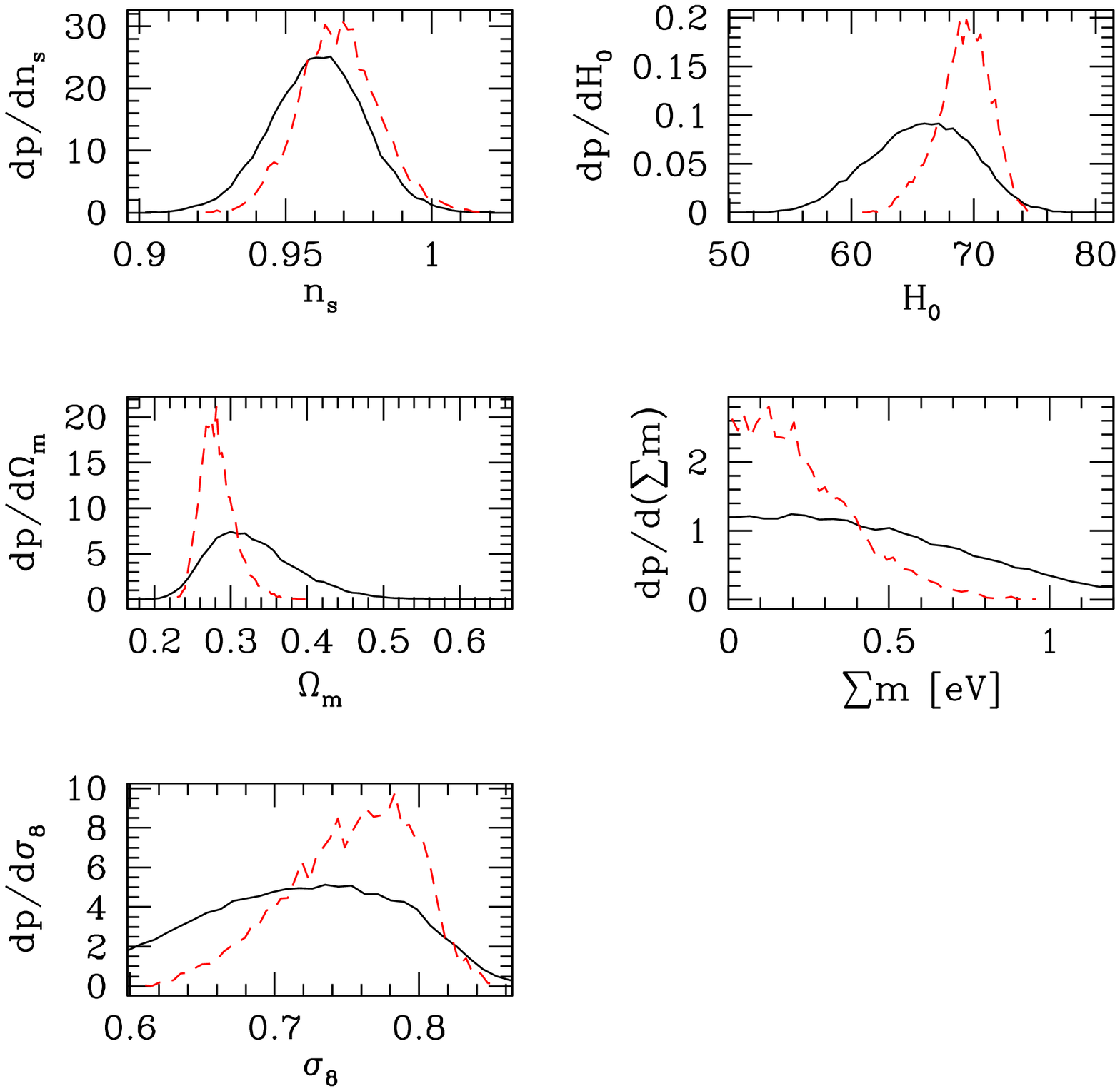}
\end{center}
\caption{1D probability distributions for flat $\Lambda$CDM fits with massive neutrinos.  Line styles are as in
  Fig.~\ref{F:wmap-simple-1d}. \label{F:wmap-neutrino-1d}}
\end{figure}

Finally, we revert to the assumption of flat $\Lambda$CDM, but we allow for
the presence of massive neutrinos using the formalism in
Sec.~\ref{SS:powspec}.  Fig.~\ref{F:wmap-neutrino-1d} shows
1D posterior probability distributions and
Table~\ref{T:fits-wmap} gives best-fitting parameter
constraints, again with WMAP7 alone (fit 12) and with our data included
(fit 13).  In addition to tightening constraints on $\sigma_8$,
$\Omega_m$, and $H_0$ as before, our data help to rule out models with
neutrino masses on the higher end of those allowed by WMAP7, such that the one-sided 95 per cent upper limit on
neutrino mass goes down from 1.1 eV with WMAP7 alone to
0.56 eV with WMAP7 and our data. In general, the presence of
massive neutrinos can significantly broaden the allowed parameter
space for $\Omega_m$ and $H_0$ from CMB alone, and our data rule out some of the allowed high-$\Omega_m$
or low-$H_0$ values. 

\subsection{Comparison with previous work}\label{S:compprev}

The results of the previous section illustrate that our data provide
cosmological parameter constraints that are consistent with and
complementary to those from WMAP7.  Here, we compare our constraints
with those from other cosmological probes.

We do not compare our results against those from a pure clustering
analysis of the shape and amplitude of the galaxy power spectrum, due
to systematic uncertainties in treatments of non-linear galaxy bias,
redshift-space distortions, and other issues. 
  However, it is valuable to compare against measurements of baryonic
  acoustic oscillations (BAO), a measure of the expansion history of
  the universe rather than the growth of structure, since it is
  significantly less prone to such uncertainties.  The most recent measurement of BAO in the SDSS DR7 
\citep{2012MNRAS.427.2168M,2012MNRAS.427.2132P,2012MNRAS.427.2146X}
represents the first use of the `reconstruction' technique
\citep{2007ApJ...664..675E} to reduce the effects of non-linear
evolution of the density field in smoothing the BAO peak.
\cite{2012MNRAS.427.2168M} demonstrates that when combining the BAO
peak position with CMB data, they find $\Omega_m=0.280\pm 0.014$ (in
the context of flat $\Lambda$CDM). This result is fully consistent
with our data when fitting for $\sigma_8$ and $\Omega_m$, which
yielded $\Omega_m=0.257^{+0.038}_{-0.034}$
(Sec.~\ref{SS:paramconstraints-alone}).  This consistency is
non-trivial, given that the BAO results use an identical sample to
make a fully geometric constraint on cosmology, whereas we measure the
amplitude of clustering well below the BAO peak.  

We can also compare against the BAO results from the Baryon
Oscillation Spectroscopic Survey (BOSS) presented in
\cite{2012MNRAS.427.3435A}. Assuming flat $\Lambda$CDM, when using
WMAP7 and the BAO results \refresp{for two galaxy samples, they find 
$\Omega_m=0.293\pm 0.012$ and $H_0=68.8\pm 1.0$~km~s$^{-1}$
Mpc$^{-1}$.  We compare this with our fit 7 (combining our data with
WMAP7), which gave $\Omega_m=0.261\pm 0.014$ and $H_0=71.1\pm
1.5$~km~s$^{-1}$.  A naive comparison of the results -- neglecting the
fact that both measurements include WMAP7 -- suggests a $1.7\sigma$
and $1.3\sigma$ discepancy for $\Omega_m$ and $H_0$, respectively.
While these are not very significant, the tension is in fact worse
since a significant part of the constraining power comes from the CMB
data, which is the same for both measurements.  We defer exploration
of this possible tension between the BAO and our lensing constraints
to future work; however, we note that fit 13 suggests that including
the effects of massive neutrinos would help to reduce this tension.}

We also compare our results against those from other lensing analyses,
particularly cosmic shear.  First, we compare against those from the COSMOS survey,
including the original analysis from \cite{2007ApJS..172..239M} and a
re-analysis in \cite{2010A&A...516A..63S}.  The results from
\cite{2007ApJS..172..239M} used a 3D analysis to infer $\sigma_8
(\Omega_m/0.3)^{0.44}=0.866^{+0.085}_{-0.068}$ (68 per cent CL,
stat. $+$ sys.).  We can compare this result against our result when
fitting for $\sigma_8$ and $\Omega_m$, 
$\sigma_8 (\Omega_m/0.25)^{0.57}=0.80 \pm 0.05$ marginalised over
nuisance parameters.  The COSMOS results are 
$\sim 1.6\sigma$ above ours, 
giving a higher amplitude of clustering.  The 3D COSMOS lensing 
analysis in \cite{2010A&A...516A..63S} gives, for flat $\Lambda$CDM, a
value $\sigma_8 (\Omega_m/0.3)^{0.51} = 0.75\pm 0.08$, 
consistent with our results at the $\sim 0.2\sigma$ level. Part of the reason for the lower quoted
clustering amplitude in \cite{2010A&A...516A..63S} is a different
treatment of the non-linear power spectrum (more consistent with ours): if they use the same
treatment as \cite{2007ApJS..172..239M}, they find $0.79\pm 0.09$,
higher by 5 per cent.  Other differences in clustering amplitude
between the two COSMOS results could come from the different treatment of PSF
estimation, charge-transfer inefficiency, the availability of more
photometric data to improve the photometric redshifts, or differences
in analysis methods (scales used and so on).  In short, the
COSMOS lensing results are consistent with ours, with the exact comparison 
depending on the method of analysis and the treatment of systematic errors.

We can also compare against cosmic shear results from stripe 82 of the
SDSS itself.  There are two such results that use largely independent
analysis methods on the same data, by \cite{2011arXiv1112.3143H} and
\cite{2012ApJ...761...15L}.  The work in \cite{2011arXiv1112.3143H}
used the same PSF correction technique as we have used, and also the
same simulation method relying on space-based training data
\citep{2012MNRAS.420.1518M} to calibrate the shape measurements, so
its systematic errors may not be fully independent from ours.  However, given
that the area of stripe 82 is $\sim 3$
per cent of the area used here, we can consider those
results to be statistically independent of ours.
With fixed $\Omega_m$ close to our value, \cite{2011arXiv1112.3143H} find a relatively low
amplitude of matter fluctuations, $\sigma_8=0.636^{+0.109}_{-0.154}$,
which can be compared with our $0.80\pm 0.05$.
Assuming completely independent errors, this represents a $1.4\sigma$
discrepancy, which is not statistically significant mainly due to the
small size of stripe 82 and the limited source number density due to
the SDSS seeing.  Comparing with   \cite{2012ApJ...761...15L}, they
find for flat $\Lambda$CDM that $\sigma_8
\Omega_m^{0.7}=0.252^{+0.032}_{-0.052}$.  For our value of
$\Omega_m=0.25$, that constraint becomes
$\sigma_8=0.665^{+0.084}_{-0.137}$, quite similar to that from
\cite{2011arXiv1112.3143H}.  In both cases we therefore see a slight
tension with our results, but only at the $\sim 1.4\sigma$ level.  

\refrespt{We also compare against the most recent cosmic shear results
  from the Canada-France-Hawaii Telescope Legacy Survey (CFHTLS),
  presented in \cite{2013arXiv1303.1808H}.  The sample in this
  analysis covers 154 deg$^2$ and has a median redshift of $z_{\rm
    med}=0.70$.  After computing tomographic cosmic shear signals and
  marginalizing over a model for intrinsic alignments, they find a
  best-constrained parameter combination of
  $\sigma_8(\Omega_m/0.27)^{0.46\pm 0.02} = 0.774^{+0.032}_{-0.041}$
  before combining with any external datasets.  This result is
  completely consistent with our findings of $\sigma_8
  (\Omega_m/0.25)^{0.57}=0.80 \pm 0.05$.}

\refresp{Finally, we compare our results with those from analyses that used small-scale
  lensing and clustering measurements, despite our caveats from
  Sec.~\ref{S:theory}.  First, \cite{2012arXiv1207.0503C} carried out
  a joint lensing and clustering analysis of SDSS `Main' sample
  galaxies to constrain cosmology.  They employed the alternate
  approach, discussed briefly in Sec.~\ref{S:theory}, of using the
  measurements to small scales, which requires use of an HOD model for
  how galaxies populate dark matter halos in order to interpret the
  measurements.  Despite use of SDSS data, that measurement is
  somewhat independent of this one because it (a) uses a subset of the
area and (b) employs smaller scales than this one.  Their results for
the flat $\Lambda$CDM model with WMAP7 priors on $n_s$, $h$, and
$\Omega_b$ are consistent with ours, $\Omega_m =
0.278^{+0.023}_{-0.026}$ and $\sigma_8=0.763^{+0.064}_{-0.049}$ (95
per cent CL).  Second, we compare with the results from
\cite{2012ApJ...745...16T}, who used the mass-to-number ratio for
galaxy clusters combined with the mass versus richness calibration
based on weak lensing, and the galaxy clustering (using an
HOD). These measurements should be somewhat but not highly 
correlated with ours, because of the different range of scales used. For
their combination of observables and modeling method, they find
$\sigma_8 \Omega_m^{0.5} = 0.465\pm 0.026$, or $\sigma_8
(\Omega_m/0.25)^{0.5} = 0.93\pm 0.05$ ($1\sigma$).  Compared with our
result of $\sigma_8
  (\Omega_m/0.25)^{0.57}=0.80\pm 0.05$, there is clearly some tension,
since the discrepancy is $2.5\sigma$ assuming independent errors, and
in fact there should be some correlation between the results.
Understanding the exact source of this tension is beyond the scope of
this work, but likely it lies in the different assumptions behind the
methods.  We also note that when combining with CMB data, they find 
$\Omega_m=0.290\pm 0.016$ and $\sigma_8=0.826\pm 0.020$, which should
be compared with our flat $\Lambda$CDM results of
$\Omega_m=0.270^{+0.030}_{-0.027}$ and $\sigma_8=0.810\pm 0.029$.
Here, the tension is less evident, presumably because of the
combination with identical CMB data.}

\section{Discussion}\label{S:discussion}

We have used updated measurements of galaxy-galaxy weak lensing and
galaxy clustering for several samples of spectroscopic galaxies in the
SDSS DR7 to place competitive constraints on the amplitude of matter
fluctuations and, by combining with WMAP7 data, the growth of
structure with time.  The novelty in comparison with previous lensing
cosmology constraints is that we have used galaxy-galaxy lensing (a
cross-correlation of lens galaxy positions and the background shear
field) rather than cosmic shear (the auto-correlation of the shear
field).  From a statistical perspective, the former is more detectable
in low-redshift surveys such as SDSS; however, even at higher
redshift, the galaxy-galaxy lensing typically has a lower systematic
error budget, because the use of cross-correlations allows us to more
easily remove several systematic errors (intrinsic alignments,
additive shear systematics) that plague cosmic shear.  To avoid
contamination from the smallest scales, where there are uncertainties
in the galaxy-mass cross-correlation due to the way that galaxies
populate halos, we have used the annular differential surface density
(ADSD) statistic $\Upsilon$, which strictly
removes contributions below some scale $R_0$ (chosen based on
comparison with simulations to be $4$\hmpc\ for the clustering
analysis, and $2$\hmpc\ for the lensing analysis). 
We apply this approach to three different non-overlapping samples extracted from SDSS DR7: an
intermediate redshift ($0.16<z<0.36$) LRG sample, high redshift
($0.36<z<0.47$) LRG sample, and a low redshift sample ($z<0.155$) with a 
fainter absolute magnitude limit and no colour selection. 
We have opted to model the signals using the non-linear
matter power spectrum along with a perturbation theory-based model for the
non-linear galaxy bias, containing parameters over which we marginalise. We 
see clear evidence for a scale-dependent bias in our LRG sample with large-scale bias 
around $2$, while there is no evidence for the scale-dependent bias 
for the lower luminosity sample with large-scale bias around $1.25$. This trend is consistent 
with theoretical expectations based on simulations and analytic 
predictions \citep{2012PhRvD..86h3540B}. 

Using our data and fixing all cosmological parameters except for
$\sigma_8$ and $\Omega_m$, we find $\sigma_8
(\Omega_m/0.25)^{0.57}=0.80\pm 0.05$ ($1\sigma$,
stat. $+$ sys.) after marginalising over the galaxy bias parameters
and a nuisance parameter for the lensing calibration.  This result is
highly consistent with that from the WMAP7 CMB analysis, and with many
other cosmological measurements as discussed in Sec.~\ref{S:compprev}.
The 6 per cent errorbar, including both statistical and systematic
contributions, indicates 
that the SDSS is a quite powerful survey for weak lensing cosmology at
low redshift (in the context of other extant lensing datasets).  Moreover, given its low effective
redshift of $\sim 0.27$, we imagine future benefits from the
combination with other lensing measurements that typically have higher
effective redshifts.

When we include WMAP7 data in the analysis, we find that for flat
$\Lambda$CDM, we are able to provide significant additional
constraining power on $\sigma_8$, $\Omega_m$, and $H_0$ due to 
orthogonal degeneracy directions, effectively halving the allowed
region in parameter space; we do not provide significant
additional constraining power on $n_s$.   When we allow the equation
of state of dark energy $w_\mathrm{de}$ to vary from $-1$ (while still
assuming it is constant in time), when we further allow the
possibility of curvature, or when we include massive neutrinos in the
context of flat $\Lambda$CDM, we likewise find that our low-redshift
constraint on the amplitude of matter fluctuations is crucial for
reducing major parameter degeneracies.  It will be interesting to
combine these results with a low-redshift constraint on the expansion
history of the universe, such as from BAO; we defer this exercise to
future work\refresp{, but note that Section~\ref{S:compprev} suggests
  that inclusion of massive neutrinos may be necessary to reduce some
  tension between constraints on $\Omega_m$ and $H_0$ from the two
  probes}.

We emphasise that these results represent an entirely new opportunity
for the field of lensing to constrain cosmological parameters in a way
that is largely independent of the details of how galaxies populate
dark matter halos, but also less sensitive than cosmic shear to
several important observational and astrophysical systematic errors.
Among these are all additive contributions such as telescope or atmosphere 
effects that induce shear-shear correlations but not shear-galaxy correlations. 
Intrinsic alignments 
also do not contribute 
to shear-galaxy correlations as long as the redshift separation between lenses and 
sources, using photometric redshift information, is effective. In contrast, the dominant intrinsic alignment 
contribution to shear-shear correlations, induced by correlations
between sheared 
galaxies in the background and 
intrinsically-aligned galaxies in the foreground, cannot be eliminated simply 
by using photometric redshift information \citep{2004PhRvD..70f3526H}. 

In addition to the intrinsic value of these cosmological constraints
in and of themselves, this work is a proof of concept for this
analysis technique for the next generation of large, wide-field
surveys that will carry out lensing
measurements, such as Hyper Suprime-Cam (HSC\footnote{\texttt{http://www.naoj.org/Projects/HSC/index.html}},
\citealt{2006SPIE.6269E...9M}), Dark Energy Survey
(DES\footnote{\texttt{https://www.darkenergysurvey.org/}},
\citealt{2005astro.ph.10346T}), the KIlo-Degree Survey
(KIDS\footnote{\texttt{http://www.astro-wise.org/projects/KIDS/}}),
the Panoramic Survey Telescope and Rapid Response System
(Pan-STARRS\footnote{\texttt{http://pan-starrs.ifa.hawaii.edu/public/}},
\citealt{2010SPIE.7733E..12K}); and even more ambitious programmes
such as the Large Synoptic Survey Telescope
(LSST\footnote{\texttt{http://www.lsst.org/lsst}},
\citealt{2009arXiv0912.0201L}),
Euclid\footnote{\texttt{http://sci.esa.int/science-e/www/area/index.cfm?fareaid\
    =102}}, and the Wide-Field Infrared Survey Telescope
(WFIRST\footnote{\texttt{http://wfirst.gsfc.nasa.gov/}}).  The ability
to make cosmological measurements with galaxy-galaxy lensing rather
than cosmic shear is particularly important for making use of early
data from upcoming surveys, when additive shear systematics will be less
well-understood.  
We expect that this approach will yield results that are competitive and 
complementary to shear-shear analysis for these
future surveys as well. 

\refresp{The data used for the cosmological parameter constraints, and
the MCMC chains, can be downloaded directly from the first author's website.}

\section*{Acknowledgements}

\refresp{We thank the anonymous referee for providing numerous
  comments that improved the presentation of this work.}  We thank Jim Gunn, Robert Lupton, Benjamin Joachimi, and David Spergel for useful discussions related to this work, and
we thank Eyal Kazin for helpful discussions on selection of the
Luminous Red Galaxy sample.  
During this project, U.~S. was supported in part by the U.S. Department of Energy under
Contract No. DE-AC02-98CH10886, the Swiss National Foundation under
contract 200021-116696/1 and the WCU grant R32-10130.
During this project, C.~H. has been supported by the US Department of
Energy (DOE-FG03-92-ER40701 and DOE.DE-SC0006624), the National
Science Foundation (NST AST-0807337), and the David \& Lucile Packard
Foundation. 
R.~E.~S. acknowledges support from a Marie Curie Reintegration
Grant and an award for Experienced Researchers from the
Alexander von Humboldt Foundation. We thank V. Springel for making
public GADGET-2 and for providing his B-FoF halo finder,
and R. Scoccimarro for making public his 2LPT code.

Funding for the Sloan Digital Sky Survey (SDSS) and SDSS-II has been
provided by the Alfred P. Sloan Foundç∂ation, the Participating
Institutions, the National Science Foundation, the U.S. Department of
Energy, the National Aeronautics and Space Administration, the
Japanese Monbukagakusho, and the Max Planck Society, and the Higher
Education Funding Council for England. The SDSS Web site is
{\texttt http://www.sdss.org/}. 

The SDSS is managed by the Astrophysical Research Consortium (ARC) for
the Participating Institutions. The Participating Institutions are the
American Museum of Natural History, Astrophysical Institute Potsdam,
University of Basel, University of Cambridge, Case Western Reserve
University, The University of Chicago, Drexel University, Fermilab,
the Institute for Advanced Study, the Japan Participation Group, The
Johns Hopkins University, the Joint Institute for Nuclear
Astrophysics, the Kavli Institute for Particle Astrophysics and
Cosmology, the Korean Scientist Group, the Chinese Academy of Sciences
(LAMOST), Los Alamos National Laboratory, the Max-Planck-Institute
for Astronomy (MPIA), the Max-Planck-Institute for Astrophysics (MPA),
New Mexico State University, Ohio State University, University of
Pittsburgh, University of Portsmouth, Princeton University, the United
States Naval Observatory and the University of Washington. 

\bibliography{cosmo,cosmo_preprints,simground,newrefs}

\appendix

\section{Calibration of the method}\label{SS:paramconstraints-sims}

As a basic test, we use the data from the simulated LRG sample
(Sec.~\ref{SS:simulations}) and check whether we can accurately recover
the input cosmology using the analysis framework from
Sec.~\ref{SS:powspec}.  Such tests with various $R_0$ 
will allow us to determine the safest choice of
$R_0$ to minimise systematic error without excessively increasing
the statistical errors.

For this test, we fixed all cosmological parameters besides
$\sigma_8$, and varied $R_0$, making sure that we can recover the true
$\sigma_8=0.8$ for the simulations\footnote{We could have fit 
  jointly for $\sigma_8$ and $\Omega_m$.  However, since those
two cosmological parameters are strongly degenerate, we could easily
(due to very small noise fluctuations in the simulations) be driven
anywhere along that degeneracy, without it being a meaningful deviation.
We therefore check that with $\Omega_m$ fixed we can infer the correct
$\sigma_8$, under the assumption that this shows we can
infer the proper degenerate combination of $\Omega_m$ and
$\sigma_8$.}. While doing this test, we allowed the lensing
calibration and the bias parameters to be free parameters. Given that moving $R_0$ above
2\hmpc\ for the galaxy-galaxy lensing severely impacts the $S/N$, we
consider only this value of $R_0$ for the lensing, but vary $R_0$ for the
clustering from 2 to 6\hmpc, in steps of 2\hmpc.  This procedure is
also theoretically motivated since, as pointed out in
Sec.~\ref{SS:r0choice}, the clustering signal should be more strongly
sensitive than the g-g lensing signal to the details of how galaxies occupy dark matter halos on
these scales. 
Shape noise was
not added to the simulated LRG 
data, \refrespt{and the simulation volume is 40 times larger than the
  cosmological volume covered by the real data.  Thus, the cosmic
  variance is substantially smaller than that in the real data, and in
any case, our dominant source of noise (shape noise) is not present.
Because of this, we can trust the simulations to reveal low-level
biases due to our fitting procedure, at the level of $\sim 0.2\sigma$
(where $\sigma$ is the statistical uncertainty in the cosmological
parameters in the fits to real data).}  
For $R_{0,\rmg\rmg}=2, 4, 6$\hmpc, the best-fitting $\sigma_8=0.763, 0.795,
0.792$ (these are the global best-fit values, not marginalised over
nuisance parameters).  The first value, for $R_{0,\rmg\rmg}=2$\hmpc, indicates a
statistically-significant bias in $\sigma_8$.  However, the
results for larger $R_0$ are consistent with no bias, so we adopt
$R_0=2$\hmpc\ and $4$\hmpc\ for the g-g lensing and galaxy clustering,
respectively. \refrespt{The fact that the most likely point agrees
  with the theory means that there is no inherent bias in the theory
  predictions with respect to our simulations.}

We can also check the effect of priors and marginalisation over 
lensing calibration bias and galaxy bias parameters.  After
marginalisation \refrespt{over the galaxy bias parameters and the
  lensing calibration}, the median of the likelihood is $0.78$, which differs from the 
input value \refrespt{and the global best-fitting value} of 0.8.  
Since our best fit model (maximum likelihood) is at the position of the input
model, the discrepancy between the median and the input value must be
due to the effect of the prior\refrespt{s and the marginalisation process}. The median is the standard value quoted in the
Markov Chain Monte Carlo (MCMC) analyses, since it can be robustly estimated. It is also invariant under a monotonic transformation of the variable, i.e. 
 the median of $\sigma_8$ is 
the same as the median of $\sigma_8^2$ assuming the same prior, which is believed to be useful 
since there is no good reason {\em a priori} to use the linear fluctuation amplitude $\sigma_8$ as opposed to the correlation 
function amplitude $\sigma_8^2$. However, this does not mean that the median is invariant under the change of the prior, 
i.e. if we start with a uniform prior on $\sigma_8$ we obtain a different median than if we use 
a uniform prior on $\sigma_8^2$. The only number that is
prior-independent is the maximum likelihood value, which is unstable
in multi-dimensional MCMC analyses, since the likelihood surface is shallow and MCMC 
has difficulties finding the absolute maximum with a finite number of steps. 
It is nevertheless convenient that the main reported 
number agrees with the input value in simulations. Since a  
prior that is uniform in $\sigma_8$ does not result in this property, and since 
there is nothing especially natural about that choice of the prior, we
will empirically choose a prior such that the median $\sigma_8$ agrees with 
the input value in simulations, and then report the median determined
in a similar way in the data. 
If we apply a flat prior on $\sigma_8^5$, the median increases to $\sigma_8=0.80$, and since this agrees 
with the input value we
adopt this prior for the rest of the paper\footnote{\refresp{For those
  who wish to make cosmological parameter constraints using these data
while exploring different choices of the prior, the MCMC chains
and the data used for the fits can be downloaded directly from the
first author's website.}}. Note that as we add more
data, such as additional SDSS datasets (we use three samples, not just
LRGs as in the tests in this section) 
and WMAP, the effect of the prior is
diminished and we converge on the true value. \refrespt{This is the
  reason to apply a prior, rather than a correction factor such as
  $0.80/0.78$, which would have an excessive impact in the case where
  we add more data.}

 The plots of simulated $\Upsilon$ are 
shown in Fig.~\ref{F:nlbiasmodel}, 
and we see that the simulated data and the best model 
agree reasonably well both for galaxy-galaxy lensing and galaxy clustering, 
with $b_2=0.25$ as the best fit value.

We also note that in the real data, when carrying out the fitting
procedure, we see a qualitatively similar trend in
$\sigma_8(R_{0,\rmg\rmg})$ increasing with $R_0$, however the
change in $\sigma_8$ is larger in amplitude (but consistent with that
given here within the errors).   As a final sanity check of our
procedure above, we carry out the analysis using 
another HOD sample, constructed 
from simulations with $\sigma_8=0.90$ and selected to mimic a
higher-luminosity and higher-mass sample. 
For this HOD sample, with our adopted values of $R_0$ for g-g lensing
and galaxy clustering and our prior on $\sigma_8$, we find
best-fitting and median $\sigma_8=0.90$ and $0.92$. The former number is  completely
consistent with the input cosmological model, reassuring us that our
choice of $R_0$ is robust to significant changes in 
the galaxy mass and input cosmology. The latter number
suggests that for large values of $\sigma_8$, the adopted prior will cause a
bias that is $\sim 40$ per cent of our statistical errors. However, we
see no evidence for such a high $\sigma_8$, and so this worst-case
scenario is not an issue in practise.

\section{Weak lensing systematics tests}\label{SS:reslensingsys}

In this Appendix, we present several tests of systematic error in the
lensing signals, focusing exclusively on those tests that were not
done for the same shape catalogue by R12.

\subsection{Calculation of $\Upsilon_{\rmg\rmm}$}\label{SSS:calcupsgm}

Since we use \upsgm\ for cosmological parameter constraints, and
it is a derived quantity that relies on determination of
$\ds(R_0)$, here we present tests illustrating the accuracy of that
determination.  

\begin{figure}
\includegraphics[width=\columnwidth]{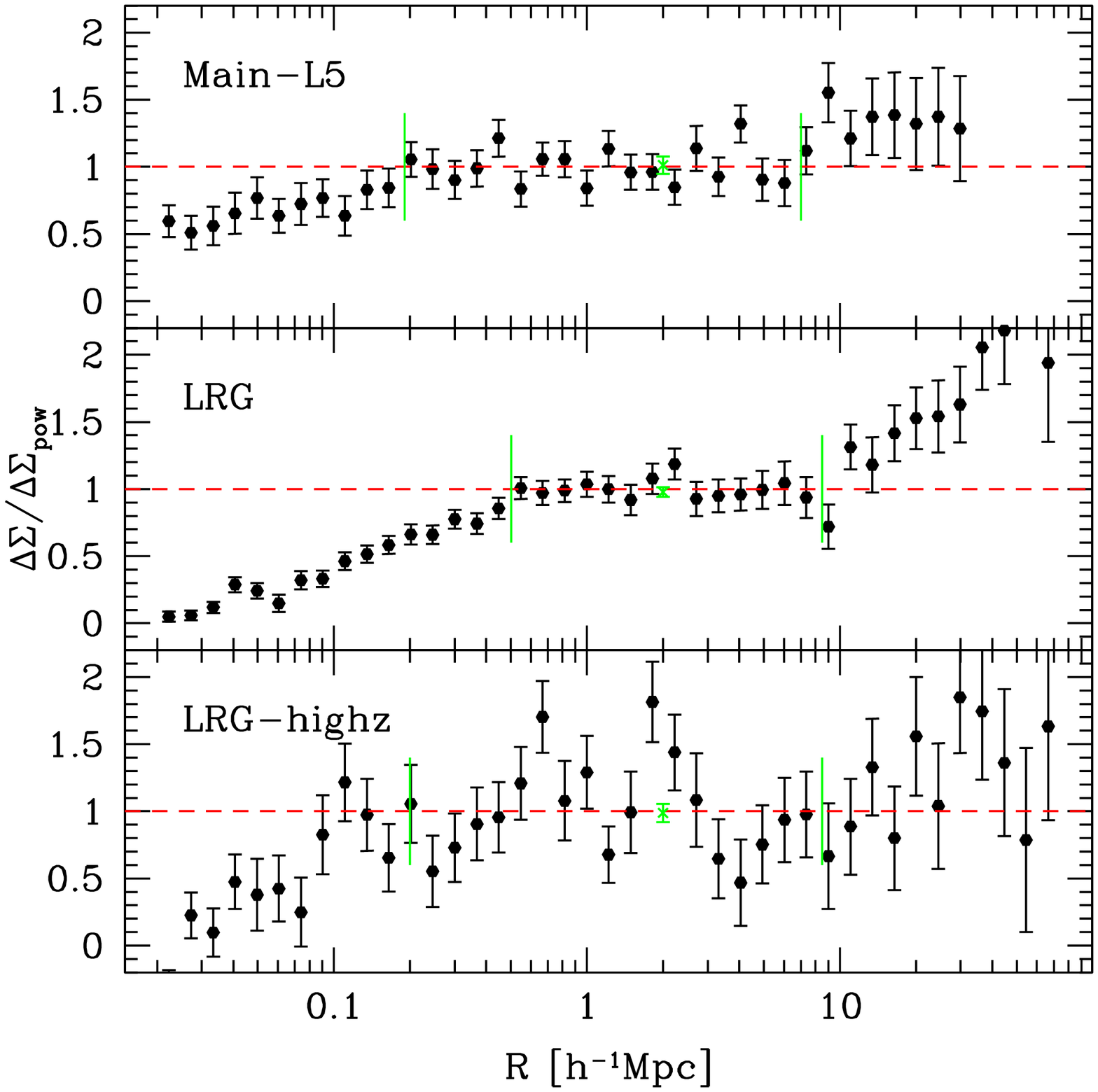}
\caption{Lensing signals for the three samples (as labelled on the
  plot) divided by the best-fit power-law using the range of scales
  indicated by vertical solid lines.  The horizontal dashed line shows
  the ideal value $\ds/\ds_\mathrm{pow}=1$.  Also, a single point
  shown as an $\times$ with its own errorbar at $R_0=2$\hmpc\ shows the
  jackknife mean 
  value that was used for $\ds(R_0)$, and its jackknife error.\label{F:dsr0}}
\end{figure}

As stated in Sec.~\ref{SS:obs-gglensing}, we determine $\ds(R_0)$ using power-law
fits over a range of scales on which $\ds$ appears
consistent with a power-law, including $R_0$ itself to avoid extrapolation.  In Fig.~\ref{F:dsr0}, we show (for all
three samples) the observed $\ds(R)$ divided by the best-fit power-law
for the chosen range of scales which are indicated by vertical lines.  This power-law is determined from a
fit to the jackknife mean $\ds$, weighted by the inverse variance
(assumed to be diagonal, which is appropriate for these scales).
Ideally, this ratio of observed signal to power-law would be consistent
with one for all scales between the vertical lines.  In addition, we show the jackknife
mean value of $\ds(R_0)$ and its jackknife errorbar, again with
respect to the best-fit power-law for the chosen range of scales and
therefore with an ideal value of $1$.  It is clear that the power-law
approximation is valid on the range of scales used, but the observed
signal deviates from it strongly outside of that range.  This
power-law fit is used {\em only} for empirical determination of
$\ds(R_0)$, not for any other purpose.

The best-fit power-laws and ranges of scales used are defined as $\ds
= \ds_0 (R/R_0)^{\alpha}$ for $R_\mathrm{pow,min}<R<R_\mathrm{pow,max}$, where $\ds_0$ is in units of
$hM_\odot/\mathrm{pc}^2$.  Given this definition, the power-laws that
went into Fig.~\ref{F:dsr0} are described in Table~\ref{T:dspowlaw}.
\begin{table}
\begin{tabular}{ccccc}
\hline
\hline
Sample & $R_\mathrm{pow,min}$ & $R_\mathrm{pow,max}$ & $\ds_0$ &
$\alpha$ \\
 & $[$\hmpc$]$ & $[$\hmpc$]$ & $[hM_\odot/\mathrm{pc}^2]$ & \\
\hline
Main-L5 & 0.2 & 7.0 & $2.49$ & $-0.97$ \\
LRG & 0.5 & 8.5 & $5.81$ & $-1.22$ \\
LRG-highz & 0.2 & 8.5 & $5.41$ & $-1.08$ \\
\hline\hline
\end{tabular}
\caption{Best-fitting power-law functions (as defined in Sec.~\ref{SSS:calcupsgm})
  to $\ds(R)$ for a limited range of scales, used to estimate
  $\ds(R_0)$ and therefore \upsgm.\label{T:dspowlaw}}
\end{table}

\subsection{Correction for physically associated sources}\label{SSS:physassoc}

\begin{figure}
\includegraphics[width=\columnwidth]{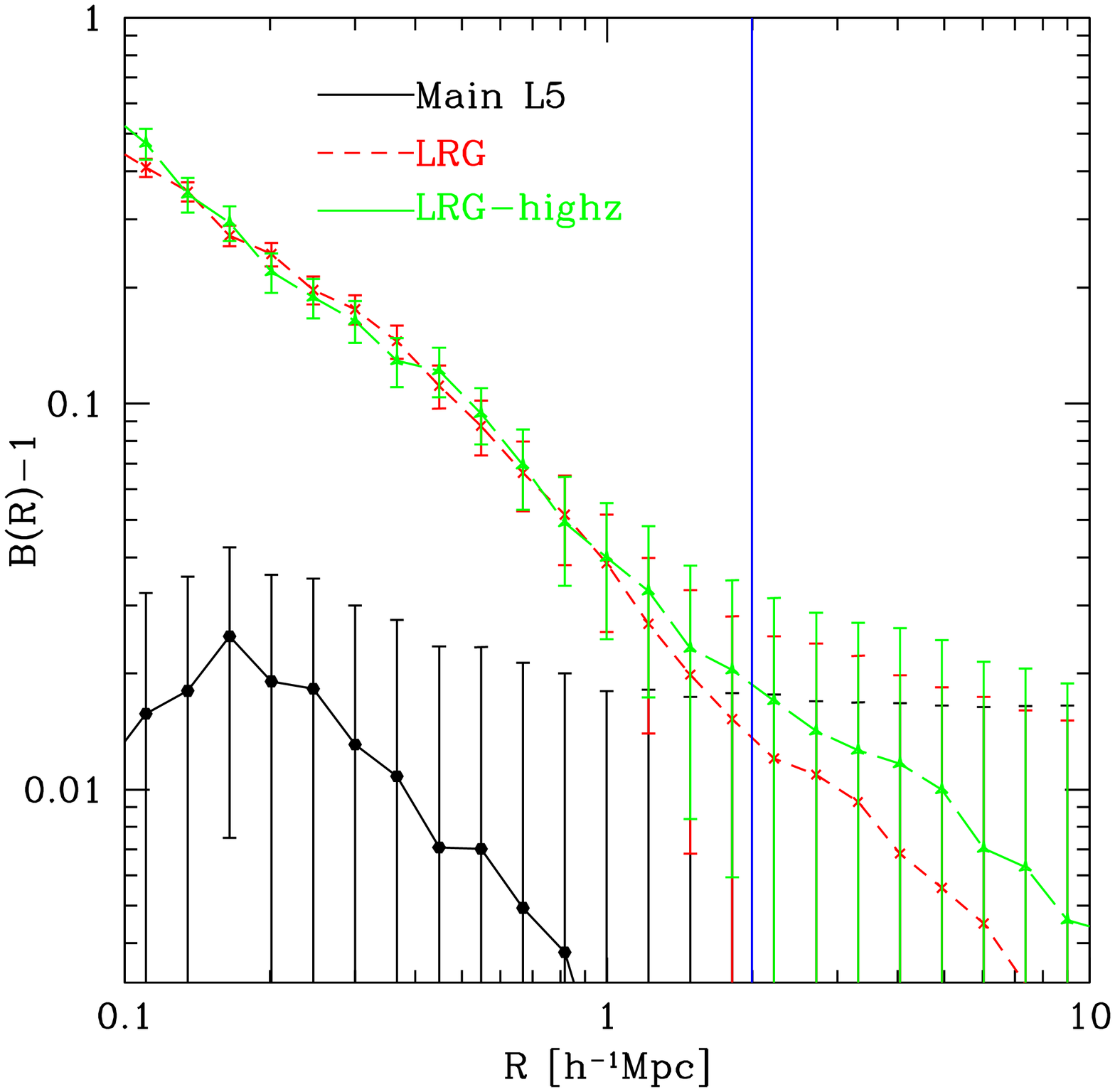}
\caption{Boost factors $B(R)-1$ as a function of separation for the three
  samples used in this analysis, as labeled on the plot.  The vertical
  solid line at $R=2h^{-1}$Mpc indicates the physical scales used for
  the cosmological constraints in this paper.\label{F:boosts}}
\end{figure}

The boost factors (Eq.~\ref{E:boostdef}) that implicitly went into those signals are shown in
Fig.~\ref{F:boosts}.  As shown, the correction is very small ($\sim$
per cent level) on the scales used for this measurement, $\gtrsim
2h^{-1}$Mpc. The size of the errorbars indicates that there are $\sim
1$ per cent-level density fluctuations in the real catalogue on large
scales,  perhaps
due to dependence of the lens number density on systematics that are not properly reproduced in the
random catalogues.  This does not bias the lensing signal, it simply
acts as a minor contribution to the statistical error budget,
subdominant to shape noise.

\subsection{Intrinsic alignments}\label{SSS:ia}

We must consider the contribution of intrinsic alignments to our
measurement.  In principle, they should only contribute if \photoz\
errors scatter galaxies that are actually physically-associated with
our lenses into the source sample, and if those physically-associated
galaxies have a tendency to align radially or tangentially with
respect to our lenses.  In principle, we expect a radial alignment for
elliptical galaxies, which should follow the linear alignment
model to some extent \citep{2004PhRvD..70f3526H}; this alignment has been
observed \citep[e.g.,][]{2007MNRAS.381.1197H} but primarily for bright red galaxies, rather than
the faint ones that we use as sources in our lensing analysis.  Disk 
galaxies are expected to align in a way that relates their disk
angular momentum to the dark matter halo angular momentum, which leads
to essentially no net radial or tangential alignment with respect to
lens galaxies \citep{2004PhRvD..70f3526H}.

Thus, our approach here is to assume that blue sources, which
contribute $\sim 70$ per cent of the weight to the lensing
measurement, contribute zero intrinsic alignment, and red sources
contribute some amount that we must empirically constrain.  A method
for carrying out this empirical constraint is presented in
\cite{2012JCAP...05..041B}; it relies on calculating shears of sources
selected in different \photoz\ bins with respect to the lenses.  That
work shows that using this method, we can constrain intrinsic
alignment contamination of the galaxy-galaxy lensing signal at
10\hmpc\ (the scale which roughly dominates our lensing constraints) to $<2$
per cent, for our LRG lens sample (which has the lowest statistical
error and therefore is the most demanding in terms of
systematics). Assuming that $\sim 30$ per cent of the source sample
has at most a 2 per cent intrinsic alignment contamination, this
translates to $<0.6$ per cent intrinsic alignment contamination of our
lensing signals.  This is far subdominant to our other assumed
systematic errors (e.g., 4-5 per cent for lensing calibration), so
we assume it is negligible for the purpose of these
cosmological constraints. 

\subsection{Ratio test}

As described in Sec.~\ref{SS:calibbias}, we apply a number of
corrections for known sources of calibration bias such as \photoz\
error.  To support the claim that we understand these effects well, we
carry out ratio tests of the signals computed using the same lens sample
and different source samples.  After correcting for known 
calibration biases that are different for each sample, these ratios
should simply be consistent with one within the errors (which are
typically $\pm 0.06$).

There are several ratio tests that we can carry out, each of which is
sensitive to our understanding of several effects.  For example, when
dividing the sample into red and blue galaxies, we are most sensitive
to differences in \photoz\ errors and to differences in the intrinsic
ellipticity distribution of the red and blue galaxy populations.
When dividing into faint versus bright galaxies, we are most sensitive
to noise rectification bias and \photoz\ errors in the former.  When
dividing into well- vs. poorly-resolved galaxies, we are most
sensitive to selection biases that couple the shear to the apparent
size.  When dividing into all sources vs. only those with
$\zphot>0.45$, we are most sensitive to \photoz\ errors in the former
(the \photoz\ errors in the latter are less important because of the
larger redshift separation between lenses and sources).  Therefore, an
ability to achieve a ratio of one in each of these four cases would
suggest that we understand the different systematic errors that are
affecting the lensing signals in each case well enough to constrain
cosmology at the $\sim 5$ per cent level.

Our findings, using the LRG lens sample, is that in three cases the
ratio is within $1\sigma$ of one.  In the final case, the split into
red versus blue sources, the ratio is $0.92\pm 0.06$; thus the
discrepancy is only a little more than $1\sigma$, and since we expect
such a case to turn up after doing several ratio tests, we conclude
that the ratio tests do not indicate any major misunderstanding of
the predominant systematic errors affecting the lensing calibration. 

\subsection{Large-scale systematic shear}\label{SSS:largescale}

In this section, we present tests of the large-scale shear
systematics.  As noted in \cite{2005MNRAS.361.1287M}, the presence of
a coherent PSF ellipticity along the scan direction in the SDSS data
can result in a non-zero tangential shear on large scales, where
lens-source pairs may be lost due to survey edge effects.  There, we
corrected for this effect by subtracting off the g-g lensing
signal around random points, but also noted that if the systematic
shear correlates in some way with the lens number density, this
correction may not be sufficient.  Thus, we must test the accuracy of
this procedure when using large-scale lensing signals to constrain cosmology.

\begin{figure}
\includegraphics[width=\columnwidth]{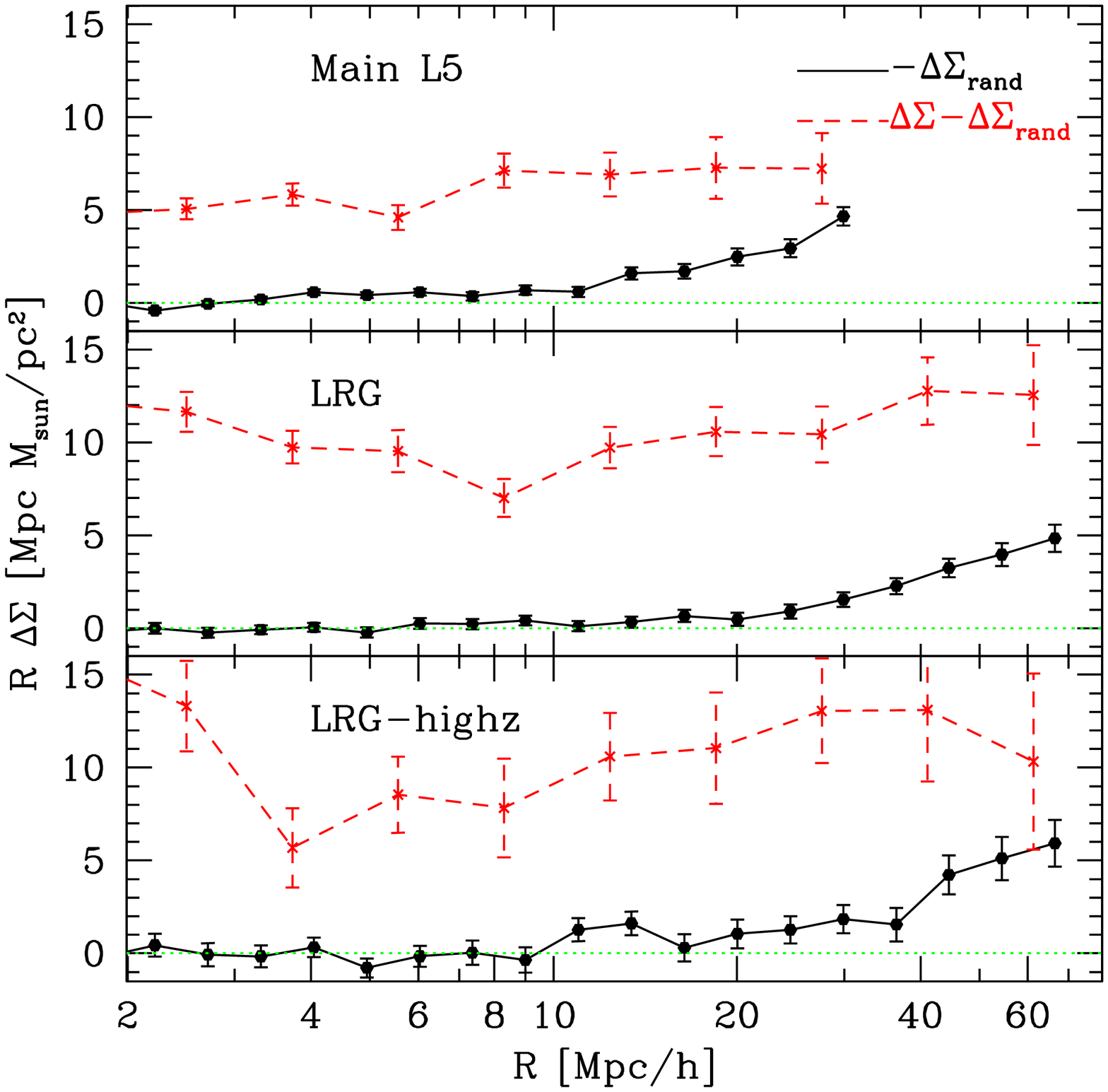}
\caption{The black points and lines show $-R \dsrand$, a measure of
  systematic shear in the source catalogue.  Each panel is a
  different sample (Main L5, LRG, LRG-highz from top to bottom).  For
  reference, the actual signal that we use for science is also shown
  as the red dashed lines.\label{F:sysshear}}
\end{figure}

In Fig.~\ref{F:sysshear}, we show the g-g lensing signal around
random points for our three lens samples.  As shown, it becomes
significantly 
inconsistent with zero for scales above $10$, $20$, and $15h^{-1}$Mpc 
for Main-L5, LRG, and high-$z$ LRG,
respectively.  The reason for the different scales is that it is a
systematic associated with an angular scale, which becomes different
physical scales when we convert angular separation $\theta$ to $R$ at
the different typical lens redshifts.  However, its magnitude also 
depends on the apparent magnitude and resolution factors
of sources used, as we have confirmed explicitly by dividing our
source sample by source properties\footnote{This result is expected,
  because systematic errors in determination of the shear generically
  depend on both the $S/N$ and apparent size of the source compared to
  the PSF; e.g., \cite{2010MNRAS.405.2044B} and \cite{2012MNRAS.423.3163K}.}.  These two effects go in competing directions,
driving the typical scale of the systematic error to increasing $R$ as
$z$ increases, then at a certain point above $z\sim 0.3$, the typical
scale starts to decrease again.  For reference, this figure also
shows the observed signal around real lenses in the same form ($R\Delta\Sigma$), to show at which
point the systematic correction is comparable in size.  Generally, on
the maximum scale used for science, the systematic correction ranges
from $1/3$ to $1/2$ of the real signal, and is $1$--$2$ times larger
than the statistical error.  

Given the significance of this systematics signal on the largest
scales used for our measurements, we must assess whether the
assumptions behind the correction for it are correct
(c.f. Sec.~\ref{SS:lensingsys}).  In order to do so, we rely on the
following facts: First, if there is some coherent systematic shear,
then depending on the survey geometry, it can also show up as a signal
in the other shear component $\gamma_\times$ on large scales.  Second,
there is no gravitational lensing signal in that shear component.
Thus, we measure $\dsx$ around real lenses and
$\dsxrand$ around random points, and check to see
whether they are consistent.  An inconsistency would call into
question the validity 
of this correction using random points.

\begin{figure}
\includegraphics[width=\columnwidth]{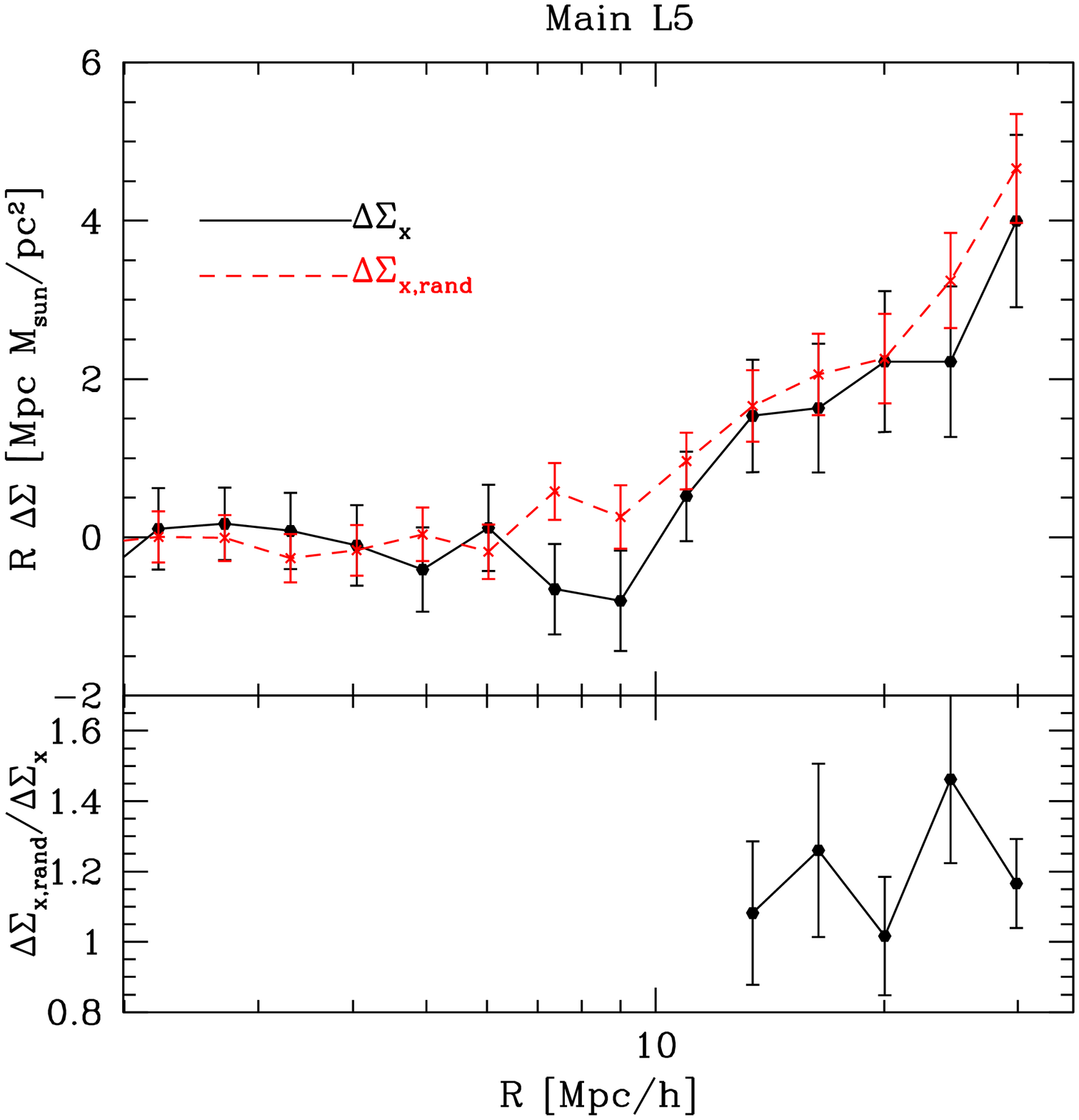}
\caption{{\em Top:} The black solid and red dashed lines show $R\dsx$
  and $R\dsxrand$, respectively, for the Main L5 sample.  {\em
    Bottom:} For the scales that have a significant systematic shear
  signal, we show the ratio $\dsxrand/\dsx$.\label{F:sysshear45}}
\end{figure}

As shown in Fig.~\ref{F:sysshear45} for
one of the lens samples, we find that there is a significant signal in the $\times$
shear component on the same scales as for the $+$ component, and
$\dsxrand \approx 1.25 \dsx$ (actually it is
$\sim 1.2\dsx$ on this Figure; $1.25$ is the result if we
average over all the lens samples to reduce the noise).  Taken at face
value, this suggests that the systematic shear correlates with some
factor that determines the effective lens number density, and that correlation
is {\em not} properly taken into account by our procedure for
producing random catalogues.  This finding is also consistent with the
results in Sec.~\ref{SSS:physassoc}.  There is likely a simple
physical explanation for this issue related to the data processing; for
example, very slight correlations of lens selection probability with the
PSF FWHM, extinction, or sky level should also correlate with
fluctuations in the systematic shear in the sources.

We consider the implications for the LRG sample, as an example.  That
sample has $\ds\sim 0.30\pm 0.03$ and
$\ds_\mathrm{rand}\sim -0.10$  at 50\hmpc.  If we assume that
$\ds_\mathrm{rand}$ is overestimating the true systematic shear by a
factor of 1.25, then that means we should have been subtracting
$-0.08$, not $-0.10$, from the original signal to get our final
$\ds$.  Thus, our original $\ds$ should have been $0.28\pm 0.03$, a
shift of $(2/3)\sigma$.  The size of this correction compared to the
statistical error suggests that we should impose the correction by
simply dividing $\ds_\mathrm{rand}$ by 1.25 before subtracting it to
get our estimate of the lensing signal.  Indeed, we applied
corrections in this way to all signals used 
for science and shown in all  plots in this paper.

As an additional test, we computed the signal while excluding all LRGs
within $60$\hmpc\ of a survey edge.  The resulting systematic shear
signal was consistent with zero, and the value of \ds\ was unchanged
on average on small scales, and decreased at the expected level on the
largest scales.  However, we merely present this test to validate our
correction scheme; we do not use the signal with survey edges removed
for science, because the statistical errors on a few \hmpc\ scales
increase significantly (20 per cent).     

Finally, we note that a non-negligible fraction of this systematic
shear signal ($\sim 30$--$40$ per cent) is due to an error in the SDSS
PSF
model identified in R12, which resulted in all PSF models in the $r$ band
in one of the camera columns having some spurious ellipticity.  The
remainder is due to the inadequacy of the adopted PSF correction
method at removing the PSF ellipticity from the galaxy shapes.

\section{Galaxy clustering systematics tests}\label{SS:resclustsys}

\subsection{Calculation of $\Upsilon_{\rmg\rmg}$}\label{SSS:calcupsgg}

Since we use \upsgg\ for our cosmological parameter constraints, and
it is a derived quantity that relies on determination of
$\wgg(R_0)$, here we present tests illustrating the accuracy of that
determination.  

\begin{figure}
\includegraphics[width=\columnwidth]{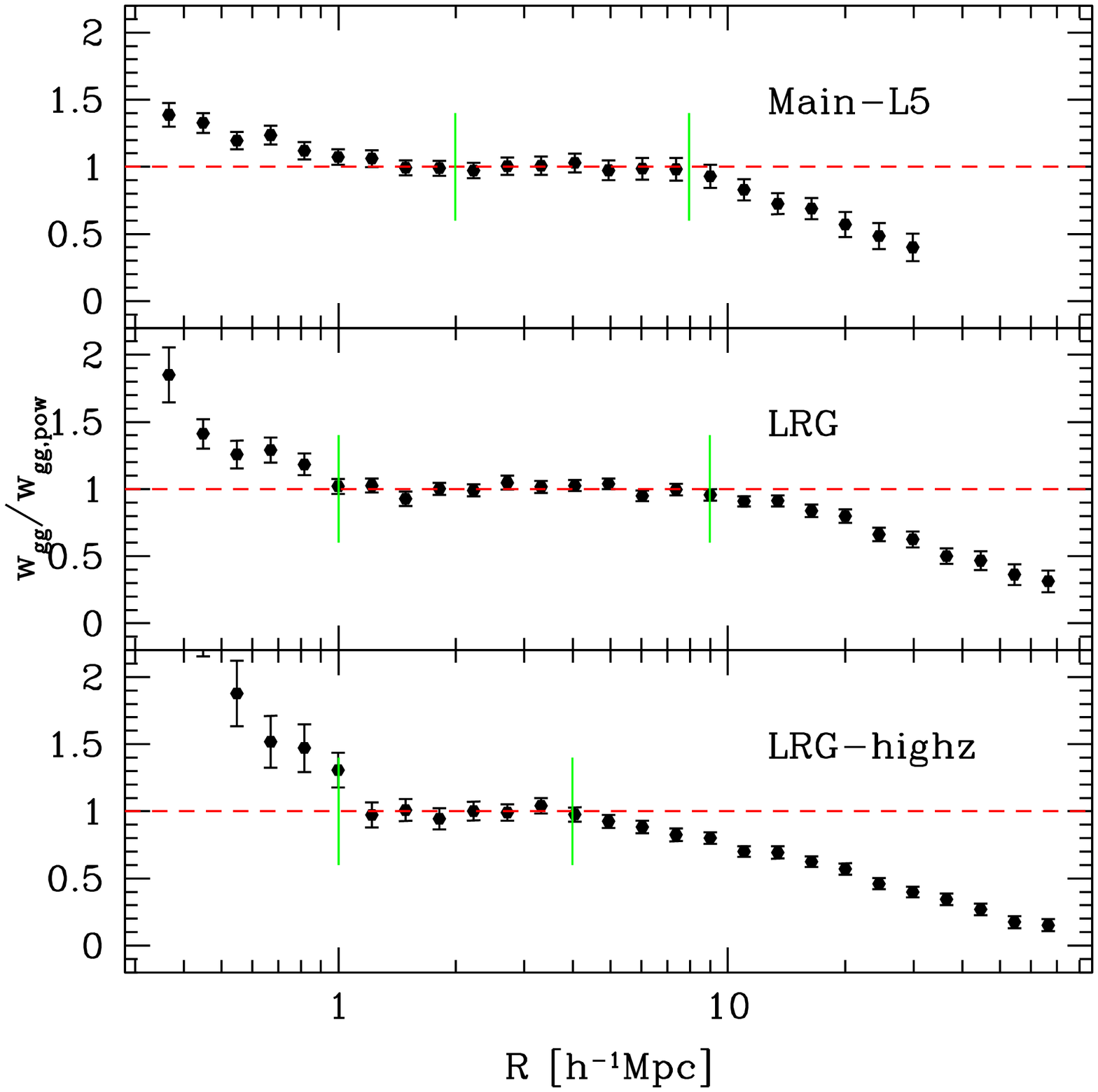}
\caption{Clustering signals for the three samples (as labelled on the
  plot) divided by the best-fit power-law using the range of scales
  indicated by vertical solid lines.  The horizontal dashed line shows
  the ideal value $\wgg/w_\mathrm{gg,pow}=1$.  \label{F:w0}}
\end{figure}

As for the lensing signal \ds, we determine $\wgg (R_0)$ using power-law
fits over a range of scales including $R_0$ on which \wgg\ appears
consistent with a power-law.  In Fig.~\ref{F:w0}, we show (for all
three samples) the observed $\wgg(R)$ divided by the best-fit power-law
for the chosen range of scales which are indicated by vertical lines.  This power-law is determined from a
fit to the jackknife mean $\wgg$, weighted by the inverse variance
(assumed to be diagonal on these scales).  Ideally, this ratio of
observed signal to best-fit power law would be consistent
with one between the vertical lines.   It is clear that the power-law
approximation is valid on the range of scales used, but the observed
signal deviates from it strongly outside of that range, which is what
we expect for $\Lambda$CDM and typical models of scale-dependent bias.

The best-fit power-laws and ranges of scales used are defined as $\wgg
= w_0 (R/R_0)^{\beta}$ for $R_\mathrm{pow,min}<R<R_\mathrm{pow,max}$, where $w_0$ is in units of
\hmpc.  Given this definition, the power-laws that
went into Fig.~\ref{F:w0} are given in Table~\ref{T:wpowlaw}.
\begin{table}
\begin{tabular}{ccccc}
\hline
\hline
Sample & $R_\mathrm{pow,min}$ & $R_\mathrm{pow,max}$ & $w_0$ &
$\beta$ \\
 & $[$\hmpc$]$ & $[$\hmpc$]$ & $[$\hmpc$]$ & \\
\hline
Main-L5 & 2.0 & 8.0 & $35.4$ & $-0.71$ \\
LRG & 1.0 & 9.0 & $75.7$ & $-0.83$ \\
LRG-highz & 1.0 & 4.0 & $88.0$ & $-0.64$ \\
\hline\hline
\end{tabular}
\caption{Best-fitting power-law functions (as defined in Sec.~\ref{SSS:calcupsgg})
  to $\wgg(R)$ for a limited range of scales, used to estimate
  $\wgg(R_0)$ and therefore \upsgg.\label{T:wpowlaw}}
\end{table}

\end{document}